\documentclass[journal,twoside]{IEEEtran}

\usepackage[]{amsmath}
\usepackage{amsfonts,amssymb,amsthm}
\usepackage{mathtools}
\usepackage{color}
\usepackage{xcolor}
\usepackage{colortbl}
\usepackage{graphics,graphicx}
\usepackage{algorithm,algpseudocode}
\usepackage{multicol,multirow,booktabs}
\usepackage{cite}
\usepackage{array}
\usepackage{float,floatpag}
\usepackage{enumerate}
\usepackage{caption}
\usepackage{subfig}
\usepackage{listings}
\usepackage{pifont,textcomp}
\usepackage{tabularx,longtable}
\usepackage{makecell}
\usepackage{cuted}
\usepackage{balance}

\stripsep.sp

\theoremstyle{definition}

\newtheorem{proposition}{Proposition}
\newtheorem{remark}{Remark}
\newtheorem{lemma}{Lemma}

\makeatletter
\renewenvironment{proof}[1][\proofname]{
\par\pushQED{\qed}\normalfont
\topsep6\p@\@plus6\p@\relax
\trivlist\item[\hskip\labelsep\bfseries#1\@addpunct{.}]
\ignorespaces
}{\popQED\endtrivlist\@endpefalse}
\makeatother

\newcounter{tempcount}

\definecolor{thegreen}{rgb}{0.00,0.60,0.30}
\definecolor{theblue} {rgb}{0.20,0.45,0.75}
\definecolor{thered}  {rgb}{0.90,0.30,0.30}

\usepackage{hyperref}
\hypersetup{colorlinks=true,urlcolor=theblue, citecolor=theblue, linkcolor=theblue}

\newcommand{\bm}[1]{\boldsymbol{#1}}

\newcommand{\tone}  [1]{\parbox[c]{28mm}{\vspace*{2.5pt}#1\vspace*{1.5pt}}}
\newcommand{\ttwo}  [1]{\parbox[c]{57mm}{\vspace*{2.5pt}#1\vspace*{1.5pt}}}
\newcommand{\tthree}[1]{\parbox[c]{28mm}{\vspace*{2.5pt}#1\vspace*{1.5pt}}}
\newcommand{\tfour} [1]{\parbox[c]{40mm}{\vspace*{2.5pt}#1\vspace*{1.5pt}}}

\newsavebox\ltmcbox
\newenvironment{longtab}
{\setbox\ltmcbox\vbox\bgroup
\csname @twocolumnfalse\endcsname
\csname col@number\endcsname\csname @ne\endcsname}
{\unskip\unpenalty\unpenalty\egroup\unvbox\ltmcbox}

\begin{document}

\title{EcoMobiFog -- Design and Dynamic Optimization of a 5G Mobile-Fog-Cloud Multi-Tier Ecosystem for the Real-Time Distributed Execution of Stream Applications}

\author{Enzo~Baccarelli, 
        Michele~Scarpiniti,~\IEEEmembership{Senio~Member,~IEEE,}
        and~Alireza~Momenzadeh,
\thanks{Manuscript received March 24, 2019; accepted April 17, 2019, date of publication April 29, 2019.}%
\thanks{Authors are with the Department of Information Engineering, Electronics and Telecommunications (DIET), ``Sapienza'' University of Rome, Via Eudossiana 18, Rome 00184, Italy, e-mail: michele.scarpiniti@uniroma1.it.}
\thanks{Digital Object Identifier 0.1109/ACCESS.2019.2913564}
}

\markboth{IEEE Access,~Vol.~7,~pp.~55565--55608,~May~2019}%
{Baccarelli \MakeLowercase{\textit{et al.}}: A 5G Mobile-Fog-Cloud multi-tier ecosystem}



\maketitle

\begin{abstract}
The emerging 5G paradigm will enable multi-radio smartphones to run high-rate stream applications. However, since current smartphones remain resource and battery-limited, the 5G era opens new challenges on how to actually support these applications. In principle, the service orchestration capability of the Fog and Cloud Computing paradigms could be an effective means of dynamically providing resource-augmentation to smartphones. Motivated by these considerations, the \textit{peculiar} focus of this paper is on the \textit{joint} and \textit{adaptive} optimization of the resource and task allocations of mobile stream applications in 5G-supported \textit{ multi-tier} Mobile-Fog-Cloud virtualized ecosystems. The objective is the \textit{minimization} of the computing-plus-network energy of the overall ecosystem under \textit{hard constraints} on the minimum streaming rate and the maximum computing-plus-networking resources. To this end, (i) we model the target ecosystem energy by explicitly accounting for the virtualized and multi-core nature of the Fog/Cloud servers; (ii) since the resulting problem is non-convex and involves both continuous and discrete variables, we develop an optimality-preserving decomposition into the cascade of a (continuous) resource allocation sub-problem and a (discrete) task-allocation sub-problem; (iii) we numerically solve the first sub-problem through a suitably designed set of gradient-based \textit{adaptive} iterations, while we approach the solution of the second sub-problem by resorting to an ad-hoc-developed \textit{elitary} Genetic algorithm. Finally, we design the main blocks of \textit{EcoMobiFog}, a technological virtualized platform for supporting the developed solver. Extensive numerical tests confirm that the energy-delay performance of the proposed solving framework is typically within a \textit{few per-cent} the benchmark one of the exhaustive search-based solution.
\end{abstract}

\begin{IEEEkeywords}
Multi-tier Mobile-Fog-Cloud ecosystems, multi-radio 5G, service models, real-time mobile stream applications, adaptive joint resource and task allocation.
\end{IEEEkeywords}


\section{Introduction}
\label{sec:introduction}

\IEEEPARstart{W}{ith} smartphones becoming our symbiotic personal assistant, high-quality mobile applications are playing an important role in our life. This is mainly due to the fact that current smartphones are more and more being equipped with an increasing number of heterogeneous sensors and wireless Network Interface Cards (NICs), that make today feasible to support multimedia mobile stream applications \cite{andrade2014}. These applications usually exploit video cameras and/or other native sensors, in order to carry out in real-time perception-based jobs, like, for example, object and/or gesture recognition and augmented-reality immersive experiences, just to cite a few. However, these applications share two main features that make them hard to be supported by current stand-alone smartphones. First, by definition, they require the continuous high-throughput processing of the data streams generated by high-data-rate sensors, in order to guarantee accuracy \cite{andrade2014}. For example, low-resolution video streams may miss/veil object poses or human gestures and, then, may give rise to a low Quality of Service (QoS). Second, the mining and/or machine learning-based algorithms used to extract from the acquired data streams useful information are typically computation intensive. Hence, since the computing and battery capacities of current smartphones are still limited at a large extent, it could be appealing to resort to the so-called Mobile Cloud Computing (MCC) paradigm and, then, offload computation-intensive tasks to remote (i.e., distant) resource-rich Cloud data centers for their execution \cite{khan2014}. However, due to the delay and throughput-sensitive features of typical mobile stream applications, this solution would increase both the network traffic to be sustained by the Mobile-Cloud backhaul network and the overall service latency \cite{khan2014}. In principle, a more performing approach could be to allow the smartphones to leverage both their native multi-radio capability and the ultra-short latencies guaranteed by the emerging Fifth Generation (5G) network technology \cite{agiwal2016}, in order to suitably allocate the offloaded application tasks over both the remote Cloud and proximate virtualized servers, generally referred to as Fog nodes \cite{mahmud2018}. An examination of Table \ref{table:01} unveils why the integration of the three pillar paradigms of Fog Computing (FC), Cloud Computing (CC) and Multi-Radio 5G (MR-5G) could improve both the energy performance of smartphones and the throughput (i.e., processing rate) performance of the supported stream applications.

\begin{table*}[htbp]
\caption{Native features and synergic interplay of the pillar FC, CC and MR-5G paradigms.}
\label{table:01}
\centering
\small
\resizebox{\textwidth}{!}{
\renewcommand{\arraystretch}{1.5}
\begin{tabular}{
>{\columncolor[HTML]{80DBFC}}p{2.1cm}
>{\columncolor[HTML]{80DBFC}}p{3.0cm}
>{\columncolor[HTML]{F5E1BE}}p{1.9cm}
>{\columncolor[HTML]{F5E1BE}}p{3.1cm}
>{\columncolor[HTML]{9FDDD3}}p{1.7cm}
>{\columncolor[HTML]{9FDDD3}}p{3.3cm}}
\toprule
\rowcolor[HTML]{7BB6F2}
\multicolumn{2}{c}{\textbf{FOG COMPUTING}}  &  \multicolumn{2}{c}{\textbf{CLOUD COMPUTING}}  &  \multicolumn{2}{c}{\textbf{MULTI-RADIO 5G}}  \\
%
\midrule
\textbf{Pervasive deployment} & Fog servers are pervasively deployed at the network edge, in order to limit the network delay &
\textbf{Centralized deployment} & Cloud datacenters sit in the backbone network and their access delays are high  & 
\textbf{Support for multi-radio technologies} & Multiple short/long-range  radios are simultaneously supported and dynamically turned ON-OFF by 5G  smartphones \\
%
\midrule
\textbf{Light virtualization} & Virtualized clones of the served smartphones are hosted by Fog servers. Container-based virtualization technologies are employed for reducing the resulting virtualization overhead & 
\textbf{Heavy virtualization} & Clones of the served devices are statically deployed by resorting to large-size (i.e., heavy) Virtual Machines  & 
\textbf{Dynamic bandwidth provisioning} & Wireless bandwidth is dynamically provided to the requiring multi-radio smartphones on a per-radio basis \\
%
\midrule
\textbf{Support for throughput-sensitive stream applications}  & Fog nodes exploit low-latency short-range links for enabling fast task offloading from smartphones & 
\textbf{Support for delay-tolerant applications} & Being natively equipped with a large number of powerful servers, Cloud data centers
may execute computing-intensive (but delay-tolerant) tasks offloaded by remote devices 
 & 
\textbf{Bandwidth aggregation} & The simultaneous utilization of multiple radios allows the aggregation of the wireless bandwidth through bandwidth pooling  \\
\midrule
\textbf{Energy saving} & Resource-limited smartphones may save energy by leveraging proximate Fog servers as computing clones & 
\textbf{Energy wasting} & The access to remote Clouds by Mobile devices requires the utilization of energy-wasting multi-hop cellular links & 
\textbf{Ultra-low access delay} & Sub-millisecond access delays are achieved by the synergic utilization of 5G-enabled multiple radios \\
\bottomrule
\end{tabular}
}
\renewcommand{\arraystretch}{1.0}
\end{table*}

Fog Computing is a quite novel computing paradigm \cite{mahmud2018}. By definition, it enables pervasive \textit{local} access to virtualized small-size pools of computing resources that can be \textit{quickly} provisioned, \textit{dynamically} scaled up/down and released on an on-demand basis. Proximate resource-limited mobile devices may access these resources by establishing \textit{single-hop} communication links. The first column of Table \ref{table:01} points out the native features of the FC paradigm.

Somewhat complementary features are retained by the (more traditional) Cloud Computing paradigm (see the second column of Table \ref{table:01}). In fact, by definition, the CC paradigm enables \textit{ubiquitous} access to \textit{large-size} pools of virtualized computing resources by establishing (typically) multi-hop cellular-type communication paths. Resource provisioning/releasing entails no negligible bootstrapping delays, and resource scaling embraces latencies of tens of milliseconds. Hence, offloading of computing-intensive but delay-tolerant and communication-light tasks well matches the native feature of the CC paradigm.

Thanks to its ultra-short latencies and support of multi-radio terminals, the forthcoming 5G paradigm is expected to be an ideal ``glue'' for enabling the synergic integration of the Fog and Cloud paradigms (see the third column of Table \ref{table:01}). In fact, by design, 5G provides a multi-radio network platform that hosts existing 2G, 3G and 4G cellular technologies. It is envisioned that 5G may also integrate other short/long-range communication technologies (like, for example, WiFi, mobile satellite system, digital video broadcasting) by resorting to multi-tier spatial coverage based on the overlay of macro, pico, femto and other types of cells \cite{agiwal2016}.

\subsection{Why the convergence of Fog-Cloud-5G? Some motivating use cases}
\label{ssec:WhyFogCloud5G}

In order to appreciate the potential impact of the synergic integration of the three pillar paradigms of Fog Computing, Cloud Computing and Multi-Radio 5G, let us consider the general mobile operative scenario in which a user equipped with a smartphone desires to process a stream of frames of a given application. This last is composed of a number of inter-connected tasks (i.e., sub-routines, methods or threads) and it is described by the corresponding application Directed Acyclic Graph (DAG) \cite{andrade2014}. Since the smartphone is energy limited and equipped with limited computing resources, the corresponding operative system may decide to execute each task of the current frame locally or offload it to a connected Fog or Cloud node by leveraging the 5G-enabled multi-radio capability of the smartphone.

In the sequel, we shortly review (few) emerging use cases that fit the aforementioned general scenario, so to illustrate the supporting role played by the underlying Fog-Cloud-5G integrated system (see, for example, \cite{rahmani2018} for a detailed presentation of a spectrum of Fog-supported use cases).

\medskip\noindent\textit{Object recognition applications} -- Let us consider a mobile user who desires to quickly detect the presence/absence of a specific object from the real-time video stream captured by the camera of his/her smartphone. Since the underlying object recognition algorithm must operate on a per-frame basis, it may be too complex to be fully executed by the smartphone during an inter-frame interval. Therefore, the operative system hosted by the smartphone splits the overall algorithm into three main components, namely the object-detection, feature-extraction and object-recognition components. Afterwards, during each inter-frame interval, the first component may be executed locally by the smartphone, while the second and third ones may be executed by a proximate Fog server and a remote Cloud server, respectively. The required Mobile-Fog-Cloud data exchange is supported by the underlying WiFi and Cellular parallel connections managed by the multi-radio interfaces that equip the smartphone.

\medskip\noindent\textit{Augmented reality and immersive mobile applications} -- The development of near-to-eye display technologies (like, for example, Google Glasses) is opening the doors to new types of immersive applications that exploit the so-called Augmented Reality (AR) paradigm. Just as an example, let us consider a museum, where a network of local Fog servers is strategically deployed along the visiting tour. In this scenario, beside to listen explanations through headphones, visitors may be guided in real-time through a stream of visual annotations by exploiting the WiFi connections sustained by the local Fog servers. So doing, a stream of scenes can come to life right before the visitors' eyes immersing them in ancient history. Furthermore, specific queries by the visitors may be addressed by streaming the required information from a central archive hosted by a (possibly, distant) Cloud server.

\medskip\noindent\textit{Smart shopping centers} -- Let us consider a multi-floor shopping center where a number of local Fog servers collectively forms an integrated multimedia database about the offered products. In this scenario, Fog servers at different floors store floor-related information that, in turn, is periodically updated by a central (possibly, remote) Cloud server. By exploiting WiFi connections, the Fog servers can collectively stream radio-navigation services to smartphone-equipped mobile users, in order to interactively guide them through the mall and annotate in real-time all visited products on their shopping lists.

\medskip
The common feature of all these real-world applications is that they require the real-time stream execution of similar programs, like, for example, radio-positioning programs, object recognition programs, and 3D visual rendering programs, just to name a few. Although these programs are already available at a large extent \cite{andrade2014}, their computational complexities are typically high, so that current smartphones are not still capable of supporting their complete execution in a standing-alone way \cite{rahmani2018}.

\subsection{The considered multi-tier multi-radio ecosystem}
\label{ssec:ConsideredMulti-tier}

\begin{figure*}[htb]
\centering
\includegraphics[width=0.9\textwidth]{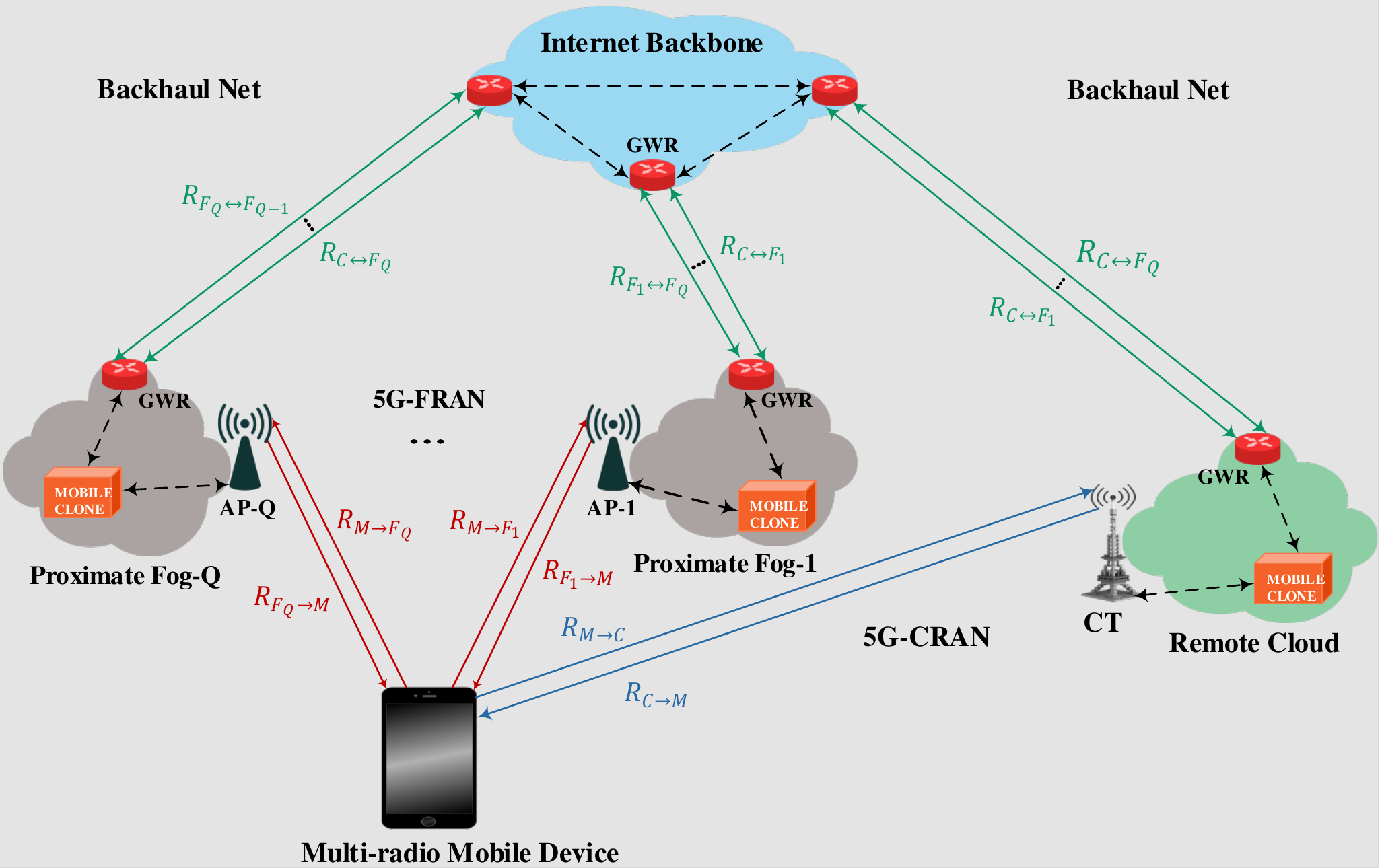}
\caption{The considered 5G-supported Mobile-Fog-Cloud multi-tier ecosystem. Single (resp. double)-arrow paths indicate one-way (resp. two-way) TCP/IP connections. AP:= Access Point; GWR:= GateWay Router; CT:= Cellular Tower; FRAN:= Fog Radio Access Network; CRAN:= Cloud Radio Access Network.}
\label{fig:multi-tier}
\end{figure*}

Motivated by this consideration, in Fig.\ \ref{fig:multi-tier}, we sketch the main building blocks of the considered networked multi-tier multi-radio virtualized ecosystem for the support of task offloading from a resource-limited Mobile device. The ecosystem is composed by a Mobile device (i.e., a smartphone), a number $ Q \geq 1 $ of proximate Fog nodes and a remote Cloud node. A 5G-based FRAN (resp., CRAN) supports the Mobile-Fog (resp., Mobile-Cloud) wireless up/down single-hop TCP/IP connections, while a (possibly wired and/or multi-hop) Backhaul network guarantees the inter-Fog and Cloud-Fog TCP/IP connectivity.

In the considered framework of Fig.\ \ref{fig:multi-tier}, the Mobile device may be equipped with multiple wireless NICs, in order to process in parallel multiple transmit-receive wireless streams. For this purpose, it is assumed that the Transport-layer of the protocol stack at the Mobile device hosts the Multi-Path TCP, i.e., MPTCP (see, for example, the contributions in \cite{peng2016,baccarelli2018}, and references therein for extensive performance analysis of MPTCP and related implementation aspects).

Virtualization is employed in 5G-supported Fog/Cloud data centers, in order to \cite{rahmani2018}: (i) dynamically multiplex the available physical computing, storage and networking resources over the spectrum of the served mobile devices; (ii) provide homogeneous interfaces atop (possibly) heterogeneous 5G mobile devices; and, (iii) isolate the applications running atop a same physical server, so as to provide trustworthiness. Hence, according to the considered virtualized environment, the Fog and Cloud nodes of Fig.\ \ref{fig:multi-tier} are equipped with (software) clones of the Mobile device. Each clone acts as a (virtual) multi-core ``server'' processor and provides resource augmentation to the ``client'' Mobile device by processing workload on behalf of it. For this purpose, each clone is run by a container that is instantiated atop the host computing node \cite{pahl2017}. So doing, the clone is capable of exploiting (through resource multiplexing) a slice of the physical computing and network resources of the host computing node.

\begin{figure*}[htb]
\centering
\subfloat[]
{\label{fig:VirtNodeArchitec-a}
\includegraphics[width=.47\textwidth]{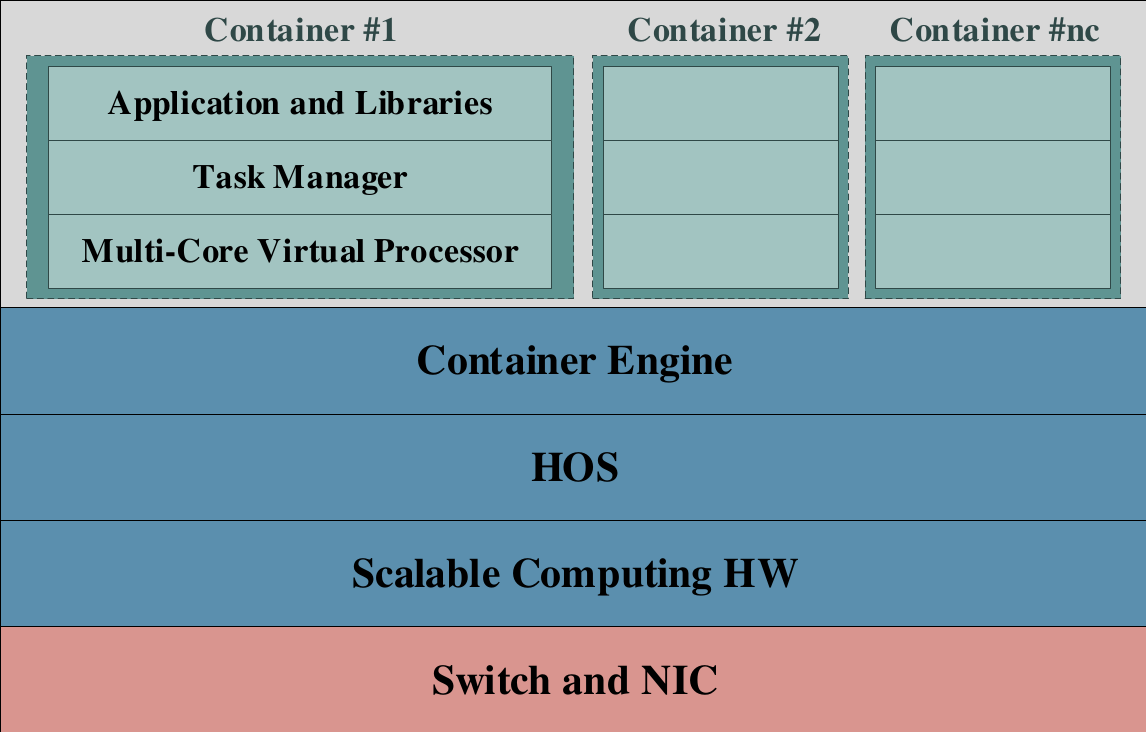}}
\hspace{1em}
\subfloat[]
{\label{fig:VirtNodeArchitec-b}
\includegraphics[width=.41\textwidth]{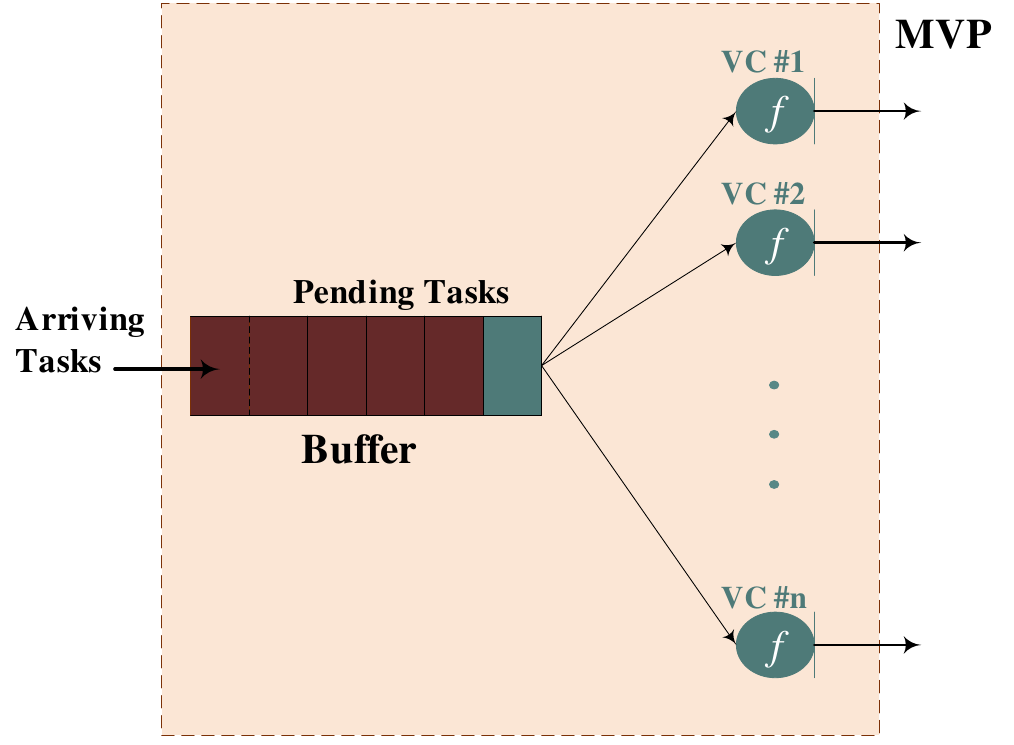}}	
\caption{Container-based virtualization of a physical server equipping the Mobile, Fog, and Cloud nodes. (a) Virtualized server architecture; (b) Architecture of a multi-core virtual processor. HW:= CPU HardWare; NIC:= Network Interface Card; HOS:= Host Operating System; MVP:= Multi-core Virtual Processor; VC:= Virtual Core; n:= Number of virtual cores; f:= Per-core processing frequency.}
\label{fig:VirtNodeArchitec}
\end{figure*}

Fig.\ \ref{fig:VirtNodeArchitec} reports the basic elements of the container-based virtualized architecture that equips the Mobile device and each computing node of Fig.\ \ref{fig:multi-tier}.

Specifically, according to Fig.\ \ref{fig:VirtNodeArchitec-a}, each server at the Mobile, Fog and Cloud nodes hosts a number $ nc \geq 1 $ of containers. All the containers hosted by the same physical server share: (i) the server's Host Operating System (HOS); and, (ii) the pool of computing (i.e., CPU cycles) and networking (i.e., I/O bandwidth) physical resources done available by the CPU and NICs that equip the host server. Job of the Container Engine of Fig.\ \ref{fig:VirtNodeArchitec-a} is to dynamically allocate to the requiring containers the bandwidth and computing resources done available by the host server. In order to execute the allocated workload on behalf of the Mobile device, each container is equipped with a Multi-core Virtual Processor (MVP). This last comprises (see Fig.\ \ref{fig:VirtNodeArchitec-b}): (i) a buffer that stores the currently offloaded application tasks; and, (ii) a number $ n \geq 1 $ of (typically, homogeneous) Virtual Cores (VCs), that run at the processing frequency $ f $ dictated by the Container Engine. Therefore, goal of the Task Manager of Fig.\ \ref{fig:VirtNodeArchitec-a} is to allocate the pending application tasks over the set of virtual cores of Fig.\ \ref{fig:VirtNodeArchitec-b}. This is done according to the actually implemented service discipline (see Section \ref{sec:ServiceScheduling}).

\subsection{Main contributions and roadmap of the paper}
\label{ssec:MainContributions}

On the basis of an overview of the related work carried out in Section \ref{sec:RelatedWork}, we anticipate that the main contributions of our paper may be summarized as follows:
\begin{enumerate}[1.]
\item we carefully model both the computing and networking energy of the multi-tier ecosystem of Fig.\ \ref{fig:multi-tier} by explicitly accounting for its virtualized multi-core and multi-radio features;
\item we develop a solving approach for the delay-constrained minimization of the overall computing-plus-networking energy consumed by a stream application by performing task offloading and allocation of the per-core computing frequencies and per-connection network throughput of the ecosystem of Fig.\ \ref{fig:multi-tier} in a \textit{joint} and \textit{adaptive} way. Interestingly enough, the developed solving approach allows us to account for: (i) the minimum required application throughput (i.e., the minimum rate at which the stream application must be executed); (ii) the task service and scheduling disciplines actually implemented by the computing nodes of Fig.\ \ref{fig:multi-tier}; (iii) the maximum allowed per-connection network throughput and per-core processing frequencies; and, (iv) the specific service model enforced by the Service Provider who manages the platform of Fig.\ \ref{fig:multi-tier}. For this purpose, the proposed solving approach suitably combines gradient-based adaptive iterations with a Genetic-based elitary meta-heuristic, in order to simultaneously attain adaptive resource and task allocation;
\item we design the main building blocks and define the supported services of \textit{EcoMobiFog}, i.e., the proposed virtualized technological platform for the actual support of the developed solving framework; and, finally,
\item we carry out extensive numerical tests for the evaluation and comparison of the energy-vs.-delay performance of the designed solving framework under a number of operative scenarios and application DAGs. In particular, (i) we compare the performance-vs.-computational complexity trade-off of the proposed solver with respect to the corresponding ones of five benchmark solvers, namely the \textit{Only-Task Allocation}, \textit{Only-Fog}, \textit{Only-Mobile}, \textit{Only-Cloud} and \textit{Exhaustive-Search} solvers; and, (ii) we numerically test the sensitivity of the energy-delay performance of designed solver on two pillar service models, namely the \textit{Eco-centric} and the \textit{Mobile-centric} service models. All the reported numerical results have been carried out by the recently developed \textit{VirtFogSim} toolbox\footnote{Available online at: \url{https://github.com/mscarpiniti/VirtFogSim}}.
\end{enumerate}

The roadmap of the remaining part of the paper is as follows. After reviewing the related work in Section \ref{sec:RelatedWork}, Section \ref{sec:modeling} formally introduces the main features of DAGs for mobile stream applications, while Sections \ref{sec:ServiceScheduling} and \ref{sec:ModelingEnergyConsumption} are devoted to formally characterize the service/scheduling disciplines at the computing nodes and the models of the computing and network energy, respectively. Section \ref{sec:ConsideredJOP} introduces the afforded \textit{Joint Optimization Problem} \textit{(JOP)}, as well as its decomposition in the cascade of a \textit{Resource Allocation Problem} \textit{(RAP)} and a \textit{Task Allocation Problem} \textit{(TAP)}. Afterwards, Sections \ref{sec:RAP} and \ref{sec:TAP} present the proposed solving approaches of the \textit{RAP} and \textit{TAP}, together with the analysis of the associated computational complexities. In Section \ref{sec:Technologicalplatform}, we detail the architecture of \textit{EcoMobiFog}, i.e., the proposed technological platform for the actual support of the developed \textit{JOP} solver. Afterwards, in Section \ref{sec:Performancetests}, we numerically test and compare the actual energy-vs.-delay performance of the proposed solving framework under a number of application scenarios and benchmark DAGs. Conclusive Section \ref{sec:Conclusion} recaps the main results of our work and provides some hints for future research. Appendix \ref{appendix:A} reports the main taxonomy of the paper, together with the meanings/roles of the main used symbols/parameters, their measuring units and simulated values. Final Appendixes \ref{appendix:B}, \ref{appendix:C}, \ref{appendix:D}, and \ref{appendix:E} present the analytical proofs of the main formal results of the paper.

Regarding the adopted notation, we point out that the arrowed subscript: $ \vec{x} $ indicates a row vector, $ \left|\mathcal{V} \right| $ is the size (i.e., the cardinality) of the set $ \mathcal{V} $, $ \left\lceil . \right\rceil $ (resp., $ \left\lfloor . \right\rfloor $) is the ceil (resp. floor) function, while $ \left[ M \right] \overset{\mathrm{def}}{=} \left[ m_{i,j} \right]_{i,j=1}^{L} $ denotes an $ \left(L \times L \right) $ matrix, whose $ \left( i,j \right) $-th element is $ m_{i,j} $. Furthermore, the symbol $ u_{-1} \left(y\right) $ indicates the unit-step Heaviside function (i.e., $ u_{-1}\left(y\right) \overset{\mathrm{def}}{=} 0 $ for $ y \leq 0 $, and $ u_{-1}\left(y\right) \overset{\mathrm{def}}{=} 1 $, otherwise), while $ \delta \left(y\right) $ is the Kronecker's delta function (i.e., $ \delta \left(y\right) \overset{\mathrm{def}}{=} 1 $ for $ y=0 $, and $ \delta \left(y\right) \overset{\mathrm{def}}{=} 0 $, otherwise)

Finally, formal assumptions are marked by bullets.

\section{Related work}
\label{sec:RelatedWork}

An overview of the large body of literature related to the broad topic of MCC points out that Mobile Edge Computing (MEC) is another computing model that is sometimes (mis)understood as a synonymous of Fog Computing \cite{garcia2015,satyanarayanan2015,yousefpour2019}. In this paper, we distinguish these two paradigms. The main reason is that, in the MEC paradigm, proximate network nodes are exploited for \textit{only} providing resource augmentation to Mobile devices by exploiting single-hop connections. As a consequence, the resulting MEC computing infrastructure is inherently composed by \textit{only} two tiers of entities, i.e., the ``client'' Mobile devices and the ``server'' edge nodes. In contrast, Fog Computing aims to harness computing across the \textit{full path} followed by the data to be processed, and this path may include \textit{multiple} (possibly, hierarchically organized) tiers of intermediate server nodes, as well as a remote Cloud data center (see Fig.\ \ref{fig:multi-tier}). Therefore, Fog infrastructures are natively composed of three or more tiers of nodes. So doing, the computational needs of mobile devices and edge nodes can be supported by cloud-like proximate resources or, alternatively, processed data can be transported from the remote cloud node to the edge of the network \cite{yousefpour2019}. 

According to this observation, we note that a first (rich) branch of research work on task placement considers two-tier MEC scenarios that involve only two computing nodes, i.e., a first node hosted by the Cloud or a proximate MEC data center, and a second node running atop the mobile device (see the recent tutorial on MEC in \cite{mao2017}). A second (substantial) research branch focuses on the so-called task migration/allocation problem, where two physical computing nodes (like, for example, a Mobile device and a cloud or MEC node) are still involved \cite{mach2017,secci2016,kwak2015}.

However, to date, less considerable work seems to be available on the core problem tackled by this paper, i.e., the dynamically optimized placement of application tasks over \textit{three} or \textit{higher order-tier} networked computing platforms. In this regard, an overview of this last set of work shows that, roughly speaking, the related on-going research is moving along three main lines, namely: (i) the optimized placement of multi-task applications in multi-tier data-centers; (ii) the design of multi-tier computing architectures for MCC and related management protocols; and, (iii) the design of task offloading and resource allocation algorithms for multi-tier mobile computing environments.

A first group of contributions in \cite{chowdhury2012,bansal:2011,dutta2012,kuo2016,lukovszki2015} focuses on the optimized placement of multi-task applications in multi-tier data-centers. In this regard, the authors of \cite{chowdhury2012} develop an algorithm for the minimization of the total cost of task placement under load balancing constraint. The proposed algorithm is based on Linear Program (LP) relaxation, its computational complexity scales linearly with the number of tasks to be allocated and does not allow resource sharing of the computational resources of the underlying physical nodes. In order to address this last point, the authors of \cite{bansal:2011} propose an algorithm for mapping application DAGs with tree topologies onto the physical graphs of networked computing nodes. The goal is still the minimization of the total cost of the performed mapping under constraints on the maximum utilization of each link of the underlying physical graph. Being the afforded problem NP-hard, a suboptimal low-complexity online version is also developed in \cite{bansal:2011} by relying on a suitable linear relaxation of the afforded problem. The paper in \cite{dutta2012} proposes an LP-based algorithm for the (dual) problem of the offline mapping of DAG paths onto data centers with tree topology. The goal is the minimization of link congestion so that load balancing of the computing nodes is not included by the adopted objective function. Furthermore, the constraints considered in \cite{dutta2012} force the DAG tasks to be only mapped into the leaves of the tree-shaped graph of the underlying data center. The contributions in \cite{kuo2016} and \cite{lukovszki2015} focus on the (quite recent) problem of the embedding of the service chains. Triggered by the emerging trend of Network Function Virtualization (NFV), the common goal of these contributions is to map a linear application DAG (i.e., a DAG with chain topology) onto the physical path joining fixed source and sink computing nodes, so that a sequential chain of operations may be performed on the data packets moving from the source node to the destination one. However, the topic of link placement optimization is not considered by these papers. Overall, like our contribution, these first set of papers consider the general problem of optimized task placement onto multi-tier networked computing infrastructures. Nevertheless, \textit{unlike} our contribution, their solutions are not adaptive.

A second group of contributions in \cite{cuervo2010,kosta2012,bahl2012,rahimi2013,cheng2015,yang2013,zhou2017} tackles with the design aspects and architectures of multi-tier computing technological platforms for the support of MCC applications. For this purpose, the authors of \cite{cuervo2010} propose a code offloading framework, i.e., \textit{MAUI}, that supports method-level energy-aware offloading for mobile applications described by DAGs. The developed framework allows to annotate methods and retrieves information from a set of profilers, in order to take decisions on whether to offload.  The remarkable feature shared by the \textit{Thinkair}, \textit{Cloudlet} and \textit{Music} frameworks in \cite{kosta2012,bahl2012,rahimi2013} is that they rely on the virtualization of the served mobile devices, in order to enable them to offload computing-intensive tasks to their clones running on distant nodes. As our contribution, all these frameworks consider virtualized multi-tier offloading technological platforms. However, \textit{unlike} our framework, all these papers subsume stable (i.e., static) operative environments, which may be over-optimistic under failure-prone networking scenarios. Later, a number of works proposes to consider other types of resources for offloading. For example, the authors of \cite{cheng2015} develop an architecture composed by wearable devices, mobile devices and cloud for code offloading. As our contribution, the goal of \cite{cheng2015} is to allow the execution of computing-intensive applications on wearable devices through task offloading towards proximate/remote server nodes. However, \textit{unlike} our contribution, in \cite{cheng2015}, the impact on the offloading performance of the (possibly, time-varying) feature of the underlying wireless connections is not considered. The focus of \cite{yang2013} is on the design of a system architecture (i.e., \textit{StreamCloud}) that supports fine-grained offloading of tasks of stream applications from a mobile device towards distant serving nodes. In order to solve the underlying decision process, this paper presents a Genetic-based meta-heuristic, that is capable to maximize the application throughput under the constraint on the available maximum wireless bandwidths. Hence, like our work, also this contribution considers the application throughput as a pivotal performance metric for stream applications and resorts to the Genetic paradigm as solving approach. However, \textit{unlike} our work, the paper in \cite{yang2013}: (i) does not perform dynamic optimization of the network and/or computing resources; and, (ii) does not consider the network and/or computing energy consumption as target objectives to be minimized. These aspects are, indeed, addressed at some extent by the so-called \textit{mCloud} framework recently proposed in \cite{zhou2017}. Specifically, the authors of this last contribution develop a technological platform for task offloading from a mobile device to remote Clouds and/or nearby Fog nodes. The target is the minimization of the task execution times by leveraging context-awareness, in order to dynamically select the most energy-saving wireless connection over the ones simultaneously managed by the Mobile device. Hence, like our work, the resulting offloading framework of \cite{zhou2017} accounts for: (i) the presence of a multi-tier networked computing infrastructure, that is capable to provide resource augmentation to resource-limited mobile devices; and, (ii) the time-varying and heterogeneous power-vs.-delay profiles of the wireless connections managed by the mobile device. However, \textit{unlike} our contribution, the \textit{mCloud} framework: (i) does not perform dynamic scaling of the computing resources available at the Mobile and Cloud/Fog nodes; (ii) the tasks to be offloaded are considered mutually independent, i.e., no precedence constraints are assumed to be enforced by the underlying application DAG; (iii) the impact of the service discipline at the computing nodes is not modeled; and, (iv) all processing nodes are assumed single-core.

A third set of contributions in \cite{chen2013,hung2015,lin2013,jia2014,kao2015,mahmoodi2016,zhang2015,wang2017} affords the (broad) topic of the optimized design and performance evaluation of task offloading and resource allocation algorithms for mobile multi-tier computing environments. In this regard, the authors of \cite{chen2013} develop a semi Markovian-based framework for triggering the offloading decision, that aims at attaining a good trade-off between the contrasting requirements of low DAG execution times and low energy consumption. However, the developed decision framework assumes a stable network condition, that is quite over-optimistic in mobile environments. In order to reduce the decision-delays that inherently affect the solving approaches based on the Markov Decision Process, the contribution in \cite{hung2015} resorts to profile and cache the already computed offloading planes, while \cite{lin2013} proposes a proactive approach that exploits location-awareness for performing mobility prediction. The common feature shared by the contributions in \cite{jia2014,kao2015,mahmoodi2016,zhang2015,wang2017} is that they model the underlying application as a weighted DAG, in order to account for the task dependencies and related computing/communication workloads. Specifically, \cite{jia2014} pursues the criterion of workload balancing between mobile device and distant servers, in order to develop a heuristic DAG-partitioning algorithm for the reduction of the resulting execution time. The paper in \cite{kao2015} investigates the latency of DAG executions under constraints on the available computing/communication resources, and develops a polynomial-complexity approximate solution with guaranteed performance. The goal of \cite{mahmoodi2016} is the minimization of the energy consumed by the mobile device through task offloading. For this purpose, both the scheduling and offloading decisions are jointly optimized by numerically solving a suitable integer program. In order to efficiently cope with the fading phenomena impairing the mobile channels supporting task offloading, the paper in \cite{zhang2015} performs the delay-constrained minimization of the average energy consumption of the mobile device. To this end, in \cite{zhang2015}, the afforded problem is turned into a stochastic shortest path problem, that is solved through suitable \textit{one-climb} offloading policies. Finally, the recent contribution in \cite{wang2017} develops online approximate algorithms with poly-log competitive ratios for the load-balanced mapping of an application DAG onto a networked computing graph under constraints on the link utilization.

In summary, on the basis of the carried out research overview, we may conclude that the peculiar feature of our contribution is as follows. It aims at maximizing the energy efficiency of task offloading of \textit{throughput-constrained} mobile stream applications. To this end, a \textit{dynamic} framework that leverages the \textit{virtualization} of the available computing/networking resources by \textit{jointly} optimizing resource and task allocation. This is done by accounting for: (i) an ecosystem of (possibly, heterogeneous) offloading destinations, that inter-communicate through (possibly, heterogeneous) TCP/IP 5G connections; (ii) the dynamically changing network conditions; and, (iii) the service policy actually enforced by the involved Service Providers.

\section{Modeling stream applications}
\label{sec:modeling}

Real-life applications are composed by a number of basic tasks that can exhibit arbitrary sets of inter-dependencies. In general, a suitable description of these last may be exploited by the mobile device of Fig.\ \ref{fig:multi-tier}, in order to improve the energy performance of the carried out task-offloading process.

For this purpose, application Component Dependency Graphs (also referred to as application Task-Call Graphs) may be utilized \cite{andrade2014}. From a formal point of view, an application component dependency graph is a DAG: $ \mathcal{G}_{APP} \overset{\mathrm{def}}{=} \left(\mathcal{V} , E \right) $, whose node set: $ \mathcal{V} \overset{\mathrm{def}}{=} \left\{ i: i=1,\dots,V \overset{\mathrm{def}}{=} \left|\mathcal{V} \right| \right\} $ represents the application tasks, while the set: $ E \overset{\mathrm{def}}{=} \left\{ e_{ij}, i \in \mathcal{V}, j \in \mathcal{V} \right\} $ of the directed edge captures the inter-task dependencies \cite{andrade2014}. Being a weighted graph, an application DAG is formally characterized by \cite{andrade2014}:
\begin{enumerate}[i.]
	\item the binary-valued matrix: $ \left[ A \right]  \overset{\mathrm{def}}{=}  \left[ a_{ij} \right]_{i,j=1}^{V} $ of the pair-wise task adjacencies;
	\item the real-valued matrix: $ \left[ D_A \right]  \overset{\mathrm{def}}{=}  \left[ d_{ij} \right]_{i,j=1}^{V} $ of the edge weights, with $ d_{ij} $ being the weight (measured in $ \mathrm{(bit)} $) of the $ \left(i,j\right) $-th edge $ e_{ij} \in E $; and,
	\item the vector: $ \vec{s} \overset{\mathrm{def}}{=}  \left[ s_1 , \dots , s_V \right] $ of the task sizes, with $ s_i $ (measured in $ \mathrm{(bit)} $) being the workload to be sustained for the execution of the $ i $-th task of the DAG.
\end{enumerate}

Furthermore, according to, for instance, \cite{mao2017,mach2017}, in the sequel, we assume that every application DAG describing a mobile application retains the following six defining properties:

\begin{itemize}
	\item each node $ i \in \mathcal{V} $, with $ i \neq 1 $ and $ i \neq V $, has an in-degree and an out-degree of (at least) one. The in-degree of the first node vanishes, while the out-degree of the last node is zero;
	\item each intermediate node $ i \in \mathcal{V} $, with $ i \neq 1 $ and $ i \neq V $, has at least one directed path from the first task and at least one directed path to the last task, so that the first and last tasks are the root and the sink of the considered application DAG, respectively;
	\item the DAG is loop-free, so that each root-to-sink directed path is of finite length;
	\item each task must be processed by one and only one computing node of the ecosystem of Fig.\ \ref{fig:multi-tier};
	\item both the first and last nodes cannot be offloaded, and, then, must be executed by the Mobile device;
	\item a clone of the Mobile device is already deployed at the Cloud node and at each Fog node of Fig.\ \ref{fig:multi-tier}. Each clone has the same software stack as its associated Mobile device and is equipped with the application DAG to be executed.  
\end{itemize}

Regarding the rationales behind the above assumptions, five main explicative remarks are in order.

First, the first three assumptions are (at least) necessary, in order to have finite DAG execution times.

Second, depending on the more or less fine granularity of the considered DAG, in our framework, a task may represent a routine, a method or a thread. Hence, the fourth assumption is compliant with the atomic nature of these entities.

Third, the fifth assumption reflects the fact that the executions of first and last tasks of real-life mobile applications typically require the utilization of input/output \textit{hardware} cards (like, screens, keyboards, sensors, photo/video-cameras, microphones and similar) that are hosted by the Mobile device, so that these tasks are not off-loadable \cite{andrade2014,mao2017}.

Fourth, in practice, the sixth assumption may be actuated by performing suitable real-time migrations of the underlying containers. Since this specific topic has been recently addressed, for example, in \cite{baccarelli2018}, we limit to note that, under this last assumption, the Mobile device only needs to transmit a small volume of signaling data to the mobile clones, in order to perform synchronization and indicate the tasks to be remotely executed. Therefore, in the sequel, we do not consider the time and energy consumption induced by the transmission of signaling data.

Finally, according to a network-oriented point of view, in our framework, the tasks sizes are assumed to be expressed in $ \mathrm{bit} $. In practical, the CPU workload: $ \hat{s}_i $ (measured in (CPU cycle)) required by the execution of the $ i $-th task is related to the corresponding task size: $ s_i $ (bit) through the following product formula \cite{kwak2015}:
\begin{align}
\SwapAboveDisplaySkip
\hat{s}_i = pd \times s_i .
\label{eq:CPUworkload}
\end{align}
In the above relationship, $ pd $ (expressed in (CPU cycle/bit)) is the so-called processing density of the considered application. Since it measures the average number of CPU cycles that are required for the processing of a single application bit, its actual value depends on the computing load of the underlying application. For illustrative purposes, Table \ref{table:02} reports the values of the processing densities of some benchmark mobile stream applications.

\begin{table*}[ht!]
\caption{Profiled processing densities of some classes of real-life mobile stream applications \cite{kwak2015}.}
\label{table:02}
\centering
\setlength{\extrarowheight}{0.5ex}
\renewcommand{\arraystretch}{1.2}
\begin{tabular}{lcc}
\toprule
\rowcolor[HTML]{BABFC6}
\bfseries Application class & & {\bfseries Application density} (CPU cycle/bit)  \\[0.6ex]
\midrule
\rowcolor[HTML]{F2F5F9}
Video transcoding  & & $ 200-1200 $ \\[0.4ex]
\rowcolor[HTML]{DCE1E8}
Processing of $ 400 $ frames of video game  & &  $ 2400-1200 $ \\[0.4ex]
\rowcolor[HTML]{F2F5F9}
Gesture recognition from high-resolution photos  & & $ 2500-32000 $ \\[0.4ex]
\rowcolor[HTML]{DCE1E8}
Virus scanning  & & $ 33000-37000 $ \\[0.4ex]
\bottomrule
\end{tabular}
\renewcommand{\arraystretch}{1.0}
\end{table*}

\section{The considered service and scheduling disciplines}
\label{sec:ServiceScheduling}

The goal of this section is to introduce the basic formal notation and operative assumptions about the networked ecosystem of Fig.\ \ref{fig:multi-tier}. In this regard, after indicating by:
\begin{equation}
\mathcal{A} \overset{\mathrm{def}}{=} \left\{ M , F_1 , \dots , F_Q  , C \right\} ,
\label{eq:MFC}
\end{equation}
and
\begin{equation}
{B\!H\!S} \overset{\mathrm{def}}{=}  \left\{ F_1 , \dots , F_Q  , C \right\} \equiv \mathcal{A} \setminus \left\{ M \right\},
\label{eq:BHS}
\end{equation}
the set of the available computing nodes and that of the corresponding backhaul segment respectively (see Fig.\ \ref{fig:multi-tier}), let:
\begin{equation}
{R_{N_1 \to N_2}} \; \text{(bit/s)}, \quad {N_1 \neq N_2}  , \text{with }  {N_1,N_2 \in \mathcal{A}},
\label{eq:onewayTCPIP}
\end{equation}
and
\begin{equation}
{R_{N_1 \leftrightarrow N_2}} \; \text{(bit/s)}, \quad {N_1 \neq N_2}  , \text{with }  {N_1,N_2 \in \mathcal{A}},
\label{eq:twowayTCPIP}
\end{equation}
indicate the throughput (i.e., the transport rate) of the one-way TCP/IP transport connection from node $ N_1 $  to node $ N_2 $, and that of the corresponding two-way symmetric full-duplex connection. Furthermore, since each task must be processed by only one computing node, let:
\begin{equation}
\vec{x} = \left[ x_1 , \dots , x_V \right],
\label{eq:taskallocation}
\end{equation}
be the $ V $-dimensional task allocation (row) vector, whose $ i $-th discrete-valued scalar component:
\begin{equation}
x_i \in \mathcal{A},  \quad  {i =1,\dots,V} ,
\label{eq:xcomponent}
\end{equation}
indicates the computing node that must process the $ i $-th task of the assigned application DAG.

\subsection{Service disciplines at the computing nodes and related service times}
\label{ssec:Servicedisciplines}

According to the virtualized node architecture reported in Fig.\ \ref{fig:VirtNodeArchitec}, let: $f_{i,N}$ (bit/s), $ i = 1, \dots, V $, $ N \in \mathcal{A} $ be the processing frequency that the Container Engine of Fig.\ \ref{fig:VirtNodeArchitec} allocates for the execution of the $i$-th task of size $ s_i $ (bit). Hence, by definition, the resulting service time: $ T_{i,N}^{\left(SER\right)}$ (s) measures the processing time of the $ i $-th task, and, then, it is formally defined as follows \cite{kumar2004}:
\begin{equation}
T_{i,N}^{\left(SER\right)}  \overset{\mathrm{def}}{=}  s_i / f_{i,N}, \quad i = 1,\dots, V ; \; N \in \mathcal{A}.
\label{eq:ServiceTime}
\end{equation}
Since the form assumed by $ f_{i,N} $ depends on the specific service discipline adopted by node $ N $ for the processing of the assigned tasks, in the sequel, we limit to assume that \cite{kumar2004}:
\begin{itemize}
\item $f_{i,N}$ is proportional to the total computing capacity $ n_N f_N $ (bit/s) available at node $ N $ (see Fig.\ \ref{fig:VirtNodeArchitec}). 
\end{itemize}

In this regard, we note that two examples of service disciplines of practical relevance that meet the above assumption are the \textit{SEQuential (SEQ)} service discipline and the \textit{Weighted Processors Sharing (WPS)} one (see, for example, \cite[Chapter 4]{kumar2004}).

By design, we shortly note that, under the SEQ service discipline, each task is \textit{individually} processed according to a specified sequential ordering. Hence, $f_{i,N}$ equates, by design, the full per-node processing capability (that is, $f_{i,N} \equiv n_N f_N$), and, then, (see Eq.\ \eqref{eq:ServiceTime})
\begin{equation}
T_{i,N}^{\left(SER;SEQ\right)}  \overset{\mathrm{def}}{=}  s_i / \left(n_{N}f_{N}\right), \quad  i = 1,\dots, V.
\label{eq:ServTimeSequen}
\end{equation}

Under the WPS service discipline, the tasks assigned to the computing node $ N $ are processed in parallel by following a weighted round-robin task scheduling \cite{kumar2004}. Specifically, the computing frequency at which the $ i $-th task is processed equates to:
\begin{equation}
f_{i,N} \equiv \left(  \frac {\phi_{i}} {\sum\limits_{j=1}^{V} \phi_{j}  \delta \left( x_j - N \right) }  \right)  \times n_N f_N,  
\label{eq:iTaskFreq}
\end{equation}
for $i = 1,\dots, V$, where \cite{kumar2004}: (i) the (dimensionless and positive) weight coefficient $ {\phi_{i}} $ ($i = 1,\dots, V $), fixes the relative priority level of the $i$-th task; and, (ii) the delta-terms at the denominator of the above equation assure that the processing capability $n_{N}f_{N}$ available at node $ N $ is shared only by the tasks that are actually to be processed by node $ N $.

\medskip\noindent\textit{Per-node total service times} -- The resulting total service time $ T_{N}^{\left(SER\right)} $ (s) at node $ N \in \mathcal{A} $ is defined as the total time spent by the node for processing \textit{all} the assigned tasks, under the assumption that all the data needed for the task processing is \textit{already available} at node $ N $ (i.e., by definition, $ T_{N}^{\left(SER\right)} $ does \textit{not} account for the inter-node network-induced transport delays \cite{kumar2004}). 

Since also the analytical expression of $ T_{N}^{\left(SER\right)} $ heavily depends on the adopted service discipline, in the sequel, we limit to assume that \cite{kumar2004}: 

\begin{itemize}
\item $T_{N}^{\left(SER\right)}$ is proportional to: $ 1 / \left( n_N f_N \right) $; and,
\item $T_{N}^{\left(SER\right)}$ does not decrease when at least one size of the assigned tasks increases.
\end{itemize}

Just as illustrative examples, the expression assumed by the cumulative service time under the (aforementioned) \textit{SEQ} service discipline is sum-like, i.e.,
\begin{flalign}
T_{N}^{\left(SER;SEQ\right)} & = {\sum\limits_{i=1}^{V} T_{i,N}^{\left(SER;SEQ\right)}  \, \delta \left( x_i - N \right) }  \nonumber \\
& \equiv  \frac{1}{n_N f_N} \times \left(   {\sum\limits_{i=1}^{V}  s_{i} \, \delta \left( x_i - N \right) }  \right), 
\label{eq:T_N-SERSEQ}
\end{flalign}
while the following $ \max $-type formula holds for the \textit{WPS} case:
\begin{flalign}
T_{N}^{\left(SER;WPS\right)} & = \! \underset{1 \leq i \leq V}{\max} \! \left\{ T_{i,N}^{^{\left(SER;WPS\right)}} \, \delta \left( x_i - N \right) \right\}   \nonumber \\
& \equiv  \frac{1}{n_N f_N} \times \underset{1 \leq i \leq V}{\max} \left\{  \frac{s_i \, \delta \left( x_i - N \right)}{ \left( \frac{\phi_i}{{\sum\nolimits_{j=1}^{V}  \phi_{j} \, \delta \left( x_j - N \right) }}\right) }  \right\} . 
\label{eq:T_N-SERWPS}
\end{flalign}

\subsection{Per-task execution times}
\label{ssec:executiontimes}

The impact of the network-induced delays on the performance of the ecosystem of Fig.\ \ref{fig:multi-tier} is accounted for by the corresponding execution time $ T_{i,N}^{\left(EXE\right)} $ (s) of the $ i $-th task at node $ N \in \mathcal{A} $. It is formally defined as the summation \cite{kumar2004}:
\begin{equation}
T_{i,N}^{\left(EXE\right)}  \overset{\mathrm{def}}{=}   T_{i,N}^{\left(SER\right)}   +   T_{i,N}^{\left(NET\right)} , \;\,  i = 1,\dots,V ;   \;  N \in \mathcal{A} ,
\label{eq:T_i,N-EXE}
\end{equation}
of the (already introduced) service time $ T_{i,N}^{\left(SER\right)} $ and the network time $ T_{i,N}^{\left(NET\right)} $, needed for the transport to node $ N $ of \textit{all} data that is required by the execution of the $ i $-th task. Hence, directly from the (previously reported) definition of the throughput of the one-way connections, the following relationship holds:
\begin{equation}
T_{i,N}^{\left(NET\right)}  =   \sum\nolimits_
{\begin{subarray}{l}	N_1 \in \mathcal{A}   \\   N_1 \ne N \end{subarray}}
\left(  \frac{vl_i^{\left(N_1 \to N\right)}}{R_{N_1 \to N}}  \right), \quad  i = 1,\dots,V .
\label{eq:T_i,N-NET}
\end{equation}
In the above equation, $ vl_i^{\left(N_1 \to N\right)} $ (bit) indicates the volume of data that must be transported from node $ N_1 $ to node $ N $ for the execution of the $ i $-th task at node $ N $. Hence, from the definitions of the adjacency and edge weight matrices of the considered application DAG, the following relationship holds for the computation of $ vl_i^{\left(N_1 \to N\right)} $, $ i = 1,\dots,V $:
\begin{equation}
vl_i^{\left(N_1 \to N\right)}  \overset{\mathrm{def}}{=}
\left( 1 + \overline{NF}_{N_1 \to N} \right) \times
\sum_{j=1}^{V}  { a_{ji} d_{ji} \delta \left( x_j - N_1 \right) } ,
\label{eq:vl_i}
\end{equation}
where $ N_1 \ne N $, and $ N_1,N \in \mathcal{A} $.

Before proceeding, three main explicative remarks about the relationships in \eqref{eq:T_i,N-NET} and \eqref{eq:vl_i} are in order.

First, since the (non-negative and dimensionless) term $ \overline{NF}_{N_1 \to N} $ is the average failure rate of the one-way connection: $ N_1 \to N $, the first factor present in \eqref{eq:vl_i} is the average overhead in the volume of the transported data that is induced by connection-failure phenomena.

Second, the sum-form of the expression of $ T_{i,N}^{\left(N\!ET\right)} $ in \eqref{eq:T_i,N-NET} applies when the multiple data streams arriving at the receiving node $ N $ are processed in a sequential way, so that their network delays are added up. However, in our framework, the mobile device is equipped by multiple NICs that, in principle, could operate in parallel. In this case, the sum-type expression in \eqref{eq:T_i,N-NET} should be replaced by the following max-type one \cite{kumar2004}:
\begin{equation}
T_{i,N}^{\left(NET\right)}     =    \max_
{\begin{subarray}{l}	N_1 \in \mathcal{A}   \\   N_1 \ne N \end{subarray}}
\left\{  \frac{vl_i^{\left(N_1 \to N\right)}}{R_{N_1 \to N}}  \right\} .
\label{eq:T_i,N-NET,maxtype}
\end{equation}
Since the point-wise maximum of convex function is still a convex function \cite{bazaraa2017}, we anticipate that the convex/nonconvex nature of the constrained optimization problem to be afforded remains unchanged under both cases. Furthermore, the parallel processing of multiple received streams may induce out-of-order phenomena, that, in turn, introduce additional queue delays at the Transport Layer of the receiving nodes \cite{baccarelli2018}. Hence, on the basis of these considerations, without substantial loss of generality, in the sequel, we assume that $ T_{i,N}^{\left(NET\right)} $ is given by the sum-type expression in \eqref{eq:T_i,N-NET}.

Finally we note that, since each virtual processor is equipped with a buffer that is reserved for the temporary storage of the assigned tasks (see Fig.\ \ref{fig:VirtNodeArchitec}), the definition in \eqref{eq:T_i,N-EXE} of the per-task execution time  $ T_{i,N}^{\left(EXE\right)} $ automatically accounts for the queue delay at the underlying computing node $N$ \cite{kumar2004}.

\subsection{Per-DAG execution times and inter-node task scheduling disciplines}
\label{ssec:Per-DAGexecutiontimes}

By definition, the DAG execution time $ T_{DAG} $ (s) is the time interval between the instant at which begins the execution of the first task and the instant at which ends the execution of the last task. From a formal point of view, $ T_{DAG} $ is a function \cite{kumar2004}:
\begin{equation}
T_{DAG}     \!  =  \!     \mathcal{X} \!
\left( \! \left\{
T_{i,N}^{\left(EXE\right)}  \delta \left(x_i-N\right) \! ,
i \! = \!  1,\dots,V    ;    N \! \in \! \mathcal{A}
\right\} \! \right) ,
\label{eq:T_DAG}
\end{equation}
of both the set of the (previously modeled) per-task execution times and task allocation vector $ \vec{x} $ in \eqref{eq:taskallocation}. Furthermore, the specific form of the $ \mathcal{X} \left(.\right) $ function depends on the actually adopted inter-node Task Scheduling Discipline (TSD). By definition, it dictates the (statically or dynamically configured) ordering in which the computing nodes process the sets of the assigned tasks \cite{kumar2004}. Since a number of TSDs have been even recently considered in the open literature \cite{mahmoodi2016}, in the sequel, we assume the adopted TSD to be assigned and, then, we limit to point out two general assumptions on the resulting $ T_{DAG} $ that are typically guaranteed by the TSDs of practical interest. Specifically, we assume that \cite{kumar2004}:
\begin{itemize}
\item $ T_{DAG} $ is a non-decreasing function of each per-task execution time $ T_{i,N}^{\left(EXE\right)} $, $ i=1,\dots,V $, $ N \in \mathcal{A} $; and,
\item $ T_{DAG} $ is a jointly convex function of the per-task execution times $ \left\{ T_{i,N}^{\left(EXE\right)}, \, i=1,\dots,V, \,  N \in \mathcal{A} \right\} $.
\end{itemize}

Just as practical examples of TDSs that meet the above assumptions, we shortly address the \textit{Sequential Task Scheduling} (STS) and the \textit{Parallel Task Scheduling} (PTS) disciplines \cite{kumar2004}.

By definition, the STS discipline forces the nodes to perform their computation in a \textit{sequential} way. Hence, this discipline does \textit{not} allow inter-node parallel task executions and, then, it is applied when the sets of tasks assigned to the various computing nodes cannot be processed in parallel. As a matter of fact, the corresponding DAG execution time assumes the following sum-type expression:
\begin{equation}
T_{DAG}^{\left(STS\right)}     =  
\sum_{i=1}^{V}
\sum_{N \in \mathcal{A}}
T_{i,N}^{\left(EXE\right)} \, \delta \left( x_i - N \right),
\label{eq:T_DAG-STS}
\end{equation}
where the delta terms in the inner summation guarantee that the $ i $-th task is executed by a single computing node. 

By definition, the PTS discipline forces the sets of tasks assigned to the computing nodes to be processed in \textit{parallel}, so that it may be applied when \textit{no} interdependence is present. As a consequence, the resulting DAG execution time is given by the following max-type expression \cite{kumar2004}:
\begin{equation}
T_{DAG}^{\left(PTS\right)}     =  
\underset{1 \leq i \leq V}{\max}
\left\{
\sum_{N \in \mathcal{A}}
T_{i,N}^{\left(EXE\right)} \, \delta \left( x_i - N \right)
\right\} .
\label{eq:T_DAG-PTS}
\end{equation}
%
By direct inspection, it can be viewed that both the above expressions meet the previously reported general assumptions on $ T_{DAG} $.

\section{Modeling the computing and networking energy consumption}
\label{sec:ModelingEnergyConsumption}

The goal of this section is threefold. First, we introduce the cost parameters used to formally feature the Software-as-a-Service (SaaS) policy enforced by the Cloud and Fog Service Providers that manage the ecosystem of Fig.\ \ref{fig:multi-tier}. Second, we develop the formal models to profile the power and energy consumption of the virtualized multi-core processors that equip the device clones at the computing nodes (see Fig.\ \ref{fig:VirtNodeArchitec}). Third, we pass to model the companion power and energy models of the TCP/IP-based 5G network connections that support the FRAN, CRAN and Backhaul segments of the overall network infrastructure of Fig.\ \ref{fig:multi-tier}. Interestingly enough, we anticipate that the models developed for both the computing and network power/energy explicitly account for the specific SaaS server policy applied by the Service Providers.

\subsection{SaaS policies for virtualized Mobile-Fog-Cloud networked ecosystems}

In this regard, we note that, since the ecosystem of Fig.\ \ref{fig:multi-tier} relies on container and 5G-based virtualization technologies for multiplexing of the underlying physical computing and networking resources, each device clone runs atop an isolated environment and this makes the SaaS model be applicable \cite{zhao2016}. According to this service model, the user who manages the mobile device may be charged on the basis of the networking and computing resources that are actually wasted by the corresponding device clones of Fig.\ \ref{fig:multi-tier}. Specifically, since the final goal of the ecosystem of Fig.\ \ref{fig:multi-tier} is to save energy by suitably pricing it, the Cloud and Service Providers may enforce one of the following three SaaS-based policies \cite{zhao2016}: 
\begin{enumerate}[i.]
\item the virtualized computing/networking resources utilized by the device clones at the proximate Fog nodes are priced or they are for free;
\item the virtualized computing/networking resources wasted by the device clone running at the remote Cloud node are subject to ``flat'' pricing policies or they are metered and priced on a per-usage basis; and,
\item the user is whether or not interested to minimize the energy consumption of his/her own mobile device.
\end{enumerate}
Since, in our framework, Fog Providers, Cloud Providers and mobile users act as independent actors, any combination of the aforementioned three basic pricing policies could be applied. We anticipate that, in order to account for this consideration, we introduce three binary-valued parameters, i.e.:
\begin{equation}
\theta_M   \in   \left\{0,1\right\} ,  \quad
\theta_F   \in   \left\{0,1\right\} ,  \quad
\theta_C   \in   \left\{0,1\right\} ,
\label{eq:theta}
\end{equation}
whose settings are dictated by the SaaS policies actually implemented by the three mentioned actors (see Remark \ref{remark:remark2} of Section \ref{ssec:DesignTarget} for the formal description of the role played by the theta parameters in \eqref{eq:theta}).

\subsection{Modeling the computing energy in multi-core virtualized execution environments}
\label{ssec:ModelingComputingEnergy}

According to a number of (even recent) contributions (see, for example, \cite{basmadjian2012} and references therein), the computing energy $ \mathcal{E}_N $ (Joule) wasted by the device clone at node $ N \in \mathcal{A} $ for the execution of the assigned tasks is the summation of a static $ \mathcal{E}_N^{\left(STA\right)} $ (Joule) part and a dynamic $ \mathcal{E}_N^{\left(DYN\right)} $ (Joule) part.

Specifically, the static part accounts for the energy wasted by the clone in the idle state (i.e., the clone is turned ON but it is not running). As a consequence, when the clone is actually turned ON for DAG execution, $\mathcal{E}_N^{\left(STA\right)}$ equates the following product: $\mathcal{E}_N^{\left(STA\right)}   \overset{\mathrm{def}}{=}  \mathcal{P}_N^{\left(STA\right)} \times T_{DAG} $, where $ T_{DAG} $ is the (previously introduced) DAG execution time, and $ \mathcal{P}_N^{\left(STA\right)} $ (Watt) is the corresponding per-clone computing static power. After considering that each physical server at node $ N $ consumes: $ \mathcal{P}_{CPU-N}^{\left(IDLE\right)} $ (Watt) unit of power for sustaining $ {nc}_{N} \geq 1$ containers in the idle state, (see Fig.\ \ref{fig:VirtNodeArchitec} ), the per-clone computing static power may be, in turn, modeled as \cite{basmadjian2012}: $ \mathcal{P}_N^{\left(STA\right)} = \mathcal{P}_{CPU-N}^{\left(IDLE\right)} / {nc}_{N}$, $ N \in \mathcal{A} $.

Passing to model the dynamic part of the computing energy, we note that, by definition, a clone processes the assigned tasks over a time interval equal to the (previously defined) service time. Therefore,  the dynamic component: $\mathcal{E}_N^{\left(DYN\right)}$ of the per-clone computing energy equates the product: $ \mathcal{E}_N^{\left(DYN\right)} \overset{\mathrm{def}}{=}  \mathcal{P}_N^{\left(DYN\right)} \times {T}_N^{\left(SER\right)}$, where $ \mathcal{P}_N^{\left(DYN\right)} $ (Watt) is the dynamic power consumed by the clone for computing purpose. By definition, this last power depends on three main factors, namely \cite{basmadjian2012}:
\begin{enumerate}
\item the number of cores $ n_N $ equipping the virtual processor at node $ N \in \mathcal{A} $ (see Fig.\ \ref{fig:VirtNodeArchitec});
\item the corresponding computing frequency $ f_N $ (bit/s) at which the Container Engine of Fig.\ \ref{fig:VirtNodeArchitec} does run the available cores; and,
\item the fraction $ r_N \in \left[ 0,1 \right]  $ of the overall dynamic power: $ \mathcal{P}_N^{\left(DYN\right)} $ that is shared by all cores for common operation.
\end{enumerate}

Hence, according to, for example, \cite{basmadjian2012}, $ \mathcal{P}_N^{\left(DYN\right)} $ may be formally profiled through the following power-like expression: $ \mathcal{P}_N^{\left(DYN\right)} = n_N \left( 1-r_N \right) k_N \left( f_N \right)^{\gamma_{N}} $, where: (i) $ {\gamma_{N}} $ is a dimensionless shaping exponent (i.e., typically $ \gamma_N \geq 2 $); and, (ii) the positive constant $ k_N $ $\left( \text{Watt}/(\text{bit/s})^{\gamma_{N}} \right)$ accounts for the common power profiles of the utilized (homogeneous) cores \cite{basmadjian2012}.

Overall, on the basis of the above considerations, the computing energy wasted by the mobile clone at node $ N \in \mathcal{A} $ may be analytically profiled as follows:
\begin{align}
\mathcal{E}_N  \! = &
\underbrace{
\overbrace{\left( {\mathcal{P}_{CPU-N}^{\left(IDLE\right)}}/{nc_N} \right)  }^{\mathcal{P}_N^{\left(STA\right)}}
\! \times T_{DAG} \! \times \! u_{-1}
\left(
\sum_{i=1}^{V}{\delta \left( x_i-N \right)}
\right)
}_{\mathcal{E}_N^{\left(STA\right)}} \nonumber \\
& + 
\underbrace{
\overbrace{n_N \left(1-r_N\right) k_N \left(f_N\right)^{\gamma_{N}}}^{\mathcal{P}_N^{\left(DYN\right)}}
\times T_{N}^{\left(SER\right)}
}_{\mathcal{E}_N^{\left(DYN\right)}} \, ,
\label{eq:E_N}
\end{align}
where the unit step-size factor: $ u_{-1} \, \left(  \sum_{i=1}^{V}{\delta \left( x_i-N \right)}\right) $ accounts for the fact that the device clone at node $ N $ is actually turned ON only when it must process at least a pending task.

\subsection{Modeling the networking energy of the inter-node connections}
\label{ssec:ModelingEnergyInternode}

Let $\mathcal{E}_{N_1 \leftrightarrow N_2} $ (Joule) be the network energy consumed by the two-way (i.e., bi-directional) end-to-end TCP/IP Transport-layer connection between the computing nodes $ N_{1} $ and $ N_{2} $, with $ N_1 \ne N_2 $, and $ N_1,N_2 \in \mathcal{A} $. Since it equates to the summation:
\begin{equation}
\mathcal{E}_{N_1 \leftrightarrow N_2}  =
\mathcal{E}_{N_1 \to N_2}  +
\mathcal{E}_{N_2 \to N_1}  ,
\label{eq:E_N1N2}
\end{equation}
of the corresponding energy $ \mathcal{E}_{N_1 \to N_2} $ (Joule) and $ \mathcal{E}_{N_2 \to N_1}$ of the underlying one-way (i.e., directed) Transport connections: $ {N_1 \to N_2} $ and $ {N_2 \to N_1} $ respectively, we may directly focus on the modeling of $ \mathcal{E}_{N_1 \to N_2}$. According to the analytical and experimental models reported, for example, in \cite{altamimi2015}, it may be profiled as the summation of a static part and dynamic one as in:
\begin{equation}
\mathcal{E}_{N_1 \to N_2}  =
\mathcal{E}_{N_1 \to N_2}^{\left(STA\right)}  +
\mathcal{E}_{N_1 \to N_2}^{\left(DYN\right)}  .
\label{eq:E_N1toN2}
\end{equation}
The static component: $\mathcal{E}_{N_1 \to N_2}^{\left(STA\right)} \mathrm{(Joule)}$ may be expressed, in turn, as in:
\begin{align}
\mathcal{E}_{N_1 \to N_2}^{\left(STA\right)}  = &
\mathcal{P}_{N_1 \to N_2}^{\left(STA\right)}
\times  T_{DAG}     \nonumber \\
& \times u_{-1} \left(
\sum_{i=1}^{V}{
\sum_{j=1}^{V}
{ a_{ij} \delta \left(x_i-N_1\right)  \delta \left(x_j-N_2\right)
}} \right) ,
\label{eq:E_N1toN2static}
\end{align}
with
\begin{equation}
\mathcal{P}_{N_1 \to N_2}^{\left(STA\right)}  =
\theta_{N_{1}}  \mathcal{P}_{NET-N_{1}}^{\left(IDLE\right)}     +
\theta_{N_{2}}  \mathcal{P}_{NET-N_{2}}^{\left(IDLE\right)}     .
\label{eq:P_N1toN2static}
\end{equation}
In the above formulas, we have that: (i) $ \mathcal{P}_{NET-N_{1}}^{\left(IDLE\right)} $ (resp., $ \mathcal{P}_{NET-N_{2}}^{\left(IDLE\right)} $) is the power (measured in $ \mathrm{(Watt)} $) consumed in the idle state by the NIC equipping node $ N_1 $ (resp., $ N_2 $); and, (ii) the unit step-size factor in \eqref{eq:E_N1toN2static} accounts for the fact that the connection: $ N_1 \to N_2 $  is turned ON if there is at least a task assigned to $ N_1 $ whose output data is required for the execution of at least a task assigned to $ N_2 $.

Since the involved NICs remain turned ON only during the time needed for the transport of data from node $ N_1 $ to node $ N_2 $, the dynamic component in \eqref{eq:E_N1toN2} of the per-connection network energy equates, by definition, the following product: 
\begin{equation}
\mathcal{E}_{N_1 \to N_2}^{\left(DYN\right)}  =
\mathcal{P}_{N_1 \to N_2}^{\left(DYN\right)}
\times
T_{N_1 \to N_2}  .
\label{eq:E_N1toN2dynamic}
\end{equation}
In the above relationship, we have that: (i) $ \mathcal{P}_{N_1 \to N_2}^{\left(DYN\right)}  \mathrm{(Watt)} $ is the dynamic part of the network power needed for sustaining the connection: $ {N_1 \to N_2} $; and, (ii) $ T_{N_1 \to N_2} \, \mathrm{(s)}$  is the total time needed for the transport of the required data from $ N_1 $ to $ N_2 $. By design, it is given by:
\begin{align}
T_{N_1 \to N_2}  
\overset{\mathrm{def}}{=} &
\; \frac{vl_{N_1 \to N_2}}{R_{N_1 \to N_2}}   \equiv
\frac{\left( 1 + \overline{NF}_{N_1 \to N_2} \right)}{R_{N_1 \to N_2}}     \nonumber \\
%
& \times
\sum_{i=1}^{V}{
\sum_{j=1}^{V}{
a_{ij}d_{ij} \delta  \left( x_i - N_1 \right) \, \delta  \left( x_j - N_2 \right)
}}  ,
\label{eq:T_N1toN2}
\end{align}
where (see \eqref{eq:vl_i}) $ vl_{N_1 \to N_2} \mathrm{(bit)} $ is the total volume of data to be transported from node $ N_1 $ to node $ N_2 $.

After swapping the indexes $ N_1 $ and $ N_2 $, the same formulas also hold for modeling the energy $ \mathcal{E}_{N_2 \to N_1} $ consumed by the one-way connection $ {N_2 \to N_1} $ in \eqref{eq:E_N1N2}.

\medskip\noindent\textit{Modeling the dynamic energy wasted by throughput-adaptive up-down wireless connections} -- The above formulas for the network energy apply to the evaluation of the set of energy: $ \left\{ \mathcal{E}_{M \leftrightarrow N}, N \in BHS \right\} $ of the two-way up/down single-hop connections: $ \left\{ {M \leftrightarrow N}, N \in BHS \right\} $ embraced by the FRAN and CRAN of Fig.\ \ref{fig:multi-tier}. For this purpose, it suffices that the corresponding dynamic components $ \left\{ \mathcal{P}_{M \to N}^{\left(DYN\right)} \right\} $ of the one-way network power in \eqref{eq:E_N1toN2dynamic} are suitably modeled, in order to account for the dependence of the transmit and receive dynamic power $ \mathcal{P}_{M \to N}^{\left(DYN;Tx\right)} $ and: $ \mathcal{P}_{M \to N}^{\left(DYN;Rx\right)} $ on the underlying transport throughput: $ R_{\left(M \to N\right)} $ (bit/s). In this regard, the results reported, for example, in \cite{altamimi2015}, support the adoption of the following quite general power-like model: 
\begin{align}
\mathcal{P}_{M \to N}^{\left(DYN\right)}  = &
\underbrace{
\theta_M \Omega_{\left(M,N\right)}^{\left(Tx\right)} \times \left( R_{M \to N} \right) ^ {\xi_{\left(M,N\right)}^{\left(Tx\right)}}
}_{\mathcal{P}_{M \to N}^{\left(DYN;Tx\right)}} \nonumber \\
& + 
\underbrace{
\theta_N \Omega_{\left(M,N\right)}^{\left(Rx\right)} \times \left( R_{M \to N} \right) ^ {\xi_{\left(M,N\right)}^{\left(Rx\right)}}
}_{\mathcal{P}_{M \to N}^{\left(DYN;Rx\right)}}   .
\label{eq:P_MtoNdynamic}
\end{align}
In the above equation, the (non-negative dimensionless) transmit/receive exponents: $\, \xi_{\left(M,N\right)}^{\left(Tx\right)} \geq \xi_{\left(M,N\right)}^{\left(Rx\right)} \geq 2 \, $ depend on the (short or long-range) single-hop wireless communication technology adopted to sustain the connection $ M \to N $. Furthermore, the (positive) transmit/receive coefficients: $ \Omega_{\left(M,N\right)}^{\left(Tx\right)} $ and $ \Omega_{\left(M,N\right)}^{\left(Rx\right)} $ (measured in $ \mathrm{\left(\text{Watt}/\left((\text{bit/s})^{\xi} \times \mathrm{\left(\text{s}\right)}^\eta \right) \right)} $) 
account for the effects of the Round-Trip-Time (RTT), spatial range and power profile of the considered connection. According, for example, to \cite{agiwal2016} and references therein, they may be modeled as follows:
\begin{align}
&\Omega_{\left(M,N\right)}^{\left(Tx\right)} = 
\frac
{ \left(RTT_{\left(M,N\right)}\right)^{\eta} \chi_{\left(M,N\right)}^{\left(Tx\right)} }
{ \left(1+\left(\ell_{\left(M,N\right)}\right)^{\alpha}\right) }
, \; \text{and} \nonumber \\
&\Omega_{\left(M,N\right)}^{\left(Rx\right)} = 
\frac
{ \left(RTT_{\left(M,N\right)}\right)^{\eta} \chi_{\left(M,N\right)}^{\left(Rx\right)} }
{ \left(1+\left(\ell_{\left(M,N\right)}\right)^{\alpha}\right) }  ,
\label{eq:Omega_MN}
\end{align}
where: (i) $ RTT_{\left(M,N\right)} $ (s) is the round-trip-time of the sustained TCP/IP connection; (ii) $ \eta $ is a dimensionless non-negative shaping exponent (i.e., typically, $ \eta \cong 0.6 $); (iii) $ \ell_{\left(M,N\right)} $ (m) is the physical length (i.e., the spatial range) spanned by  the considered connection; (iv) $ \alpha $ (with $ 2 < \alpha \leq 4 $) is the fading-induced loss exponent; and, (v) the positive coefficients: $ \chi_{\left(M,N\right)}^{\left(Tx\right)} $ and $ \chi_{\left(M,N\right)}^{\left(Rx\right)} $ account for the transmit/receive power efficiency of the adopted wireless communication technology. As a consequence, their actual values depend on a number of communication parameters \cite{baccarelli2005a,agiwal2016}, like, number of transmit/receive antennas \cite{baccarelli2004a,baccarelli2005b}, antenna gains, coding gains, implemented carrier tracker and interleaving depth \cite{baccarelli1998}, just to cite a few.

\medskip\noindent\textit{Energy models for the two-way backhaul connections} -- The general formulas reported in \eqref{eq:E_N1N2}-\eqref{eq:T_N1toN2} apply verbatim to model the energy $ \mathcal{E}_{N_1 \leftrightarrow N_2} $ consumed by the end-to-end two-way Transport-layer connection between any pair of backhaul nodes $ N_1, N_2 \in BHS $. Hence, in order to complete the corresponding model, it suffices to detail the expressions assumed by the involved dynamic network power $ \mathcal{P}_{N_1 \to N_2}^{\left(DYN\right)} $ in \eqref{eq:E_N1toN2dynamic} and the related backhaul transport throughput $ R_{N_1 \to N_2} $.

In this regard, three main remarks are in order. First, each backhaul connection is symmetric, full-duplex and it may be multi-hop. Second, since it typically uses a broadband Ethernet-type technology at the Data Link Layer, its steady-state transport capacity may be considered nearly constant over long time intervals, so that the dependence of $ \mathcal{P}_{N_1 \to N_2}^{\left(DYN\right)} $ on $ R_{N_1 \to N_2} $ is not a critical issue \cite{kumar2004}. As a consequence, $ \mathcal{P}_{N_1 \to N_2}^{\left(DYN\right)} $ may be considered fixed and given by the following product formula: 
\begin{align}
\mathcal{P}_{N_1 \to N_2}^{\left(DYN\right)} \equiv &
\mathcal{P}_{N_2 \to N_1}^{\left(DYN\right)} = 
no_{HOP}^{\left(N_1,N_2\right)} \times \mathcal{P}_{HOP}^{\left(N_1 \to N_2\right)}
\nonumber \\
& \times \max \left\{ \theta_C, \theta_F \right\}, \,
N_1 \ne N_2 ; \;  N_1,N_2 \in BHS,
\label{eq:P_N1toN2dynamic}
\end{align}
where: (i) $ no_{HOP}^{\left(N_1,N_2\right)} $ is the (integer-valued and positive) number of hops of the considered backhaul connection; (ii) $ \mathcal{P}_{HOP}^{\left(N_1 \to N_2\right)} $ (Watt) is the one-way per-hop consumed power; and, (iii) the $ \max $ factor in \eqref{eq:P_N1toN2dynamic} accounts for the fact that the Cloud and Fog Providers may apply different pricing policies, so that, in general, the mobile user is charged by the resources wasted by the utilized backhaul connection if at least one Provider prices them \cite{zhao2016}. Third, in general, a TCP/IP-based backhaul connection is not so affected by the mobility of the served Mobile device and, then, it tends to operate in the Congestion Avoidance state (i.e., in the steady-state) over large time intervals \cite{kumar2004}. As a consequence, its transport throughput may be considered nearly constant and given by the following (quite usual) formula (see, for example, \cite[Chapter 7]{kumar2004}): 
\begin{align}
R_{N_1 \to N_2}  \equiv 
R_{N_2 \to N_1} =
\frac
{1.22 \times MSS^{\left(N_1,N_2\right)}}
{RTT_{\left(N_1,N_2\right)} \times \sqrt{Pr_{LOSS}^{\left(N_1,N_2\right)}}},
\label{eq:R_N1N2}
\end{align}
where: (i) $ MSS^{\left(N_1,N_2\right)} $ (bit) is the maximum segment size of the considered TCP/IP connection; and, (ii) $ RTT_{\left(N_1,N_2\right)} $ (s) and $ Pr_{LOSS}^{\left(N_1,N_2\right)} $ are the corresponding round-trip-time and average segment loss probability.

\section{The considered Joint Optimization Problem (JOP)}
\label{sec:ConsideredJOP}

The goal of this section is threefold. First, by referring to the peculiar features of the mobile stream applications, we discuss the objective design targets to be pursued and the main constraints to be considered. Second, on the basis of them, we formally introduce the optimization problem to be tackled with, and, then, we consider its feasibility. Third, after pointing out the challenges presented by its solution, we develop an (optimality-preserving) decomposition of the stated optimization problem into the cascade of two inter-related sub-problems that are simpler to manage.

\subsection{Design targets, related constraints and afforded problem}
\label{ssec:DesignTarget}

By design, mobile stream applications refer to execution environments where \cite{andrade2014}: (i) the applications are described by DAGs that are submitted for the execution in an online way (i.e., their submissions are sequential over the time); (ii) forecasting about future submissions is not available; (iii) the most relevant performance metric is the application throughput: $ TH_{DAG} \overset{\mathrm{def}}{=} \left(1/T_{DAG}\right) $ (app/sec), that measures the rate at which the sequence of input DAGs is processed by the networked computing infrastructure of Fig.\ \ref{fig:multi-tier}; and, (iv) the energy consumption of the overall ecosystem of Fig.\ \ref{fig:multi-tier} must be as low as possible.

On the basis of this native features, we identify four main design targets to be met.

First, we must guarantee that the application throughput $ TH_{DAG} $ of the underlying stream application does not fall \textit{below} an assigned minimum value \cite{andrade2014}: $ TH_{0}^{\left(MIN\right)} \overset{\mathrm{def}}{=} \left(1/T_{DAG}^{\left(MAX\right)}\right) $ (app/sec), that, in turn,  is dictated by the QoS level to be provided to the mobile user.

Second, since Cloud and Fog providers aim to maximize the number of the users who are concurrently served and the container-based virtualization technology allows isolation of the computing resources on a per-user basis \cite{pahl2017}, the per-core computing frequency available at each computing node $ N $ must be considered limited up to a maximum value: $ f_{N}^{\left(MAX\right)} $ (bit/sec), that, in turn, is dictated by service policy of the Service Providers. 

Third, since 5G technology allows virtualization and isolation of the wireless network resources done available to each user by the FRAN and CRAN of Fig.\ \ref{fig:multi-tier} (see Table \ref{table:01}), both the per-connection up and down network throughput must be considered upper limited up to: $ R_{M \to N}^{\left(MAX\right)} $ (bit/sec), and $ R_{N \to M}^{\left(MAX\right)} $ (bit/sec), $ N \in BHS $, respectively.

Fourth, according to the so-called Green Wave pursued by both the Fog and 5G paradigms \cite{mahmud2018}, the ultimate goal is the minimization of the total energy $ E_{TOT} \mathrm{\left(Juole\right)} $ consumed by the overall ecosystem of Fig.\ \ref{fig:multi-tier} for carrying out the needed computing and network operations.

Hence, by leveraging the models detailed in Section \ref{sec:ModelingEnergyConsumption}, this total energy is formally defined by the following relationship: 
\begin{align}
\mathcal{E}_{TOT}  \overset{\mathrm{def}}{=} &
\underbrace{
\theta_M \mathcal{E}_M + \theta_C \mathcal{E}_C + \theta_F \mathcal{E}_{FOG} 
}_{\mathcal{E}_{CMP}} \nonumber \\
& + 
\underbrace{
\mathcal{E}_{SR-WNET} + \mathcal{E}_{LR-WNET} + \mathcal{E}_{BH-NET}
}_{\mathcal{E}_{NET}}  .
\label{eq:E_Tot}
\end{align}
In \eqref{eq:E_Tot}, we have that (see Section \ref{sec:ModelingEnergyConsumption}):
\begin{enumerate}[i.]
\item $ \mathcal{E}_{CMP} $ (resp., $ \mathcal{E}_{NET} $) is the total energy consumed by the overall ecosystem of Fig.\ \ref{fig:multi-tier} for sustaining the computing (resp., networking) operations needed for a single execution of the considered application DAG;
\item $ \mathcal{E}_M $ and $ \mathcal{E}_C $ are the computing energy wasted by the mobile device and cloud clone, respectively (see Section \ref{ssec:ModelingComputingEnergy} for their formal models);
\item $ \mathcal{E}_{FOG} $ (Juole) is the summation of the computing energy wasted by the device clones deployed at the Fog nodes. Hence, it is formally defined as follows:
\begin{equation}
\mathcal{E}_{FOG}
\overset{\mathrm{def}}{=}
\sum\nolimits_{l=1}^{Q} {\mathcal{E}_{F_{l}}},
\label{eq:E_Fog}
\end{equation}
where $ \mathcal{E}_{F_{l}} $ (Juole), $ l = 1,\dots,Q $, is the computing energy wasted by the device clone at the $l$-th Fog node (see Section \ref{ssec:ModelingComputingEnergy} for its formal model); 
\item $ \mathcal{E}_{SR-WNET} $ (Juole) is the total network energy wasted by the two-way (i.e., up/down) Short Range (SR) single-hop wireless connections embraced by the FRAN of Fig.\ \ref{fig:multi-tier}. It is formally defined as:
\begin{equation}
\mathcal{E}_{SR-WNET}
\overset{\mathrm{def}}{=}
\sum\nolimits_{l=1}^{Q} {\mathcal{E}_{M \leftrightarrow F_l}},
\label{eq:E_SRWNET}
\end{equation}
where: $ \mathcal{E}_{M \leftrightarrow F_l} $ (Juole), $ l = 1,\dots,Q $, is the network energy consumed by the wireless up/down single-hop connection between the Mobile device and the $ l $-th Fog node (see Section \ref{ssec:ModelingEnergyInternode} for its formal model);
\item $ \mathcal{E}_{LR-WNET} \overset{\mathrm{def}}{=}  \mathcal{E}_{M \leftrightarrow C} $ (Juole) is the energy consumed by the CRAN of Fig.\ \ref{fig:multi-tier}, in order to sustain the two-way (i.e., up/down) Long-Range (LR) single-hop cellular connection between the Mobile device and the remote Cloud (see Section \ref{ssec:ModelingEnergyInternode} for its formal model); and, finally,
\item  $ \mathcal{E}_{BH-NET} $ (Juole) is the total energy wasted by all the Fog-to-Fog (F2F) and Fog-to-Cloud (F2C) two-way (possibly, multi-hop) Transport-layer connection embraced by the Backhaul network segment of Fig.\ \ref{fig:multi-tier}. It reads as:
\begin{equation}
\mathcal{E}_{BH-NET} \overset{\mathrm{def}}{=}
\underbrace{
\sum\nolimits_{l=1}^{Q}{
\sum\nolimits_{\begin{subarray}{l}	k=1   \\   k > l  \end{subarray}}^{Q}{
{\mathcal{E}_{F_l \leftrightarrow F_k}}
}}
}_{\text{F2F Network Energy}}
+ \!
\underbrace{
\sum\nolimits_{l=1}^{Q}{ \!
{\mathcal{E}_{F_l \leftrightarrow C}}
}
}_{\text{F2C Network Energy}}
\label{eq:E_BHNET}
\end{equation}
where the energy present in the above summations are modeled as reported in the last part of Section \ref{ssec:ModelingEnergyInternode}.
\end{enumerate}
Let:
\begin{equation}
\overrightarrow{RS}
\overset{\mathrm{def}}{=} \!
\left[
f_N,  N \! \! \in \! A; R_{_{M \to N}} , R_{_{N \to M}}, N \! \! \in \! {B\!H\!S}
\right]
\!  \in \! \left( \mathbb{R}_{+} \right) ^ {3Q+4} \! ,
\label{eq:RS}
\end{equation}
be the $ \left( {3Q+4} \right) $-dimensional (row) vector that collects the processing frequencies of the device clones at the computing nodes and the up-plus-down wireless transport throughput over the FRAN and CRAN of Fig.\ \ref{fig:multi-tier}.
Therefore, from the outset, it follows that the afforded \textit{Joint Optimization Problem} (JOP) may be formally defined as the following mixed-integer constrained optimization problem:
\begin{subequations}
\begin{align}
&  \underset{ \vec{x} , \overrightarrow{RS}}{\min}
\; \mathcal{E}_{TOT} \left( \vec{x} , \overrightarrow{RS} \right), \label{eq:JOP1} \\
\shortintertext{s.t.:}
& T_{DAG} \leq \frac{1}{TH_{0}^{\left(MIN\right)}}  \equiv T_{DAG}^{\left(MAX\right)} , \label{eq:JOP2} \\
& 0  \leq  f_N  \leq  f_{N}^{\left(MAX\right)} , \quad N \in \mathcal{A} ,   \label{eq:JOP3} \\
& 0  \leq  R_{M \to N}  \leq  R_{M \to N}^{\left(MAX\right)} , \quad N \in BHS ,   \label{eq:JOP4} \\
& 0  \leq  R_{N \to M}  \leq  R_{N \to M}^{\left(MAX\right)} , \quad N \in BHS ,   \label{eq:JOP5} \\
& x_1 = x_V = M ,   \label{eq:JOP6} \\
& x_i \in \left\{ M,F_1,\dots,F_Q,C  \right\}, \quad i = 2,\dots,\left(V-1\right) .   \label{eq:JOP7}
\end{align}
\label{eq:JOP}
\end{subequations}

In the reported \textit{JOP} formulation, we have that: (i) the objective function in \eqref{eq:JOP1} is given by the (weighted) summation of the computing and network energy in \eqref{eq:E_Tot}. Hence, as stressed in \eqref{eq:JOP1}, its actual value jointly depends on the task and resource allocation vectors $ \vec{x} $ and $ \overrightarrow{RS} $, that, in turn, play the role of optimization variables; (ii) the constraint in \eqref{eq:JOP2} guarantees that the per-DAG execution time meets the (previously mentioned) bound on the minimum throughput required by the processed stream application; (iii) the box constraints in \eqref{eq:JOP3}, \eqref{eq:JOP4} and \eqref{eq:JOP5} account for the (aforementioned) limitations on the allowed computing and network resources; (iv) the constraints in \eqref{eq:JOP6} account for the fact that the first and last tasks of the DAG must be executed by the Mobile device (see the assumptions of Section \ref{sec:modeling} on the application DAG); and, (v) the last group of constraints in \eqref{eq:JOP7} forces each scalar component of the task allocation vector to take only and only one value over the discrete set: $ \left\{ M,F_1, \dots,F_Q,C \right\} $.

Before proceeding, two main remarks on the practical impacts of the reported \textit{JOP} formulation and the flexibility of the developed formal framework are in order.
\begin{remark}[\textit{On the flexibility of the developed formal framework}]\label{remark:remark1}
$ $
About the flexibility of the developed system framework, two main illustrative remarks may be of practical interest.

First, the developed framework featuring the FRAN, CRAN and Backhaul network segments of the overall technological platform of Fig.\ \ref{fig:multi-tier} gives rise to a \textit{fully meshed} network topology, that, in principle, may comprise up to: $ \left(Q+1\right) \left(Q+2\right) / 2 $ end-to-end (possibly, multi-hop) Transport-layer parallel connections. Hence, \textit{any} overlay network topology of interest may be actually built up by deleting the connections that are not actually utilized. By referring to the reported \textit{JOP} formulation, this may be done by: (i) setting to zero the maximum up/down throughput $ R_{M \to N}^{\left(MAX\right)} $ and $ R_{N \to M}^{\left(MAX\right)} $, $ N \in BHS $, in the constraints of \eqref{eq:JOP4} and \eqref{eq:JOP5} that correspond to the wireless connections that are not really supported by the considered FRAN and/or CRAN; and, (ii) posing to zero the throughput: $ R_{N_1 \to N_2} \equiv R_{N_2 \to N_1}$ in \eqref{eq:R_N1N2} of the two-way backhaul connections that are not instantiated by the Backhaul network segment of Fig.\ \ref{fig:multi-tier}. So doing, \textit{any} overlay network topology connecting \textit{any number} of (possibly, hierarchically organized) Fog tiers may be built up atop the underlying networked infrastructure of Fig.\ \ref{fig:multi-tier}.

Second, the Cloud and/or Fog Providers who manage the ecosystem of Fig.\ \ref{fig:multi-tier} may enforce authorization-based policies that forbid the mobile user to access the computing resources hosted by the Cloud node and/or some Fog nodes \cite{zhao2016}. In our framework, these access limitations may be taken into account by setting to zero the maximum allowed computing frequencies $ f_N^{\left(M\!A\!X\right)} $ in the \textit{JOP} constraints of \eqref{eq:JOP3} that correspond to the forbidden computing resources.
\hfill\ensuremath{\blacksquare}
\end{remark}

\begin{remark}[\textit{Eco-vs.-Mobile centric service models}]\label{remark:remark2}
$ $
The roles played by the (binary-valued) theta parameters previously introduced in \eqref{eq:theta} are unveiled by a direct inspection of the expressions of the network energy of Section \ref{ssec:ModelingEnergyInternode} and the objective function $ \mathcal{E}_{TOT} $  in \eqref{eq:E_Tot}. It points out that, by setting $ \theta_N, N \in \left\{ M,F,C \right\} $ at zero (resp., at the unit value), both the computing and network energy wasted by node $ N $ \textit{do not} contribute to the \textit{JOP} objective function in \eqref{eq:JOP1} and, then, they are \textit{not} subject to optimization.

On the basis of these formal considerations, two SaaS models may cover practical relevance\cite{zhao2016}, namely, the Eco-centric SaaS model and the Mobile-centric one. By definition, the (emerging) Eco-centric model is defined by setting:
\begin{align}
\theta_M = \theta_F = \theta_C  \equiv 1 ,  && \text{(\textit{Eco-centric} SaaS model)},
\label{eq:EcoCentric}
\end{align}
while, under the (more usual) Mobile-centric model, we have that:
\begin{align}
\theta_M = 1 ,  \quad  \theta_F = \theta_C  \equiv 0 ,   &&  \text{(\textit{Mobile-centric} SaaS model)} .
\label{eq:MobileCentric}
\end{align}
A comparison of the above defining relationships points out that: (i) under the Eco-centric model, the computing and network energy of \textit{all} nodes of the ecosystem of Fig.\ \ref{fig:multi-tier} contribute to the objective function in \eqref{eq:JOP1}; while, (ii) under the Mobile-centric model, \textit{only} the computing and network energy wasted by the mobile device are accounted for by \textit{JOP} objective function.

As a consequence, it is reasonable to expect that both the task/resource allocations and the corresponding energy consumption returned by the \textit{JOP} optimization process could be substantially different under these two service models. We anticipate the numerical results of Section \ref{ssec:PerformanceSensitivity} confirm this expectation and support the conclusion that an eco-friendly approach is capable of (drastically) reducing the total energy consumed by the overall Mobile-Fog-Cloud platform of Fig.\ \ref{fig:multi-tier} by somewhat penalizing the corresponding energy consumption experienced by the Mobile device.
\hfill\ensuremath{\blacksquare}
\end{remark}

\subsection{JOP feasibility}

The considered \textit{JOP} is a mixed-integer non-convex optimization problem, so that its solution resists, indeed, closed-form evaluation. Due to the same reason, the derivation of closed-form analytical conditions that are both necessary and sufficient for the feasibility of the \textit{JOP} seems to be very challenging. On the basis of these considerations, in the sequel, we focus on the derivation of a sufficient condition for the \textit{JOP} feasibility that is in \textit{closed-form}, and (which is the most) does \textit{not} require the explicit pre-evaluation of the (a priori unknown) \textit{JOP} solution.

In order to formally present this sufficient condition, some dummy positions are in order. Specifically, let:
\begin{equation}
s^{\left(MAX\right)}
\overset{\mathrm{def}}{=}
\underset{1 \leq i \leq V}{\max}
\left\{ s_i\right\}  \quad  \text{(bit)} ,
\label{eq:Smax}
\end{equation}
be the maximum size of the DAG tasks, and let:
\begin{equation}
w_{IN}^{\left(MAX\right)}
\overset{\mathrm{def}}{=}
\underset{1 \leq i \leq V}{\max}
\left\{  
\sum\nolimits_{j=1}^{V} { a_{ji} d_{ji} }
\right\}  \quad  \text{(bit)} ,
\label{eq:W_INmax}
\end{equation}
be the maximum volume of data that a task receives in input from the set of its preceding tasks (i.e., its parent tasks). Furthermore, let:
\begin{equation}
\beta^{\left(MIN\right)} =
\begin{cases}
1,
& \scalebox{0.9}{\text{under the \textit{SEQ} service discipline}},\\
\displaystyle \frac{\underset{1 \leq i \leq V}{\min}\left\{\phi_i\right\}}{\sum_{j=1}^{V}{\phi_{j}}},
& \scalebox{0.9}{\text{under the \textit{WPS} service discipline}} ,
\end{cases}
\label{eq:BetaMin}
\end{equation}
indicate the minimum fraction of the per-core computing frequency that is devoted to the execution of a single task. Hence, after denoting by:
\begin{equation}
Tmin_{N}^{\left(SER\right)}
\overset{\mathrm{def}}{=}
\frac{s^{\left(MAX\right)}}{n_N f_N^{\left(MAX\right)} \beta^{\left(MIN\right)}}
, \quad N \in \mathcal{A} ,
\label{eq:Tmin_Nser}
\end{equation}
the minimum service time at node $ N $ for the task of maximum size, let:
\begin{align}
Tmax^{\left(SER\right)} &
\overset{\mathrm{def}}{=}
\underset{N \in \mathcal{A}}{\max}  \left\{ Tmin_{N}^{\left(SER\right)} \right\} \nonumber \\
& \, \equiv
\frac{s^{\left(MAX\right)}}{ \beta^{\left(MIN\right)} \times \underset{N \in \mathcal{A}}{\min} \left\{ n_N f_N^{\left(MAX\right)} \right\} },
\label{eq:Tmax_ser}
\end{align}
be the maximum of the minimum service times, and let:
\begin{equation}
Tmax^{\left(NET\right)}
\overset{\mathrm{def}}{=}
\frac
{w_{IN}^{\left(MAX\right)} \! \left( 1 + \underset{N_1,N_2 \in \mathcal{A}}{\max} \left\{ \overline{NF}_{N_1 \to N_2} \right\} \right) }
{\underset{N_1,N_2 \in \mathcal{A}}{\min} \left\{R_{N_1 \to N_2}^{\left(MAX\right)}\right\}  } ,
\label{eq:Tmax_net}
\end{equation}
indicate the maximum of the minimum network times needed to transport the (previously defined) maximum volume of input data $ w_{IN}^{\left(M\!AX\right)} $ used for the execution of a task. On the basis of the above positions, it follows that the maximum execution time: $ T_{EXE}^{\left(M\!AX\right)} $ of \textit{any} task at \textit{any} computing node equates the summation:
\begin{equation}
T_{EXE}^{\left(M\!AX\right)} = Tmax^{\left(SER\right)}  +  Tmax^{\left(NET\right)} .  
\label{eq:T_ExeMax}
\end{equation}
In the Appendix \ref{appendix:A}, it is proved that a suitable exploitation of the above relationship leads to the following sufficient condition for the \textit{JOP} feasibility.
\begin{proposition}[\textit{Sufficient condition for the JOP feasibility}]\label{proposition:proposition1}
$ $
Let the previously reported assumptions on the behavior of $ T_{DAG} $ be met. Furthermore, let:
\begin{equation}
T_{DAG}^{\left(UP\right)}
\overset{\mathrm{def}}{=}
T_{DAG}
\left(
T_i^{\left(EXE\right)}  \equiv T_{EXE}^{\left(MAX\right)} , \, i=1,\dots,V
\right) ,
\label{eq:T_DAGup}
\end{equation}
be the value assumed by the DAG execution time $ T_{DAG} $ under the (worst) case in which all the task execution times: $ T_i^{\left(EXE\right)}, \,  i=1,\dots,V $, equate to the maximum one $ T_{EXE}^{\left(MAX\right)} $ in \eqref{eq:T_ExeMax}. Hence, the satisfaction of the following inequality:
\begin{equation}
T_{DAG}^{\left(UP\right)}   \leq
\frac{1}{TH_{0}^{\left(MIN\right)}} ,
\label{eq:T_DAGupLessThan}
\end{equation}
suffices to guarantee the \textit{JOP} feasibility. 
\hfill\ensuremath{\blacksquare}
\end{proposition}

In the sequel, we formally indicate by:
\begin{equation}
\left\{ \vec{x}^{*}  ,  \overrightarrow{RS}^{*}   \right\}  ,
\label{eq:SetOfJOPsolution}
\end{equation}
the solution of a feasible \textit{JOP}.

\subsection{JOP decomposition into RAP and TAP}

Two main formal features retained by the considered \textit{JOP} make the analytical evaluation of its solution in \eqref{eq:SetOfJOPsolution} challenging. First, due to the presence of product terms that jointly involve a number of (scalar) components of both optimization variables $ \vec{x} $ and $ \overrightarrow{RS} $ (see the energy expressions in Sections \ref{ssec:ModelingComputingEnergy} and \ref{ssec:ModelingEnergyInternode}), the objective function $ \mathcal{E}_{TOT} \left(\vec{x} , \overrightarrow{RS}\right) $ in \eqref{eq:JOP1} is not jointly convex in the involved optimization variables and, then, the \textit{JOP} is a \textit{non-linear} and \textit{non-convex} optimization problem. Second, the JOP embraces both real-valued $ \overrightarrow{RS} $ and discrete-valued $ \vec{x} $ optimization variables, so that it is a \textit{mixed-integer} optimization problem.

Hence, in order to cope with these challenges, in the sequel, we develop an (optimality-preserving) approach, that aims to hierarchically decompose the reported \textit{JOP} into the cascade of two inter-depending simpler optimization sub-problems, namely, the \textit{Resource Allocation Problem} \textit{(RAP)} and the \textit{Task Allocation Problem} \textit{(TAP)}. Interestingly enough, we anticipate that the \textit{RAP} optimization involves \textit{only} the real-valued resource allocation vector $ \overrightarrow{RS} $, while the \textit{TAP} optimization is carried out over \textit{only} the discrete-valued variable $ \vec{x} $.

Specifically, under any \textit{assigned} task allocation vector $ \vec{x} $, the \textit{RAP} is formally defined as the following constrained optimization problem in the real-valued optimization variable $ \overrightarrow{RS} $:
\begin{subequations}
\begin{flalign}
&  \underset{\overrightarrow{RS}}{\min}
\; \mathcal{E}_{TOT} \left( \vec{x} ; \overrightarrow{RS} \right),  \label{eq:JOP-RAP1} \\
\text{s.t.:} &  \text{
Eqs. \eqref{eq:JOP2}, \eqref{eq:JOP3}, \eqref{eq:JOP4}, \eqref{eq:JOP5}.
}  \label{eq:JOP-RAP2}
\end{flalign}
\label{eq:JOP-RAP}
\end{subequations}
%
Let:
\begin{equation}
\widetilde{\overrightarrow{RS}} \equiv
\widetilde{\overrightarrow{RS}}  \left(\vec{x}\right), \,
\text{and }
\widetilde{\mathcal{E}}_{TOT} \equiv
\widetilde{\mathcal{E}}_{TOT}  \left(\vec{x}; \widetilde{\overrightarrow{RS}}  \left(\vec{x}\right) \right),
\label{eq:RStilde}
\end{equation}
be the ($ \vec{x} $-depending) solution of the \textit{RAP} and the corresponding ($ \vec{x} $-depending) value attained by the objective function in \eqref{eq:JOP-RAP1} at the optimum.

Hence, the \textit{TAP} is formally defined as the following constrained optimization problem in the discrete-valued optimization variable $ \vec{x} $:
\begin{subequations}
\begin{flalign}
&  \underset{\vec{x}}{\min}
\; \widetilde{\mathcal{E}}_{TOT} \!
\left( \vec{x} ; \widetilde{\overrightarrow{RS}} \left(\vec{x} \right)\right),  \label{eq:JOP-TAP1} \\
\text{s.t.:} &  \text{
Eqs. \eqref{eq:JOP6} \text{ and } \eqref{eq:JOP7} .
}  \label{eq:JOP-TAP2}
\end{flalign}
\label{eq:JOP-TAP}
\end{subequations}
%
Let:
\begin{equation}
\left\{
\vec{\hat{x}} ,  \widehat{\overrightarrow{RS}}
\overset{\mathrm{def}}{=}
\widetilde{\overrightarrow{RS}} \left( \vec{\hat{x}} \right)
   \right\}  ,
\label{eq:SetOfTAPsolution}
\end{equation}
be the solution of the \textit{TAP} in \eqref{eq:JOP-TAP1} and \eqref{eq:JOP-TAP2}. Then, the following Proposition \ref{proposition:proposition2} proves that the performed \textit{JOP} decomposition is optimality-preserving.

\begin{proposition}[\textit{On the optimality of the developed JOP decomposition}]\label{proposition:proposition2}
$ $
Let the \textit{JOP} be feasible. Hence, its solution in \eqref{eq:SetOfJOPsolution} coincides with the one in \eqref{eq:SetOfTAPsolution} which is obtained by cascading the solutions of the \textit{RAP} and \textit{TAP}, that is,
\begin{equation}
\vec{x}^{*}  \equiv \vec{\hat{x}} , \,
\text{ and } \;
\overrightarrow{RS}^{*} \equiv  \widehat{\overrightarrow{RS}} .
\label{eq:JOPsolCoincideTAPsol}
\end{equation}
\end{proposition}

\begin{proof}
$ $
The proof of \eqref{eq:JOPsolCoincideTAPsol} relies on the following three formal properties retained by the performed \textit{JOP} decomposition.

First, without loss of optimality, the joint minimum in \eqref{eq:JOP1} may be decomposed into the cascade of the following two minima \cite{bazaraa2017}:
\begin{equation}
\underset{\vec{x},\overrightarrow{RS}}{\min} \:
\mathcal{E}_{TOT} \left( \vec{x} , \overrightarrow{RS} \right)  \equiv
\underset{\vec{x}}{\min} \;
\left\{
\underset{\overrightarrow{RS}}{\min} \: \mathcal{E}_{TOT} \left( \vec{x}, \overrightarrow{RS} \right)
\right\} .
\label{eq:JOPintoTwoMinima}
\end{equation}	

Second, all the constraints of the \textit{JOP} formulation in \eqref{eq:JOP2}, \eqref{eq:JOP3}, \eqref{eq:JOP4}, and \eqref{eq:JOP5} (resp., in \eqref{eq:JOP6} and \eqref{eq:JOP7}) that involve the resource allocation vector $ \overrightarrow{RS} $ (resp., the task allocation vector $ \vec{x} $) are taken into account in the \textit{RAP} formulation (resp., the \textit{TAP} formulation).

Third, for any assigned task allocation vector $ \vec{x} $, the \textit{TAP}'s objective function: $ \widetilde{\mathcal{E}}_{TOT} \left( \vec{x} ; \widetilde{\overrightarrow{RS}} \left(\vec{x}\right) \right) $ in \eqref{eq:JOP-TAP1} explicitly accounts for the optimal solution $ \widetilde{\overrightarrow{RS}} \left(\vec{x}\right) $ returned by the \textit{RAP}.
\end{proof}

\section{The developed RAP solving approach}
\label{sec:RAP}

The goal of this section is threefold. First, we develop the formal conditions under which the \textit{RAP} is convex and feasible. Second, we develop a formal approach for computing the \textit{RAP} solution in \eqref{eq:JOP-RAP} that is adaptive and capable to self-react to the mobility-induced changes of the operating environment of the ecosystem of Fig.\ \ref{fig:multi-tier}. Third, we point out some aspects related to the implementation and implementation complexity of the developed \textit{RAP} solving approach.

Specifically, in Appendix \ref{appendix:C}, the following conditions for the convexity of the \textit{RAP} are proved.

\begin{proposition}[\textit{On the convexity of the RAP}]\label{proposition:proposition3}
$ $
Let the assumptions of Section \ref{ssec:Per-DAGexecutiontimes} on $ T_{DAG} $ be met.  Furthermore, let us assume that the exponents of the dynamic computing and network power in \eqref{eq:E_N} and \eqref{eq:P_MtoNdynamic} meet the following inequalities:
\begin{equation}
\gamma_N  \geq  2 ,  \quad  N \in \mathcal{A} ,
\label{eq:gamma_Ninequality}
\end{equation}	
and
\begin{equation}
\xi_{\left(M,N\right)}^{\left(Tx\right)}  \geq 2, \;\,
\xi_{\left(M,N\right)}^{\left(Rx\right)}  \geq 2, \quad
\, N \in {BHS} .
\label{eq:xi_MNinequality}
\end{equation}	
Then, the \textit{RAP} in \eqref{eq:JOP-RAP1}, \eqref{eq:JOP-RAP2} is \textit{convex} in the resource allocation vector $ \overrightarrow{RS} $  under any assigned task allocation vector $ \vec{x} $. 
\hfill\ensuremath{\blacksquare}
\end{proposition}

Passing to consider the \textit{RAP} feasibility, let:
\begin{align}
\overrightarrow{RS}^{\left(M\!A\!X\right)}
\! \overset{\mathrm{def}}{=} &
\!
\left[
f_N^{\left(M\!A\!X\right)},  N \in A;
\right.
\nonumber \\
&
\left.
\! R_{{M \to N}}^{\left(MAX\right)} , \! R_{{N \to M}}^{\left(MAX\right)}, \! N \! \in \! {B\!H\!S}
\right]
\! \! \in \!\! \left( \mathbb{R}_{+} \right) ^ {3Q+4} \!\! ,
\label{eq:RSmax}
\end{align}
indicate the vector of the maximal allowed resources. Hence, in Appendix \ref{appendix:D}, the following necessary and sufficient condition is proved. 

\begin{proposition}[\textit{On the feasibility of the RAP}]\label{proposition:proposition4}
$ $
Let the assumptions of Section \ref{ssec:Per-DAGexecutiontimes} on $ T_{DAG} $ be met. Then, the \textit{RAP} in \eqref{eq:JOP-RAP1}, \eqref{eq:JOP-RAP2} is feasible \textit{if and only if} the following inequality holds:
\begin{equation}
TH_{0}^{\left(MIN\right)}  \times
T_{DAG} \left( \vec{x} ; \overrightarrow{RS}^{\left(MAX\right)} \right)
- 1 \leq 0 ,
\label{eq:TH_0mintimesT_DAGinequality}
\end{equation}	
where $ T_{DAG} \left( \vec{x} ; \overrightarrow{RS}^{\left(MAX\right)} \right) $ indicates the value assumed by the DAG execution time under the task and resource  allocation vectors $ \vec{x} $ and $ \overrightarrow{RS}^{\left(MAX\right)} $, respectively.
\hfill\ensuremath{\blacksquare}
\end{proposition}

Before proceeding, two explicative remarks are in order concerning the conditions reported in the above two Propositions \ref{proposition:proposition3} and \ref{proposition:proposition4}. First, thereinafter we assume that the inequalities in \eqref{eq:gamma_Ninequality}, \eqref{eq:xi_MNinequality} are met, so that the considered \textit{RAP} is guaranteed to be convex. Second, since the satisfaction of the inequality in \eqref{eq:TH_0mintimesT_DAGinequality} may depend on the considered $ \vec{x} $, in the sequel, we label a task allocation vector $ \vec{x} $ as \textit{RAP-feasible} if it meets the \textit{RAP} feasibility condition in \eqref{eq:TH_0mintimesT_DAGinequality}.

\subsection{Featuring the RAP solution}

Let us assume that the considered \textit{RAP} is convex and feasible. Hence, the general results reported, for example, by \textit{Theorems 4.3.7} and \textit{4.3.8} of \cite{bazaraa2017} guarantee that the Karush-Kuhn-Tucker (KKT) conditions are both necessary and sufficient for the evaluation of the \textit{RAP} solution, provided that the \textit{RAP} also meets the so-called Slater's qualification condition (see, for instance, \cite[Chapter 5]{bazaraa2017}). In this regard, in Appendix \ref{appendix:E}, the following formal result is proved.  

\begin{proposition}[\textit{Slater's qualification for the RAP}]\label{proposition:proposition5}
$ $
Let us assume that the feasibility condition in \eqref{eq:TH_0mintimesT_DAGinequality} is met with the \textit{strict} inequality under the considered task allocation vector $ \vec{x} $. Then, the Slater's qualification holds for the \textit{RAP} in \eqref{eq:JOP-RAP1}, and \eqref{eq:JOP-RAP2}.
\hfill\ensuremath{\blacksquare}
\end{proposition}

Under the assumption that the condition of Proposition \ref{proposition:proposition5} is met, we resort to a suitable application of the so-called \textit{Primal-Dual Solving Approach (PDSA)}, in order to evaluate the \textit{RAP} solution (see, for example, Chapter 10 of \cite{bazaraa2017}). In this regard, we note that the \textit{RAP} convexity allows to manage the box constraints in \eqref{eq:JOP3}-\eqref{eq:JOP5} on the maximal resources as implicit ones. Hence, under any assigned task allocation vector $ \vec{x} $, the resulting Lagrangian function $ \mathcal{L} \left( \overrightarrow{RS} , \lambda \right) $ (Joule) of the \textit{RAP} reads as follows:
\begin{align}
\mathcal{L} \left( \overrightarrow{RS} , \lambda \right)
\overset{\mathrm{def}}{=} & \:
\mathcal{E}_{TOT} \left( \vec{x} ; \overrightarrow{RS} \right)  \nonumber \\
&  \! \! + \! \lambda \!
\left(
TH_{0}^{\left(M\!I\!N\right)} \! \times \!
T_{D\!AG} \left( \vec{x} ; \overrightarrow{RS} \right)
- 1 
\right) \!  ,
\label{eq:Lagrangian}
\end{align}	
where the (scalar and non-negative) parameter $ \lambda $ (Joule) is the Lagrange multiplier associated to the (convex) constraint in \eqref{eq:JOP2}. Therefore, from a formal point of view, the constrained max-min optimization problem to be solved for the evaluation of the \textit{RAP} vector solution $ \widetilde{\overrightarrow{RS}} $  in \eqref{eq:RStilde} and the corresponding optimal Lagrange multiplier $ \tilde{\lambda} $ is the following one (see, for example, \cite[Chapter 6]{bazaraa2017}):
\begin{equation}
\underset{\lambda \geq 0}{\max}
\left\{
\underset{\vec{0} \leq \overrightarrow{RS} \leq \overrightarrow{RS}^{\left(MAX\right)}}{\min}
\left\{
\mathcal{L} \left( \overrightarrow{RS} , \lambda \right)
\right\} \right\} ,
\label{eq:LambdaTildeOptimal}
\end{equation}	
where $ \overrightarrow{RS}^{\left(M\!AX\right)} $ is the vector of the maximum available resources in \eqref{eq:RSmax}.

Now, let $ \overrightarrow{\nabla} \mathcal{L} \left( \overrightarrow{RS} , \lambda \right) $ be the $ \left( 3Q+5 \right) $-dimensional gradient vector of the Lagrangian function in \eqref{eq:Lagrangian} performed with respect to the $ \left( 3Q+4 \right) $ scalar components of the resource vector $ \overrightarrow{RS} $ (i.e., the vector of the so-called primal variables) and the (scalar) Lagrange multiplier $ \lambda $ (i.e., the so-called dual variable). Hence, \textit{Theorem 6.2.6} of \cite{bazaraa2017} guarantees that the solution of the max-min problem in \eqref{eq:LambdaTildeOptimal} (that is, the so-called saddle-point of the considered Lagrangian function) may be computed by performing the orthogonal projection onto the box set: $\left[\vec{0} , \overrightarrow{RS}^{\left(MAX\right)} \right]$ of the allowed resources of the vector solution of the following  $ \left( 3Q+5 \right) $-dimensional algebraic equation system:
\begin{equation}
\overrightarrow{\nabla}
\mathcal{L}
\left( \overrightarrow{RS} , \lambda \right) = \vec{0} .
\label{eq:NablaEquations}
\end{equation}	

\subsection{Developed adaptive solving approach}
\label{ssec:AdaptiveSolving}

Two main remarks about the solution:
\begin{equation}
\left\{
\widetilde{\overrightarrow{RS}}  ,  \tilde{\lambda}
\right\}  ,
\label{eq:NablaSolutionSet}
\end{equation}	
to the system of equations in \eqref{eq:NablaEquations} are in order. First, since the \textit{RAP} objective function $ \mathcal{E}_{TOT} $ in \eqref{eq:JOP-RAP1} and the constraint in \eqref{eq:JOP2} on the allowed $ T_{DAG} $ are nonlinear functions of the optimization variable $ \overrightarrow{RS} $, the resulting system of algebraic equations in \eqref{eq:NablaEquations} is nonlinear, and, in general, its solution resists closed-form evaluation. Second, even if it would possible to solve in closed-form the algebraic equations in \eqref{eq:NablaEquations} under some specific cases, the obtained closed-form solution should be re-evaluated \textit{from scratch} when the device mobility and/or the occurrence of network congestion phenomena induce abrupt (and, typically unpredictable) changes in the operating environment of the ecosystem of Fig.\ \ref{fig:multi-tier}. 

Therefore, in the sequel, we develop an iteration-based solving approach, in order to allow the solution in \eqref{eq:NablaSolutionSet} of \eqref{eq:NablaEquations} to \textit{self-react} to environmental changes. Specifically, the pursued approach allows us to iteratively evaluate the solution of \eqref{eq:NablaEquations} through the adaptive implementation of a suitable set of projected gradient-based primal-dual scaled iterations. As pointed out in \cite{bazaraa2017}, the primal-dual algorithm is an iterative procedure that updates on a per-step basis both the primal variables (i.e., the optimization variables) and the dual ones (i.e., the Lagrange multipliers associated to the underlying constraints), in order to guide the corresponding Lagrangian function towards its saddle-point. Hence, after introducing the dummy position:
\begin{equation}
\left[z\right]_{a}^{b}
\overset{\mathrm{def}}{=}
\max
\left\{
a,\min  \left\{z,b\right\}
\right\} ,
\label{eq:z_ab}
\end{equation}	
the $ \left(m+1\right) $-th updating of the $ l $-th scalar component: $ y_l ,  l=1, \dots , \left(3Q+4\right) $, of the resource vector $ \overrightarrow{RS} $ reads as in:
\begin{equation}
y_{l}^{\left(m+1\right)}  =
\left[
y_{l}^{\left(m\right)}  -  \psi_{l}^{\left(m\right)}
\left(
\frac{\partial  \mathcal{L}\left(\overrightarrow{RS}^{\left(m\right)},\lambda^{{\left(m\right)}}\right)}{\partial y_{l}}
\right)
\right]
_{0}^{y_{l}^{\left(M\!AX\right)}} ,
\label{eq:y_l,m+1}
\end{equation}	
and the $ \left(m+1\right) $-th updating of the Lagrange multiplier is dictated by the following iteration:
\begin{equation}
\lambda^{\left(m+1\right)}  =
\left[
\lambda^{\left(m\right)}  +  \xi^{\left(m\right)}
\frac
{\partial \mathcal{L}\left(\overrightarrow{RS}^{\left(m\right)},\lambda^{{\left(m\right)}}\right)}
{\partial \lambda}
\right]
_{0}^{+\infty} .
\label{eq:Lambda,m+1}
\end{equation}	
In the above iterations, we have that: (i) $ m \geq 0 $ is an integer-valued iteration index; (ii) $ \partial \mathcal{L}(\overrightarrow{RS}^{\left(m\right)},\lambda^{{\left(m\right)}}) / \partial y_l $ (resp., $ \partial \mathcal{L}(\overrightarrow{RS}^{\left(m\right)},\lambda^{{\left(m\right)}}) / \partial \lambda $) is the partial derivative of the Lagrangian function in \eqref{eq:Lagrangian} performed with respect to the $ l $-th (scalar) component $ y_l $ of the resource vector $ \overrightarrow{RS} $ (resp., with respect to the $ \lambda $ multiplier) and evaluated at iteration $ m $; (iii) $ y_l^{\left(MAX\right)} $ in \eqref{eq:y_l,m+1} is the corresponding maximum value allowed $ y_l $ (that is, the $ l $-th scalar component of the maximal resource vector $ \overrightarrow{RS}^{\left(MAX\right)} $ in \eqref{eq:RSmax}); and, (iv) $ \left\{ \psi_{l}^{\left(m\right)} , l = 1, \dots , \left(3Q+4\right) \right\} $ and $ \xi^{\left(m\right)} $ are non-negative time-variyng (i.e., $ m $-varying) step-sizes. Furthermore, according to the $\mathrm{\max}$-$\mathrm{\min}$ saddle-point relationship in \eqref{eq:LambdaTildeOptimal}, the minus (resp., plus) sign is present in \eqref{eq:y_l,m+1} (resp., \eqref{eq:Lambda,m+1}), so that \eqref{eq:y_l,m+1} (resp., \eqref{eq:Lambda,m+1}) features descending-gradient (resp., ascending-gradient) iterations.

In Appendix \ref{appendix:F}, we detail the formulas for the computation of the derivatives of the Lagrangian function involved in the iterations of \eqref{eq:y_l,m+1} and \eqref{eq:Lambda,m+1}.

\subsection{Design of the time-varying step-sizes and convergence property}
\label{ssec:DesignTimeVarying}

The peculiar feature shared by the iterations in \eqref{eq:y_l,m+1} and \eqref{eq:Lambda,m+1} is that they resort to time-varying step-sizes, in order to guarantee \textit{fast adaptation} of the corresponding primal and dual variables in response to abrupt changes of the operating environment of the ecosystem of Fig.\ \ref{fig:multi-tier}. For this purpose, the approach quite recently reported in \cite{peng2016} may be pursued. Specifically, two main formal results of \cite{peng2016} are relevant in our context. First, \textit{Theorem 3.3} of \cite{peng2016} proves that it is sufficient to update each step-size sequence on the basis of \textit{only} the corresponding primal-dual variable, in order to guarantee the asymptotic convergence to the global optimum of \textit{all} iterations in \eqref{eq:y_l,m+1} and \eqref{eq:Lambda,m+1}. Second, a suitable choice for updating each step-size is to set it proportionally to the \textit{squared value} of the corresponding primal-dual variable at each iteration step $ m $.

Hence, motivated by these formal results, we planned to implement the following ``up/down clipped'' relationships for the updating of the step-sizes in \eqref{eq:y_l,m+1} and \eqref{eq:Lambda,m+1}: 
\begin{equation}
\psi_{l}^{\left(m\right)}   \! = \!
\max \!
\left\{
\! a_{M\!AX} , \min \!
\left\{
\!  a_{M\!AX} \times y_l^{\left(M\!AX\right)} \! , \! \left( y_l^{\left(m\right)} \right)  ^{\!\!2}
\right\} \!\!
\right\} ,
\label{eq:psi_l,m}
\end{equation}	
and
\begin{align}
\xi^{\left(m\right)}  = 
\max 
& \left\{ 
a_{M\!AX} , \min 
\left\{
a_{M\!AX} \times \underset{l}{\max} \left\{y_l^{\left(M\!AX\right)}\right\} , \right. \right. \nonumber \\
&
\left. \left.
\left( \lambda^{\left(m\right)} \right)  ^{2}
\right\} 
\right\} .
\label{eq:xi,m}
\end{align}	

Interestingly enough, the role played in \eqref{eq:psi_l,m} and \eqref{eq:xi,m} by the introduced clipping factor: $ a_{M\!AX} $ is twofold. First, (i) it allows a fast reaction in response to abrupt (possibly, unpredicted and mobility induced) environmental changes; and, (ii) it speeds up the convergence to the global optimum of the iterations in \eqref{eq:y_l,m+1}, \eqref{eq:Lambda,m+1} by forbidding too small step-size values (see the outer $ \max(.) $ in \eqref{eq:psi_l,m}, \eqref{eq:xi,m}). Second, it avoids too strong oscillations of the underlying iterations around their steady-state values by clipping the maximum value allowed by each step-size (see the inner $ \min(.) $ in \eqref{eq:psi_l,m}, \eqref{eq:xi,m}). We anticipate that the numerical results of Section \ref{ssec:AdaptiveCapability} support the actual effectiveness of the performed design and also give practical insights about the right setting of the clipping factor $ a_{M\!AX} $.

\subsection{Implementation aspects and implementation complexity of the RAP iterations}

The pseudo-code of Algorithm \ref{algorithm:alg_1} details the ordered list of steps that are needed for the software implementation of the \textit{RAP} iterations in \eqref{eq:y_l,m+1} and \eqref{eq:Lambda,m+1}. In a nutshell, the pseudo-code receives in input the task allocation vector $ \vec{x} $ to be processed, and, after checking the \textit{RAP} feasibility condition of Proposition \ref{proposition:proposition4}, it performs $ I_{MAX} $ runs of the primal-dual iterations in \eqref{eq:y_l,m+1}, \eqref{eq:Lambda,m+1}. After completing the $ I_{MAX} $-th run, it returns both the attained resource allocation vector $ \widetilde{\overrightarrow{RS}} $ and the associated consumed energy $ \widetilde{\mathcal{E}}_{TOT} $ of \eqref{eq:RStilde}. If the \textit{RAP} would be infeasible, the code of Algorithm \ref{algorithm:alg_1} sets all the returned outputs at the infinite and halts its execution (see step $2$ of Algorithm \ref{algorithm:alg_1}).

Therefore, a direct inspection of this pseudo-code leads to three main insights about the related implementation complexity. First, the implementation complexity of the developed \textit{RAP} solving approach is \textit{fully independent} of the size of the underlying application DAG. Second, since, at each run, it scales with the number of updated primal-dual variables as (see \eqref{eq:y_l,m+1} and \eqref{eq:Lambda,m+1}) $ \mathcal{O} \left(3Q+5\right) $, the overall implementation complexity of the \textit{RAP} solving approach of Algorithm \ref{algorithm:alg_1} over $ I_{MAX} $  runs scales as: 
\begin{equation}
\mathcal{O} \left(  I_{MAX} \times \left( 3Q+5 \right) \right).
\label{eq:RAPcomplexity}
\end{equation}	

Third, the above relationship points out that the scaling behavior is \textit{linear} with respect to both the number $ I_{MAX} $ of carried out runs, and the number $ \left(Q+2\right) $ of computing nodes of the ecosystem of Fig.\ \ref{fig:multi-tier}.

\begin{algorithm*}[ht!]
\caption{--- Computing the \textit{RAP} solution}
\label{algorithm:alg_1}
\textbf{Input}: Task allocation vector $ \vec{x} $.\\
\textbf{Output}: \parbox[t]{\textwidth}
{
Resource allocation vector $\widetilde{\overrightarrow{RS}}$;\\
Associated consumed energy $\widetilde{\mathcal{E}}_{TOT}$.
}
{\color{white} blank row}\Comment{Feasibility test}
\begin{algorithmic}[1]

\If{the \textit{RAP} feasibility condition in \eqref{eq:TH_0mintimesT_DAGinequality} fails}
\State Set $ \widetilde{\overrightarrow{RS}} $ and $ \widetilde{\mathcal{E}}_{TOT} $ to infinite;
\State \textbf{return} $\widetilde{\overrightarrow{RS}}$ and $\widetilde{\mathcal{E}}_{TOT}$;
\EndIf

\Comment{Iterative phase}

\For{$ m=0:(I_{MAX}-1) \;$}
\State Compute the set of Lagrangian derivatives involved by \eqref{eq:y_l,m+1} and \eqref{eq:Lambda,m+1};
\State Compute the step-sizes in \eqref{eq:psi_l,m} and \eqref{eq:xi,m};
\State Update the set of resource variables $ \left\{ y_l \in \widetilde{\overrightarrow{RS}} \right\} $ through \eqref{eq:y_l,m+1};
\State Update the Lagrange multiplier $ \lambda $ through \eqref{eq:Lambda,m+1};
\State Update $ \widetilde{\mathcal{E}}_{TOT} $ through \eqref{eq:E_Tot};
\EndFor
\State \textbf{return} $\widetilde{\overrightarrow{RS}}$ and $\widetilde{\mathcal{E}}_{TOT}$.
\end{algorithmic}
\end{algorithm*}

\section{The developed TAP solving approach}
\label{sec:TAP}

The solution in \eqref{eq:SetOfTAPsolution} of the \textit{TAP} in \eqref{eq:JOP-TAP1} and \eqref{eq:JOP-TAP2} resists closed-form evaluation due to the following three main reasons. First, the \textit{TAP} is a discrete optimization problem in the task allocation vector $ \vec{x} $ in \eqref{eq:taskallocation}, so that, unlike the \textit{RAP}, its solution cannot be approached by resorting to the (aforementioned) gradient-based KKT conditions. Second, due to the presence of product terms of unit-step functions that involve multiple scalar components of the optimization variable $ \vec{x} $, the resulting \textit{TAP} objective function: $ \widetilde{\mathcal{E}} \! \left( \vec{x}; \widetilde{\overrightarrow{RS}} \left(\vec{x}\right) \right) $ in \eqref{eq:JOP-TAP1} is not convex. Hence, numerical routines for the evaluation of the solution of convex discrete optimization problems (as, for example, CVX, i.e., the MATLAB software for disciplined Convex Programming) cannot be applied.

On the basis of these considerations, in the remaining part of this section, we address these challenges by developing a version of the Genetic Algorithm (GA) that is devised on an \textit{ad hoc} basis, tailored to the peculiar features of the \textit{TAP} to be solved.

\subsection{The proposed adaptive Genetic-based solving approach}

In the last years, the GA paradigm has been applied with good success for approaching the solutions of a number of task scheduling, resource consolidation and resource migration discrete optimization problems under various delay and energy-induced constraints (see, for example, \cite{scarpiniti2018} and \cite[Chapter 8]{pierson2015}).

Roughly speaking, the GA is a meta-heuristic for iterative search, that simulates the natural evolution process. In general, the GA requires that a population (of size $ P\!S $) of discrete-valued vectors $ \left\{ \vec{x}_k, k=1,\dots,P\!S \right\} $, representing candidate solutions, evolves towards better solutions by leveraging suitable crossover and mutation functions. The goodness of each candidate: $ \vec{x}_k $  is measured through an (application-depending) fitness function $ f\!it\left(\vec{x}_k\right) $. At each generation, the fitness of each element of the current population is evaluated and a set of elements is selected by comparing their fitness. Specifically, a fraction $ C\!F $ of the ``\textit{best}'' elements of the current population is recombined through crossover, while  the remaining ``worst'' elements are modified by randomly mutating each one over $ MN $ randomly selected positions. Therefore, after selecting the ``best'' $ P\!S $ elements from the obtained set of all crossed over and mutated elements, a new population is formed. This last is used in the next iteration of the GA. The GA algorithm halts when either a maximum number $ G_{MAX} $ of generations (i.e., iterations) is carried out, or a good fitness level is reached by the \textit{best} element $ \vec{x}_{BEST} $ of the generated populations.             

By design, the \textit{Adaptive Genetic Task Allocation Strategy (A-GTA-S)} proposed in this paper for approaching the \textit{TAP} solution retains the following three main characteristics.

First, each individual $ \vec{x} $ of a population is a $ V $-tuple task allocation vector, that, in turn,  is defined according to \eqref{eq:taskallocation}.

Second, the fitness $ f\!it\left(\vec{x}\right) $ of each individual equates to the inverse of the energy $ \widetilde{\mathcal{E}}_{TOT} \left(\vec{x}\right) \equiv \widetilde{\mathcal{E}}_{TOT} (\vec{x};\widetilde{\overrightarrow{RS}}(\vec{x}))$  in \eqref{eq:RStilde} returned by the solution of the \textit{RAP}, that is,
\begin{equation}
f\!it\left(\vec{x}\right)
\overset{\mathrm{def}}{=}
1 / \widetilde{\mathcal{E}}_{TOT} \left(\vec{x}\right), \quad \left(\text{Joule}^{-1}\right).
\label{eq:fitFunctionOfx}
\end{equation}	
This definition reflects the fact that, in our framework, the ``best'' task allocations are those that require less energy, in order to be sustained.

Third, from a formal point of view, the proposed \textit{A-GTA-S} is an example of \textit{elitary} GA \cite{pierson2015}, i.e., it guarantees that the final returned solution $ \vec{x}_{BEST} $ is the \textit{global best} over the set of all computed $ G_{MAX} $ generations.
 
\medskip\noindent\textit{Pseudo-code of the proposed A-GTA-S} -- A pseudo-code of the proposed \textit{A-GTA-S} is reported in Algorithm \ref{algorithm:alg_2}. In order to facilitate its actual software implementation, the reported pseudo-code details also the main required data structures, i.e., the (dummy) matrices $ [Poplist] $, $ [Childlist] $ and $ [Mutationlist] $ of Algorithm \ref{algorithm:alg_2}. According to the GA description reported at the beginning of this section, the input data of Algorithm \ref{algorithm:alg_2} are the (previously introduced) parameters: $ PS $, $ CF $, $ G_{MAX} $ and $ MN $, while the returned output is the triplet:
\begin{equation}
\left[ \vec{x}_{BEST}, \,
\overrightarrow{RS}_{BEST}, \,
\mathcal{E}_{BEST} \right],
\label{eq:solutionOfAGTAS}
\end{equation}	
of the task allocation vector, resource allocation vector and associated consumed energy. In the sequel, we refer to the triplet in \eqref{eq:solutionOfAGTAS} as the solution of the \textit{A-GTA-S}.

\begin{algorithm*}[ht!]
\caption{--- Pseudo-code of the proposed A-GTA-S}
\label{algorithm:alg_2}
\textbf{Input:} \parbox[t]{\textwidth}
{
Population size $ PS $;\\
Fraction $ CF $ of the crossed over population;\\
Number $ G_{MAX} $ of performed generations;\\
Number $ MN $ of mutated components.
}\\[1ex]
\textbf{Output:} \parbox[t]{\textwidth}
{
Best task allocation vector $ \vec{x}_{BEST} $;\\
Best resource allocation vector $ \overrightarrow{RS}_{BEST} $;\\
Best consumed energy $ \mathcal{E}_{BEST} $.
}
{\color{white} blank row}\Comment{Initialization phase}
\begin{algorithmic}[1]

\State Generate a random initial list: $ \left\{ \vec{x}_{k} , k=1,\dots,PS \right\} $ of task allocation vectors and store it into the (dummy) matrix $ [Poplist] $ on a per-row basis;
\For{$ k=1:PS $}
\State Compute  $ \widetilde{\overrightarrow{RS}} \left(\vec{x}_k\right) $ and: $ \widetilde{\mathcal{E}}_{TOT} \! \left( \vec{x}_k; \widetilde{\overrightarrow{RS}} \left(\vec{x}_k\right) \right) $ in \eqref{eq:RStilde} by running Algorithm \ref{algorithm:alg_1} under $ \vec{x}_k \in [Poplist] $;
\State Store the obtained $ \widetilde{\overrightarrow{RS}} \left(\vec{x}_k\right) $ and $ \widetilde{\mathcal{E}}_{TOT} \! \left( \vec{x}_k; \widetilde{\overrightarrow{RS}} \left(\vec{x}_k\right) \right) $ into the $ k $-th row of the matrix $ [Poplist] $;    
\EndFor
\State Sort the row of the matrix $ [Poplist] $ for increasing values of the energy of its elements; 
\State Set the number $ Cross \overset{\mathrm{def}}{=} [CF \times PS] $ of the elements of the matrix $ [Poplist] $ to be crossed over at each generation;

\Comment{Iterative phase}

\For{$ j=1:G_{MAX} $}
\State Perform the pair-wise crossover of the first \textit{Cross} elements of the matrix $ [Poplist] $ by calling $ (Cross/2) $ times 
\Statex{\hspace*{3em}the \textit{Crossover} function in Algorithm \ref{algorithm:alg_3}};
\State Store the obtained crossed over elements in the (dummy) matrix $ [Childlist] $ on a per-row basis;
\State Randomly mutate in $ MN $ positions the last $ (PS-Cross) $ elements of the matrix $ [Poplist] $ by calling the \textit{Mutation} 
\Statex{\hspace*{3em}function in Algorithm \ref{algorithm:alg_4} and store the mutated elements into the (dummy) matrix \textit{[Mutationlist]} on a per-row basis};
\State Compute and store the resource allocation vector and the corresponding consumed energy in \eqref{eq:RStilde} of each element of 
\Statex{\hspace*{3em}the matrices $ [Childlist] $ and $ [Mutationlist] $ through $ PS $ runs of the \textit{RAP} solver in Algorithm \ref{algorithm:alg_1}};
\State Copy the first \textit{Cross} elements of the matrix $ [Poplist] $ and the full matrices $ [Childlist] $ and $ [Mutationlist] $ into the 
\Statex{\hspace*{3em}(dummy) matrix $ [Candidatelist] $};
\State Sort the $ (PS+Cross) $ elements of the matrix $ [Candidatelist] $ for increasing values of their consumed energy;
\State Copy the first $ PS $ elements of the matrix $ [Candidatelist] $ into the matrix $ [Poplist] $;
\If{energy of the first element of the matrix $ [Poplist] $ is lower than the current value of $ \mathcal{E}_{BEST} $}
\State Copy the first element (i.e., the first row) of the matrix $ [Poplist] $ into $  \vec{x}_{BEST} $, $ \overrightarrow{RS}_{BEST} $ and $ \mathcal{E}_{BEST} $ ; 
\EndIf
\EndFor
\State \textbf{return} $  \vec{x}_{BEST} $, $ \overrightarrow{RS}_{BEST} $ and $ \mathcal{E}_{BEST} $.
\end{algorithmic}
\end{algorithm*}

Algorithm \ref{algorithm:alg_3} and Algorithm \ref{algorithm:alg_4} detail the related pseudo-codes of the \textit{Crossover} and \textit{Mutation} functions called by \textit{A-GTA-S} at steps $ 9 $ and $ 11 $ of Algorithm \ref{algorithm:alg_2}. 

Shortly, the implemented \textit{Crossover} function: (i) generates a random pointer to the location index at which the crossover is performed (see step $ 1 $ of Algorithm \ref{algorithm:alg_3}); and, then, (ii) carries out the corresponding swapping of the task allocation input vectors: $ \overrightarrow{Parent_1} $ and $ \overrightarrow{Parent_2} $ (see steps $ 2 $ and $ 3 $ of Algorithm \ref{algorithm:alg_3}), so to return the crossed-over task allocation output vectors: $ \overrightarrow{Child_1} $ and $ \overrightarrow{Child_2} $ (see step $ 4 $ of Algorithm \ref{algorithm:alg_3}). The underlying rationale is to allow the crossed over output vectors to (hopefully) inherit the ``good'' fitness properties retained by the corresponding input vectors.

\begin{algorithm*}[ht!]
\caption{--- Pseudo-code of the implemented Crossover function}
\label{algorithm:alg_3}
\textbf{Input:} Two allocation vectors: $ \overrightarrow{Parent_1} $ and $ \overrightarrow{Parent_2} $ to be crossed over.\\
\textbf{Output:} Two crossed-over allocation vectors $ \overrightarrow{Child_1} $ and $ \overrightarrow{Child_2} $. \\
{\color{white} blank row}\Comment{Initialization phase}
\begin{algorithmic}[1]
\State Generate a random integer $ I $ over the interval  $ [2,V-1] $;

\Comment {Perform the swapping operation}

\State Copy the first $ I $ elements of $ \overrightarrow{Parent_1} $ and $ \overrightarrow{Parent_2} $ into the first $ I $ positions of $ \overrightarrow{Child_1} $ and $ \overrightarrow{Child_2} $, respectively;
\State Copy the last $ (V-I) $ elements of $ \overrightarrow{Parent_1} $ and $ \overrightarrow{Parent_2} $ into the last  $ (V-I) $ positions of $ \overrightarrow{Child_1} $ and $ \overrightarrow{Child_2} $, respectively;
\State \textbf{return} $ \overrightarrow{Child_1} $ and $ \overrightarrow{Child_2} $.
\end{algorithmic}
\end{algorithm*}

The opposite goal is, indeed, pursued by the \textit{Mutation} function of Algorithm \ref{algorithm:alg_4}. In fact, after generating a random vector of pointers to the locations to be mutated (see step $ 1 $ of Algorithm \ref{algorithm:alg_4}) and a random vector of mutated values (see step $ 2 $ of Algorithm \ref{algorithm:alg_4}), this function copies the generated mutated values at the pointed locations of the input vector $ \vec{x} $ (see steps $ 4 $-$ 6 $ of Algorithm \ref{algorithm:alg_4}). Afterwards, it returns the mutated output vector $ \overrightarrow{mx} $ (see step $ 7 $ of Algorithm \ref{algorithm:alg_4}). Hence, the underlying rationale is to attempt to improve (if possible) the ``bad'' fitness of the input vector $ \vec{x} $ by randomly changing a number $ MN $ of its randomly selected  components (see step $ 4 $ of Algorithm \ref{algorithm:alg_4}).

\begin{algorithm*}[ht!]
\caption{--- Pseudo-code of the implemented Mutation function}
\label{algorithm:alg_4}
\textbf{Input:} \parbox[t]{\textwidth}
{
Task allocation vector $ \vec{x} $ to be mutated;\\
Number $ M\!N $ of the scalar elements to be mutated.\vspace*{4pt}
}
\textbf{Output:} The mutated task allocation vector $ \overrightarrow{mx} $.\\
{\color{white} blank row}\Comment{Initialization phase}
\begin{algorithmic}[1]
\State Generate an $ M\!N $-tuple random vector: $ \vec{\ell} $ that points the positions to be mutated. Each element of $ \vec{\ell} $ is an integer number over
\Statex{\hspace*{1em}the interval $ [2, \, V-1] $};
\State Generate an $ M\!N $-tuple random vector: $ \overrightarrow{o\!f\!f\!set} $ that stores the mutated values. Each scalar element of $ \overrightarrow{o\!f\!f\!set} $ takes value over 
\Statex{\hspace*{1em}the discrete set $ \mathcal{A} $ in \eqref{eq:MFC}};
\State Copy $ \vec{x} $ into $ \overrightarrow{mx} $.

\Comment {Perform the mutations}

\For{$ j=1:M\!N $}
\State Copy the $ j $-th element of $ \overrightarrow{o\!f\!f\!set} $ into the position of $ \overrightarrow{mx} $ that is pointed by the $ j $-th element of $ \vec{\ell} $;  
\EndFor
\State \textbf{return} $ \overrightarrow{mx} $.
\end{algorithmic}
\end{algorithm*}

\subsection{Peculiar features of the proposed A-GTA-S and its computational complexity}

An examination of the reported Algorithms \ref{algorithm:alg_2}, \ref{algorithm:alg_3} and \ref{algorithm:alg_4} unveils the following three main \textit{peculiar} features of the proposed \textit{A-GTA-S}.

First, since it leverages the \textit{RAP} solution for the evaluation of the fitness in \eqref{eq:fitFunctionOfx} of each candidate task allocation vector (see steps $ 3 $ and $ 12 $ of Algorithm \ref{algorithm:alg_2}), the proposed \textit{A-GTA-S} inherits, by design, the \textit{adaptive capability} natively retained by the \textit{RAP} solution (see Section \ref{ssec:AdaptiveSolving} and related remarks). This feature makes its utilization appealing in the mobile scenario of Fig.\ \ref{fig:multi-tier}, where both the throughput of the wireless connections and the computing capabilities of the discovered server nodes may undergo abrupt and unpredictable changes. 

Second, formally speaking, we cannot claim that, in general, the solution in \eqref{eq:solutionOfAGTAS} returned by the A-GTA-S coincides with the optimal one in \eqref{eq:SetOfTAPsolution} of the \textit{TAP}. However, the pursued elitary approach guarantees that the fitness of the solution returned by the \textit{A-GTA-S} does not decrease for increasing values of the product: $ PS \times G_{MAX} $ (see steps $ 16-18 $ of Algorithm \ref{algorithm:alg_2}). This formal property assures, in turn, that the \textit{A-GTA-S} solution in \eqref{eq:solutionOfAGTAS} asymptotically approaches the optimal one for growing value of the product: $ PS \times G_{MAX} $.

Third, it has been experienced that elitary GAs may be too conservative, i.e., their solutions may be trapped by local minima at finite values of the product: $ PS \times G_{MAX} $ (see, for example, Chapter $ 8 $ of \cite{pierson2015} and references therein). Hence, in order to effectively cope with this potential drawback, the proposed \textit{Mutation} function of Algorithm \ref{algorithm:alg_4} randomizes \textit{both} the locations of the elements to be mutated \textit{and} the corresponding mutated values (see steps $ 1 $, $ 2 $ and $ 5 $ of Algorithm \ref{algorithm:alg_4}). In this regard, we have numerically experienced that values of the number $ M\!N $ of the mutated element in step $ 4 $ of Algorithm \ref{algorithm:alg_4} of the order of about $ V/2 $ maximize the chances of the overall \textit{A-GTA-S} escaping local  minima and reach global (or, at least, quasi global) minima (see Section \ref{sec:Performancetests}). So doing, we anticipate that the simulation results of Section \ref{sec:Performancetests} support the conclusion that the proposed \textit{A-GTA-S} is capable of attaining quasi-optimal energy performance at a reasonable low computational complexity. In this regard, an examination of Algorithm \ref{algorithm:alg_2} points out that:
\begin{enumerate}[i.]
\item the solution of the \textit{RAP} is invocated $ ((G_{MAX}+1) \times PS) $ times (see steps $ 3 $ and $ 12 $ of Algorithm \ref{algorithm:alg_2}). Furthermore, the computational cost of each \textit{RAP} invocation is given by \eqref{eq:RAPcomplexity}; and,
\item $ (G_{MAX}+1) $ sorting operations are carried out over sets composed by $ (PS+Cross) $ elements (see step $ 14 $ of Algorithm \ref{algorithm:alg_2}). Furthermore, the computational cost of each sorting operation is of the order:
$
\mathcal{O}
\left(
\left(PS+Cross\right)
\times
\log_{2} \left(PS+Cross\right)
\right)
$.
\end{enumerate}

As a consequence, the overall computational complexity of Algorithm \ref{algorithm:alg_2} is of the order: 
\begin{align}
\mathcal{O} &
\left(
\left(G_{MAX}+1\right)  \times PS \times
\left(
\left(
\left(3Q+5\right) \times I_{MAX} \right)
\right.
\right.
\nonumber \\
&
\left.
\left.
+ \left(PS+Cross\right) \times \log_{2}\left(PS+Cross\right)
\right)
\right) ,
\label{eq:OrderGA}
\end{align}	
and, then, it scales as:
\begin{equation}
\mathcal{O} 
\left(
PS \times G_{MAX} \times  \left(3Q+5\right)  \times  I_{MAX}
\right)  ,
\label{eq:OrderGAscalesAt}
\end{equation}	
for large values of the product: $ G_{MAX} \times  I_{MAX} \times PS $. We anticipate that, in Section \ref{sec:Performancetests}, the formula in \eqref{eq:OrderGAscalesAt} is exploited, in order to investigate about the right trade-off between the two contrasting requirements of quasi-optimal energy performance and low implementation complexity of the proposed \textit{A-GTA-S}.

\section{EcoMobiFog -- The proposed technological platform}
\label{sec:Technologicalplatform}

The goal of this section is to sketch the main building blocks, offered services and control flows of \textit{EcoMobiFog}, i.e., the proposed networked computer architecture for the actual support of the developed \textit{JOP} solution of Sections \ref{sec:RAP} and \ref{sec:TAP}.

Toward this end, we begin to note that, in the ecosystem of Fig.\ \ref{fig:multi-tier}, the Mobile device queries the connected Cloud and/or Fog nodes for additional computing resources, while the last may cooperate through data exchange. This means, in turn, that, in our framework: 
\begin{enumerate}[i.]
\item the Mobile-to-Fog and Mobile-to-Cloud interactions are of \textit{Client-Server} type, with the Mobile device (resp., the Cloud and Fog nodes) that plays (resp., play) the role of Client (resp., Servers); and,
\item the Fog-to-Fog and Fog-to-Cloud interactions needed for cooperative task executions are of \textit{Peer-to-Peer} type.  
\end{enumerate}

\begin{figure*}[ht!]
\centering
\includegraphics[width=0.65\textwidth]{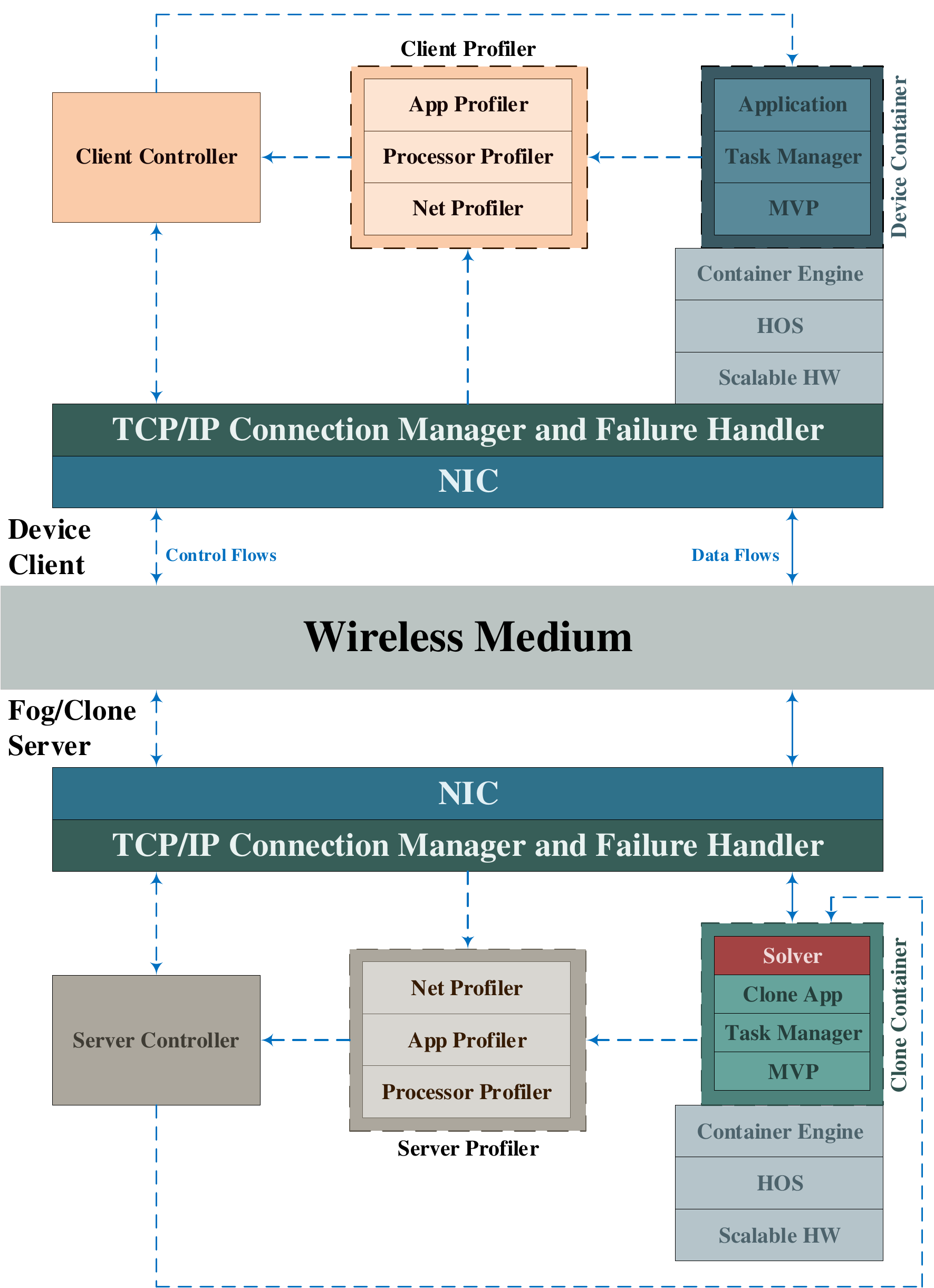}
\caption{The proposed \textit{EcoMobiFog} technological platform for the support of the developed \textit{JOP} solution. Continuous (resp., dotted) arrowed paths denote data (resp., control) TCP/IP flows.}
\label{fig:Client-ServerArch}
\end{figure*}

According to this observation, the proposed \textit{EcoMobiFog} technological platform for the support of the developed task offloading framework is composed of two main parts, i.e., a Mobile client part and a Cloud/Fog server part. Their main building blocks and exchanged control flows are sketched in Fig.\ \ref{fig:Client-ServerArch}. Specifically, according to Fig.\ \ref{fig:Client-ServerArch}, \textit{EcoMobiFog} relies on \textit{six} main agents that support the instantiated containers, namely, \textit{Profilers}, \textit{Task Managers}, \textit{Connection Managers and Failure Handlers}, \textit{Solvers}, \textit{Controllers} and \textit{Control Flows}. In the sequel, we describe the supported services, as well as their mutual interactions.

\medskip\noindent\textit{Profilers} -- Each container hosted by the Mobile, Cloud and Fog nodes is equipped with the corresponding Profiler. The function of the Profiler is to provide context-awareness for the associated container by performing in real-time the measurements of a number of context parameters, so to assist the \textit{Controller} when needed. For this purpose, each Profiler is, in turn, composed of an \textit{Application Profiler}, a \textit{Processor Profiler} and a \textit{Network Profiler}.

The \textit{Application Profiler} tracks the execution state of the tasks processed by the underlying Container Engine of Fig.\ \ref{fig:Client-ServerArch} by monitoring: the current set of tasks under execution, the related execution times $ {T_{i,N}^{EXE}}$, and the sizes $ {d_{ij}} $ of the input data required for the task executions (see Sections \ref{sec:modeling} and \ref{sec:ServiceScheduling}). 

The \textit{Processor Profiler} works on a per-virtual processor basis, and (when needed) communicates the performed measurements to the \textit{Controller}. In our framework, these measurements include: the per- virtual processor computing frequency $ f_N $, its corresponding maximum value $ f_N^{\left(MAX\right)} $, and the currently consumed computing energy $ \mathcal{E}_N $ (see Section \ref{ssec:ModelingComputingEnergy}).

The \textit{Network Profiler} collects the network state of the connections currently sustained by the active NICs, in order to detect in real-time possible network changes. Specifically, jobs of a Network Profiler is to measure the currently available set of maximum throughput:$ \{ R_{N_1 \to N_2}^{\left(MAX\right)} \} $ of the managed TCP/IP connections, as well as the actual parameters:  $ \{  \Omega^{\left(Tx\right)} , \Omega^{\left(Rx\right)} , \xi^{\left(Tx\right)} , \xi^{\left(Rx\right)} \} $ of the corresponding per-connection power profiles (see Section \ref{ssec:ModelingEnergyInternode}). From time to time, it returns to the \textit{Controller} the set $ \{ \mathcal{E}_{N_1 \to N_2} \} $ of the energy actually consumed by the managed connections. 

\medskip\noindent\textit{Task Managers} -- Each container deployed at the Mobile, Cloud and Fog nodes is equipped with the corresponding Task Manager. Its main function is to implement the adopted Task Service Discipline (like, for example, the \textit{SEQ} or \textit{WPS} ones of Section \ref{ssec:Servicedisciplines}) and, then, guarantee that the allocated tasks are executed in a compliant way.

\medskip\noindent\textit{Connection Managers and Failure Handlers} -- Each container deployed at the Mobile, Cloud and Fog nodes is equipped with the corresponding \textit{Connection Manager} and \textit{Failure Handler} module. Its goal is to manage the operations of the multiple NICs that equip the hosting node. For this purpose, the \textit{Failure Handler} module: (i) manages the time out-induced re-transmissions of the lost TCP segments; (ii) detects connection failure events and alerts the associated Network Profiler; and, (iii) measures in real-time the set: $ \{ \overline{NF}_{N_1 \to N_2} \} $ of the failure rates of the ongoing connections (see Section \ref{ssec:ModelingEnergyInternode}). In a parallel way, the \textit{Connection Manager} performs: (i) \textit{setup and tear-down} of the engaged TCP/IP connections; and, (ii) \textit{node discovery}, i.e., it monitors the strength of the signal received by each NIC, in order to discover the presence of proximate nodes. About this last functionality, we further specify that, after discovering a new node, the Connection Manager updates the information concerning the discovered node (like, connection bandwidth, IP address, computing capacity and similar), and communicates the acquired information to the Profiler. However, since node discovery through NIC monitoring can potentially increase the resulting network energy consumption, in our framework, we plan that this operation is carried out from time to time, for example, on a periodical or event-driven basis.  

\medskip\noindent\textit{Solvers} -- The goal of the \textit{Solver} module of Fig.\ \ref{fig:Client-ServerArch} is to implement the developed numerical procedure for the real-time evaluation of the solution of the \textit{JOP} (see Sections \ref{sec:RAP} and \ref{sec:TAP}).  However, since the Mobile device is resource limited and its implemented functionalities must be held at the minimum, in the proposed framework, \textit{only} the containers at the server nodes (i.e., at the Fog and Cloud nodes) are equipped with the Solver module and, then, \textit{only} these containers are capable of computing the \textit{JOP} solution in real-time. This means that, when the Controller at the mobile device decides to launch a task offloading procedure, it connects to an available Fog or Cloud node (thereinafter referred to as the \textit{solving server}), and, then, queries it to solve the underlying \textit{JOP}. After computing the \textit{JOP} solution, the solving server returns to both the Mobile device and all the other involved server nodes the computed solution in \eqref{eq:SetOfJOPsolution}. Afterward, the involved Controllers self-synchronize, and, then, start the distributed execution of the underlying application.

\medskip\noindent\textit{Controllers} -- Each container hosted by the Mobile, Cloud and Fog nodes is equipped with a corresponding \textit{Controller}. In our framework, it fulfils the following three main functions. First, it performs \textit{intra-node} synchronization, i.e., it disciplines the actions of all other modules hosted by the node. Second, on the basis of the information received by the associated Profiler and Solver modules, it decides whether and where to offload the application tasks, and, then, dispatches the tasks to the appropriate nodes. Third, it performs \textit{inter-node} synchronization, i.e., it cooperates with the Controllers hosted by the other nodes that are involved in the DAG execution, in order to guarantee the synchronized execution of the adopted inter-node Task Scheduling discipline (see the \textit{STS} and \textit{PTS} disciplines of Section \ref{ssec:Per-DAGexecutiontimes}). 

\medskip\noindent\textit{Control Flows} -- The aforementioned inter-node synchronization is supported by a set of end-to-end control flows, that connect (in a peer-to-peer fashion) all the Mobile, Cloud and Fog nodes of the ecosystem of Fig.\ \ref{fig:multi-tier}. Since these flows must be reliable, we assume that they are sustained by TCP/IP connections (see the dotted arrowed paths of Fig.\ \ref{fig:Client-ServerArch}). However, in our framework, all the server nodes are already equipped with a copy of the DAG to be executed (see the assumptions of Section \ref{sec:modeling}). As a consequence, the exchanged control data are reduce to: (i) the information data about the solution in \eqref{eq:SetOfJOPsolution} of the \textit{JOP} that has been computed by the solving server; and, (ii) the barrier-based synchronization signaling, in order to guarantee that the distributed execution of the application DAG is compliant with the adopted inter-node Task Scheduling discipline. On the basis of these design choices, it is expected that the aggregate throughput of all involved control flows remains limited and does not significantly interfere with the throughput of the corresponding data flows (see the continuous arrowed paths of Fig.\ \ref{fig:Client-ServerArch}).

\section{Performance tests and comparisons}
\label{sec:Performancetests}

The goal of this section is to numerically check and compare the adaptive capability and energy-vs.-delay performance of the proposed \textit{A-GTA-S} under multiple test operating scenarios. In order to suitably present the related multi-facet aspects, we organized this section according to the following roadmap. After describing in Section \ref{ssec:SimulatedPlatform} the simulated environment and the considered benchmark strategies, the adaptive capability of the proposed \textit{A-GTA-S} is checked in Section \ref{ssec:AdaptiveCapability}, while Section \ref{ssec:TuningTheEnergy} investigates its performance sensitivity on the population size, number of performed generations and fraction of the crossed over population. Sections \ref{ssec:TaskResourceAllocation} and \ref{ssec:PerformanceComparisons} compare the resource/task allocations and energy consumption of the proposed \textit{A-GTA-S} with respect to the corresponding ones of the considered benchmark strategies, while Section \ref{ssec:SensitivityA-GTA-S} checks the corresponding performance sensitivity on the computing-to-communication ratios of the considered benchmark DAGs. The goal of Section \ref{ssec:PerformanceSensitivity} is to check and compare the energy-vs.-delay performance of the proposed \textit{A-GTA-S} under the Eco and Mobile-centric service models in \eqref{eq:EcoCentric} and \eqref{eq:MobileCentric}. Finally, Section \ref{ssec:AverageEnergy} carries out comparative tests of the \textit{A-GTA-S} average energy performance when, due to the device mobility, the availability of the underlying Mobile-Fog connections undergoes random ON-OFF variations.

\subsection{Simulated platform and pursued comparison methodology}
\label{ssec:SimulatedPlatform}

\begin{figure*}[ht!]
\centering
\includegraphics[width=0.7\textwidth]{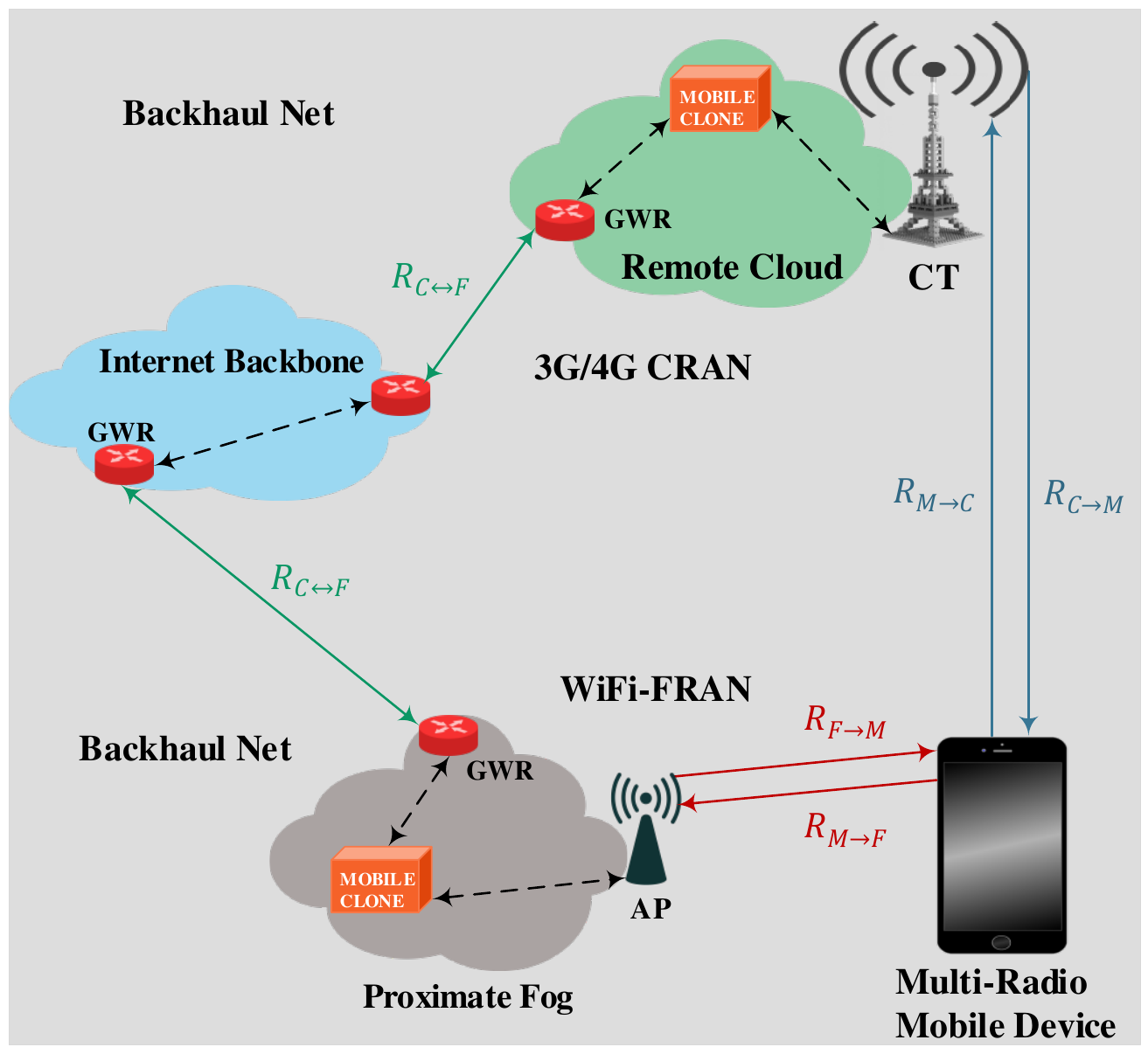}
\caption{The simulated three-tier Mobile-Fog-Cloud platform}
\label{fig:Mobile-Fog-CloudPlatform}
\end{figure*}

The most part of the numerical results and performance comparisons available in the open literature refers to three-tier Mobile-Fog-Cloud offloading systems, in which single Fog nodes are involved (see, for example, \cite{yousefpour2019} and references therein). Hence, in order to present comparative simulation results, aligned with the current literature, we have simulated the Mobile-Fog-Cloud platform sketched in Fig.\ \ref{fig:Mobile-Fog-CloudPlatform}. From a formal point of view, it is the instance of the general ecosystem of Fig.\ \ref{fig:multi-tier} that is obtained by setting $ Q=1 $ in \eqref{eq:MFC}.

Regarding the simulated platform of Fig.\ \ref{fig:Mobile-Fog-CloudPlatform}, the following two main introductory remarks are in order.

First, unless otherwise stated, in the sequel, it is understood that: (i) the settings of the main involved system parameters are the ones listed in the last column of final Table \ref{table:11}, where the communication technology sustaining the cellular Mobile-Cloud (resp., the short-range Mobile-Fog) connection is the 4G-LTE one \cite{altamimi2015} (resp., the IEEE 802.11b WiFi one \cite{xiao2014}); (ii) the reported simulated results refer to the Eco-centric model of \eqref{eq:EcoCentric} under the \textit{SEQ} and \textit{STS} service task scheduling disciplines of \eqref{eq:ServTimeSequen} and \eqref{eq:T_DAG-STS}, respectively; and, (iii) the simulated maximum allowed per-DAG execution time $ T_{DAG}^{\left(MAX\right)} $ in \eqref{eq:JOP2} ranges over the interval: $ 0.3 - 2.4 $ (s).

Second, the performed simulations have been carried out by using the recent \textit{VirtFogSim} toolbox \cite{Scarpiniti2019}, atop a hardware execution platform equipped with: (i) an Intel 10-core i9-7900X processor; (ii) a GPU ZOTAC GetForce GTX 1070; (iii) an SSD with 512 GB plus an HHD with 2 TB; and, (iv) 32 GB of RAM DDR 4. Furthermore, the simulation code exploits the release R2018a of MATLAB as software execution environment.

\medskip\noindent\textit{Pursued comparison methodology and benchmark strategies} -- Since the main peculiar features of the proposed \textit{A-GTA-S} solving approach is that it affords the two related problems of the adaptive resource \textit{and} task allocations in a \textit{joint} way. It is reasonable to pursue a methodology in which three main families of benchmark strategies are considered for comparison purpose.

Specifically, a first benchmark family of published strategies focuses on the optimization of the task allocation under \textit{fixed} (i.e., \textit{not} optimized) allocation of the underlying computing and/or networking resources. For this purpose, various versions of more or less optimized heuristic and meta-heuristic task allocation algorithms have been proposed as building blocks of Middleware technological platforms, like, for example, \textit{MAUI} \cite{cuervo2010}, \textit{CloneCloud} \cite{chun2011}, \textit{mCloud} \cite{zhou2017}, and \textit{StreamCloud} \cite{yang2013}, just to name but a few. In a nutshell, all these solutions focus on the constrained minimization of the task execution times (or the consumed energy) under \textit{fixed} resource allocation vectors.

A second family of published offloading strategies affords the dual problem of the constrained optimal allocation of the computing and/or networking resources under \textit{fixed} (i.e., not optimized) task offloading. These strategies mainly refer to single/multi-user two-tier MEC environments, and they are reviewed, for, example, in \cite[Section 3]{mao2017}.

A last family of contributions aim at pursuing (when it is possible) an optimal solving approach that directly exploits the \textit{Exhaustive Search} \textit{(ES)} \cite{mao2017}. Taking this approach, the \textit{full} space of the allowed task allocation vectors is thoroughly searched for the global best task and optimization of the allocated resources is also performed under each searched task allocation vector. Although it is guaranteed that the returned solution is the optimal one, the computational complexities of these ES-based solving approaches scale in an \textit{exponential} way with the size $ V $ of the underlying application DAGs, so that these approaches are typically applied to ``toy'' examples.

Overall, on the basis of these considerations, in the sequel, the performance of the proposed \textit{A-GTA-S} will be compared with the corresponding ones of the following five benchmark strategies:

\begin{table*}[htb]
\caption{Computational complexity of the simulated task offloading strategies; A:= Adaptive.}
\label{table:03}
\centering
\setlength{\extrarowheight}{0.5ex}
\renewcommand{\arraystretch}{1.5}
\begin{tabular}{lcl}
\toprule
\rowcolor[HTML]{BABFC6}
\textbf{Simulated Strategy} & & \textbf{Asymptotic computational complexity} \\[1ex]
\midrule
\rowcolor[HTML]{F2F5F9}
\textit{OTA-S}  &  & $ \mathcal{O} \left( PS \times G_{MAX} \right) $  \\
\rowcolor[HTML]{DCE1E8}
\textit{A-OF-S} &  & $\mathcal{O} \left( 8 \times I_{MAX} \right)$  \\
\rowcolor[HTML]{F2F5F9}
\textit{A-OC-S} & & $\mathcal{O} \left( 8 \times I_{MAX} \right)$  \\
\rowcolor[HTML]{DCE1E8}
\textit{A-OM-S} & & $\mathcal{O} \left( 8 \times I_{MAX} \right)$  \\
\rowcolor[HTML]{F2F5F9}
\textit{A-ES-S} & & $\mathcal{O} \left( 8 \times 3^{\left( V - 2 \right)}  \right)$  \\
\rowcolor[HTML]{DCE1E8}
\textit{A-GTA-S} & & $ \mathcal{O} \left( 8 \times PS \times G_{MAX} \right) $  \\
\bottomrule
\end{tabular}
\renewcommand{\arraystretch}{1.0}
\end{table*}

\begin{enumerate}[1.]
\item \textit{Only Task Allocation Strategy (OTA-S)}: by definition, the \textit{OTA-S} runs the GA of Algorithm \ref{algorithm:alg_2} under the fixed (i.e., not optimized and time-invariant) maximal resource allocation vector in \eqref{eq:RSmax}. This means, in turn, that in steps $ 3 $ and $ 12 $ of Algorithm \ref{algorithm:alg_2}, \textit{OTA-S} evaluates the energy consumed by the involved task allocation vectors by simply setting the required resource allocation vector $ \overrightarrow{RS} $ at the maximal one $ \overrightarrow{RS}^{MAX} $. In the sequel, the energy consumed under \textit{O-TAS} is indicated as $ \mathcal{E}_{OTA-S} $. \textit{O-TAS} is representative of the (aforementioned) first family of state-of-the-art offloading strategies;
\item \textit{Adaptive Only Fog Strategy (A-OF-S)}: \textit{A-OF-S} assumes that the first and last tasks of the underlying DAG are executed by the Mobile device, while all the other tasks are executed by the Fog node of Fig.\ \ref{fig:Mobile-Fog-CloudPlatform}.  Hence, by definition, \textit{A-OF-S} fixes the task allocation vector $ \vec{x} $ as in:
\begin{equation}
\vec{x} \equiv
\vec{x}_{FOG}
\overset{\mathrm{def}}{=}
\left[
M,F, \dots , F,M
\right]   ,
\label{eq:x_FOG}
\end{equation}	
and, then, invokes the \textit{RAP} solution of Algorithm \ref{algorithm:alg_1}, in order to compute the optimal resource allocation vector $ \overrightarrow{RS}_{A-OF-S}  $ and the resulting consumed energy $ \mathcal{E}_{A-OF-S} $ under $ \vec{x}_{FOG} $. \textit{A-OF-S} falls into the second family of offloading strategy;
\item \textit{Adaptive Only Cloud Strategy (A-OC-S)}: \textit{A-OC-S} assumes that the first and last tasks of the underlying DAG are executed by the Mobile device, while all the other tasks are executed by the Cloud node of Fig.\ \ref{fig:Mobile-Fog-CloudPlatform}. Hence, by definition, \textit{A-OC-S} fixes the task allocation vector $ \vec{x} $ as in:
\begin{equation}
\vec{x} \equiv
\vec{x}_{CLD}
\overset{\mathrm{def}}{=}
\left[
M,C, \dots , C,M
\right]   ,
\label{eq:x_CLD}
\end{equation}	
and, then, invokes the \textit{RAP} solution of Algorithm \ref{algorithm:alg_1}, in order to compute the optimal resource allocation vector $ \overrightarrow{RS}_{A-OC-S} $ and the resulting consumed energy $ \mathcal{E}_{A-OC-S} $ under $ \vec{x}_{CLD} $. \textit{A-OC-S} is an instance of the second family of offloading strategy;
\item \textit{Adaptive Only Mobile Strategy (A-OM-S)}: \textit{A-OM-S} assumes that all tasks are executed (when it is feasible) by the Mobile device of Fig.\ \ref{fig:Mobile-Fog-CloudPlatform}. Hence, by definition, \textit{A-OM-S} puts:
\begin{equation}
\vec{x} \equiv
\vec{x}_{MOB}
\overset{\mathrm{def}}{=}
\left[
M,M, \dots , M,M
\right]   ,
\label{eq:x_MOB}
\end{equation}	
and, then, invokes the \textit{RAP} solution of Algorithm \ref{algorithm:alg_1}, in order to compute the optimal resource allocation vector $ \overrightarrow{RS}_{A-OM-S} $ and the resulting consumed energy $ \mathcal{E}_{A-OM-S} $ under $ \vec{x}_{MOB} $. The \textit{A-OM-S} may be considered as a (limit) instance of the second  family of task allocation strategies;
\item \textit{Adaptive Exhaustive Search Strategy (A-ES-S)}: by design, \textit{A-ES-S}: (i) generates all the $ 3^{(V-2)} $ task allocation vectors; (ii) evaluates the corresponding optimal resource allocation vectors and consumed energy through $ 3^{(V-2)} $ calls of the \textit{RAP} solving procedure in Algorithm \ref{algorithm:alg_1}; and, finally: (iii) reports the task allocation vector $ \vec{x}_{A-ES-S} $, and resource allocation vector: $ \overrightarrow{RS}_{A-ES-S} $ that correspond to the minimum consumed energy: $ \mathcal{E}_{A-ES-S} $. By design, these last coincide with the solution in \eqref{eq:SetOfJOPsolution} of the \textit{JOP}, that is, it is guaranteed that:
\begin{equation}
\vec{x}_{A-ES-S} \equiv  \vec{x}^{*} ,
\overrightarrow{RS}_{A-ES-S} \equiv \overrightarrow{RS}^{*}  ,
\mathcal{E}_{A-ES-S} \equiv  \mathcal{E}^{*}  .
\label{eq:AESS}
\end{equation}	
\end{enumerate}

The computational complexities of the five considered benchmark strategies span a large range. Hence, in order to carry out fair performance-vs.-computational complexity comparisons, Table 3 reports these complexities, together with the corresponding one of the proposed \textit{A-GTA-S} (see the last row of Table \ref{table:03}). The reported formulas  account for the fact that, under the simulated scenario of Fig.\ \ref{fig:Mobile-Fog-CloudPlatform}, the computing complexity of the \textit{RAP} scales as (see \eqref{eq:RAPcomplexity} with $ Q=1 $): $ \mathcal{O} \left(8 \times I_{MAX}\right) $.

\begin{figure*}[htb]
\centering
\subfloat[]
{\label{fig:DAGs-a}
\includegraphics[width=.78\textwidth]{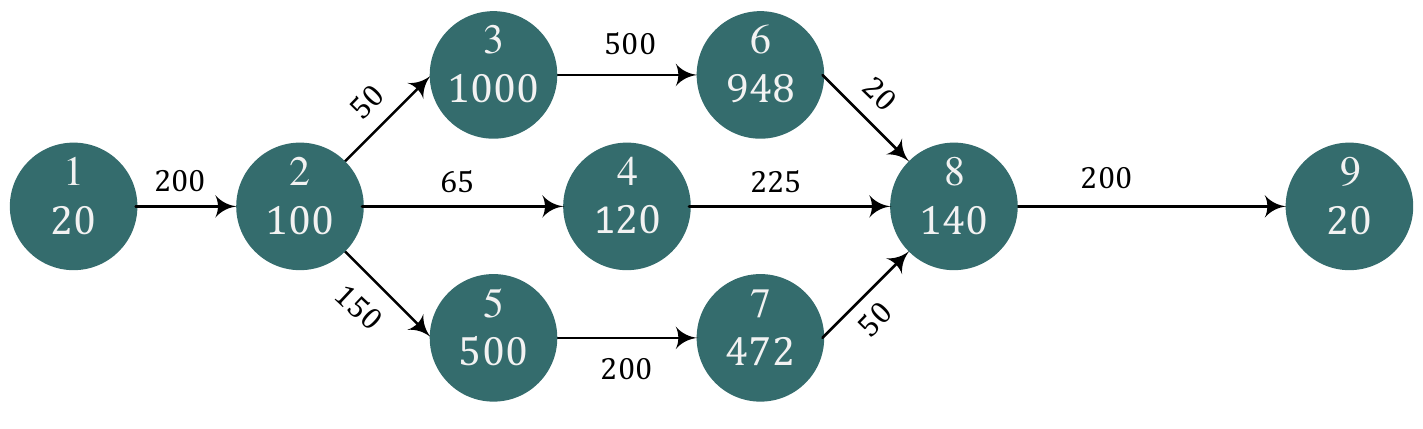}}
\hfill
\subfloat[]
{\label{fig:DAGs-b}
\includegraphics[width=.78\textwidth]{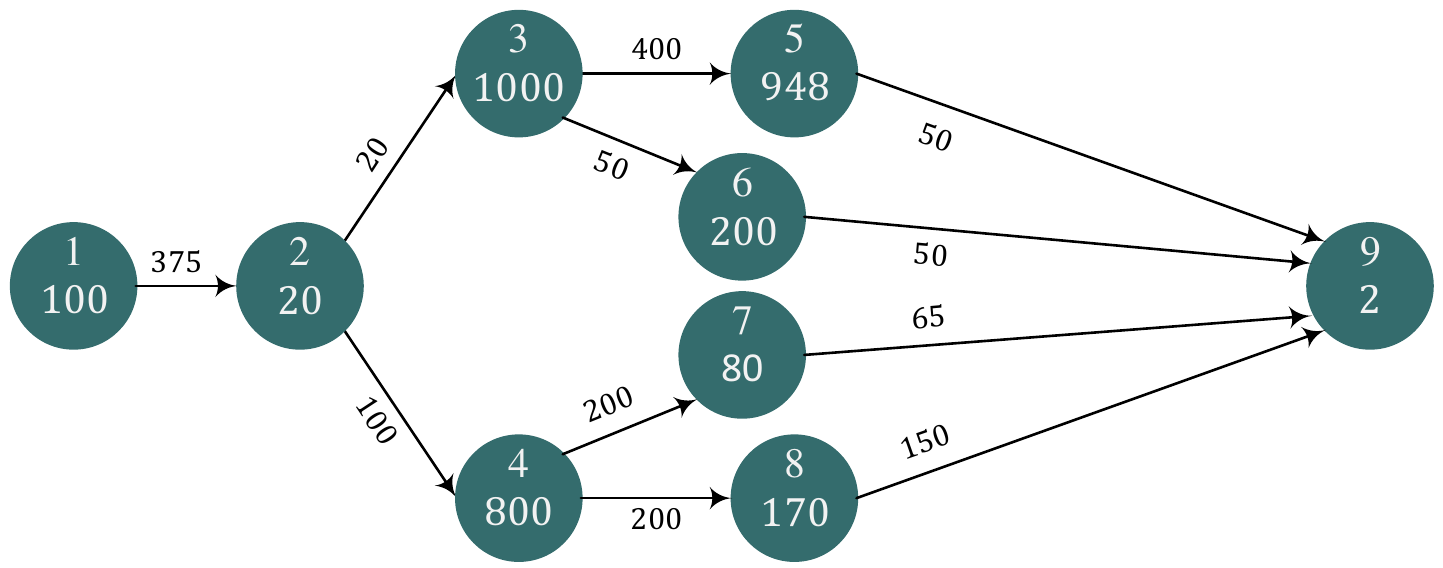}}
\hfill
\subfloat[]
{\label{fig:DAGs-c}
\includegraphics[width=.78\textwidth]{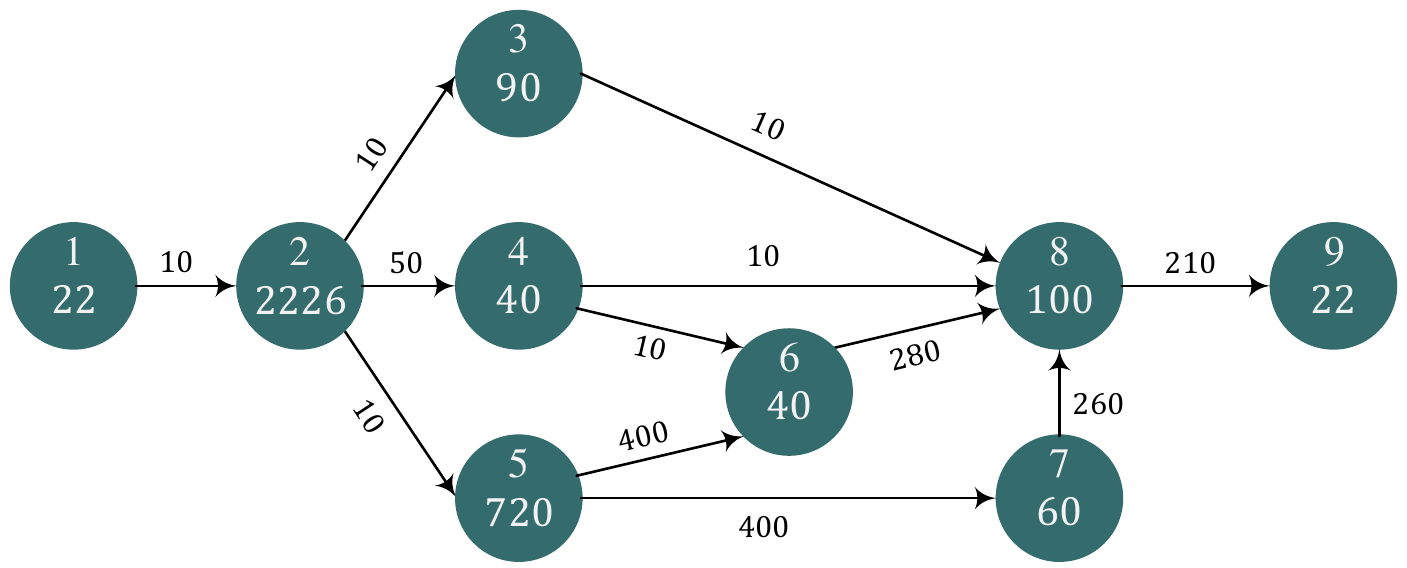}}	
\caption{A first set of test DAGs: (a) DAG1: Mesh topology; (b) DAG2: Tree topology; and, (c) DAG3: Hybrid topology. Task workloads and edge weights are in (kbit). In all cases, the summations of the task workloads and edge weights equate to $ 3.32 $ (Mbit) and $ 1.66 $ (Mbit), respectively.}
\label{fig:DAGs}
\end{figure*}

\medskip\noindent\textit{A first set of test DAGs} -- Fig.\ \ref{fig:DAGs} sketches the topology of a first set of test DAGs. The rationale behind their considerations is that they exhibit the three basic topologies (i.e., the Mesh, Tree and Hybrid topologies) typically retained by DAGs for mobile stream applications \cite{andrade2014,yang2013}. In this regard, we point out that: (i) in order to carry out fair performance comparisons, the summation of task workloads (resp., edge weights) of all DAGs of Fig.\ \ref{fig:DAGs} are normalized to $ 3.32 $ (Mbit) (resp., $ 1.66 $ (Mbit)); and, (ii) a more complex (but, more application-specific) real-world DAG will be introduced in Fig.\ \ref{fig:DAG4} and used for the final tests of Sections \ref{ssec:PerformanceSensitivity} and \ref{ssec:AverageEnergy}.

\begin{figure*}[htb]
\centering
\includegraphics[width=.75\textwidth]{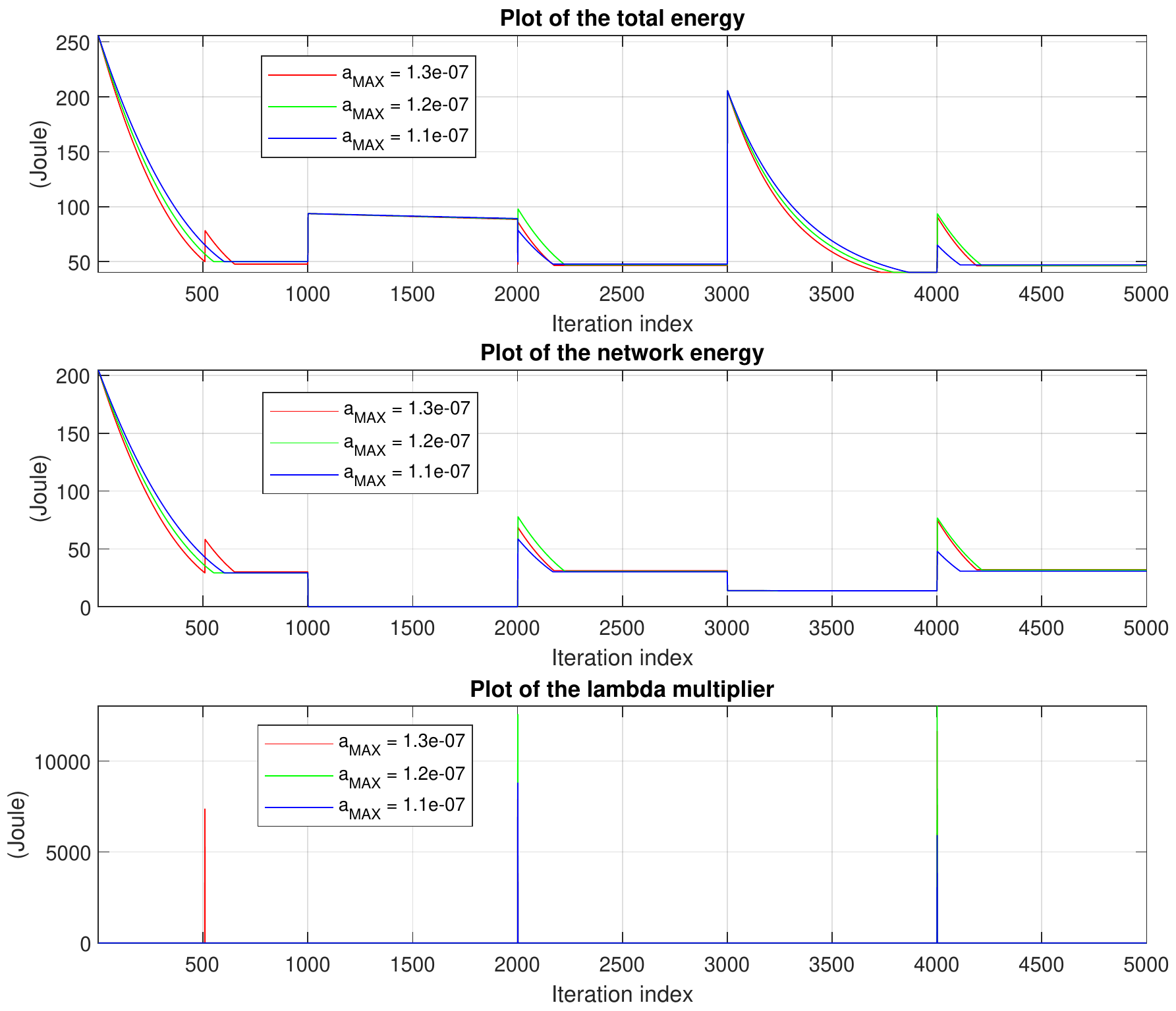}
\caption{Tracking capability of the developed \textit{RAP} function under DAG1 at $ T_{DAG}^{\left(MAX\right)} = 0.3 $ (s). (Top) Time behavior of $ \widetilde{\mathcal{E}}_{TOT} $; (middle) Time behavior of $ \widetilde{\mathcal{E}}_{NET} $; and, (bottom) Time behavior of $ \tilde{\lambda} $ multiplier.}
\label{fig:TrackingDAG1}
\end{figure*}

\subsection{Testing the adaptive capability of the developed resource allocation strategy}
\label{ssec:AdaptiveCapability}

The goal of this section is to test the sensitivity of the tracking capability of the \textit{RAP} iterations in \eqref{eq:y_l,m+1} and  \eqref{eq:Lambda,m+1} on the clipping factor $ a_{MAX} $ of \eqref{eq:psi_l,m} and \eqref{eq:xi,m}, as well as to evaluate the required convergence time $ I_{MAX} $ (in multiple of the iteration index $ m $). 

For this purpose, we have simulated a time-varying testing scenario in which, due to the mobility of the device of Fig.\ \ref{fig:Mobile-Fog-CloudPlatform}, both the Mobile-Fog up/down WiFi maximal throughput and the corresponding task allocation vectors undergo abrupt (and unpredicted) changes at the iteration indexes $ m=1,1000,2000,3000 \text{ and } 4000 $. Specifically, in the simulated setting, we have that: (i) at $ m=1 $, the up/down cellular (resp., WiFi) connections are turned ON (resp., turned OFF) and all tasks are allocated to the Cloud node, i.e. (see \eqref{eq:x_CLD}), $ \vec{x} \equiv \vec{x}_{CLD} $; (ii) at $ m=1000 $, the up/down WiFi connections are turned ON and all tasks are allocated to the Mobile node, i.e. (see \eqref{eq:x_MOB}), $ \vec{x} \equiv \vec{x}_{MOB} $; (iii)  at $ m=2000 $, the WiFi connections are still turned OFF and all tasks are re-allocated to the Cloud node, i.e., $ \vec{x} \equiv \vec{x}_{CLD} $; (iv) at $ m=3000 $, the up/down  WiFi connections are turned ON once time, and all tasks are allocated to the Fog node, i.e. (see \eqref{eq:x_FOG}), $ \vec{x} \equiv \vec{x}_{FOG} $;  and, finally, (v) at $  m=4000 $, the up/down  WiFi connections are definitively turned OFF and all tasks are re-migrated to the Cloud node, i.e., $ \vec{x} \equiv \vec{x}_{CLD} $. After each change of the setup environment, the \textit{RAP} solution is computed by running the iterations in \eqref{eq:y_l,m+1} and \eqref{eq:Lambda,m+1}, in order to properly re-allocate both the per-clone computing frequencies at the Mobile-Fog-Cloud nodes and the corresponding up/down Cellular-WiFi throughput.      

\begin{figure*}[htb]
\centering
\includegraphics[width=.75\textwidth]{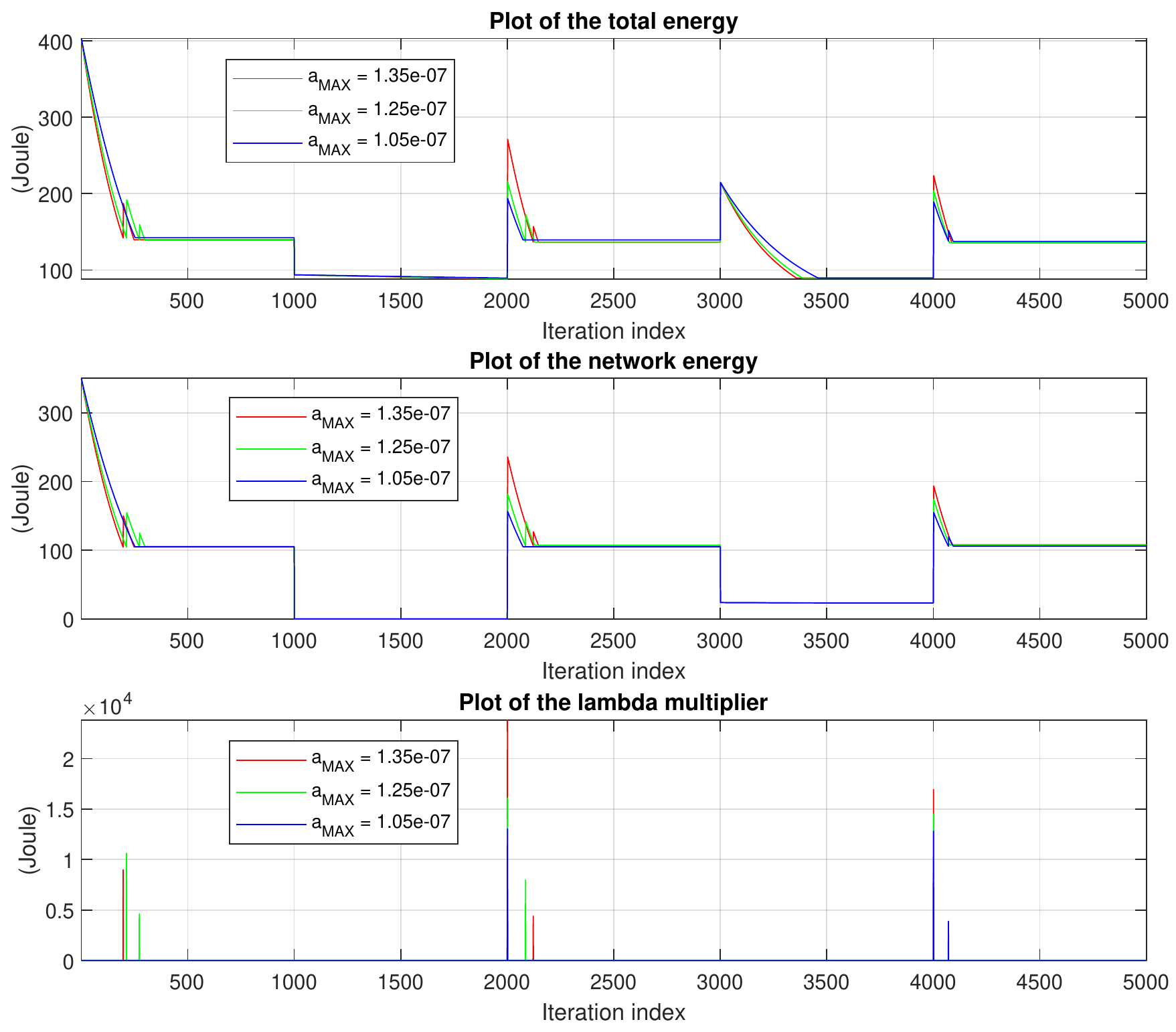}
\caption{Tracking capability of the developed \textit{RAP} solution under DAG2 at $ T_{DAG}^{\left(MAX\right)} = 0.3 $ (s). (Top) Time behavior of $ \widetilde{\mathcal{E}}_{TOT} $; (middle) Time behavior of $ \widetilde{\mathcal{E}}_{NET} $; and, (bottom) Time behavior of $ \tilde{\lambda} $ multiplier.}
\label{fig:TrackingDAG2}
\end{figure*}

\begin{figure*}[htb]
\centering
\includegraphics[width=.75\textwidth]{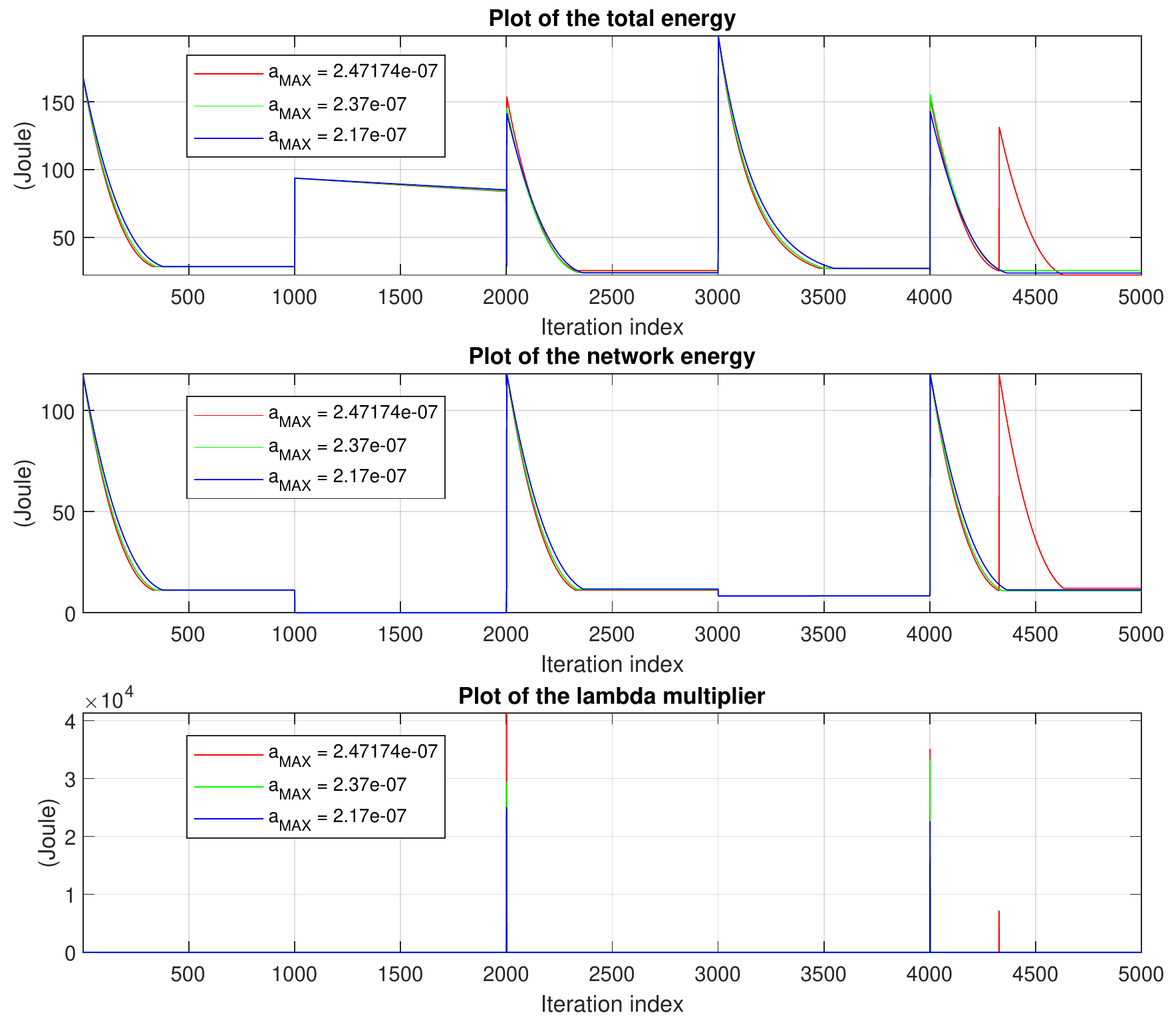}
\caption{Tracking capability of the developed \textit{RAP} function under DAG3 at $ T_{DAG}^{\left(MAX\right)} = 0.3 $ (s). (Top) Time behavior of $ \widetilde{\mathcal{E}}_{TOT} $; (middle) Time behavior of $ \widetilde{\mathcal{E}}_{NET} $; and, (bottom) Time behavior of $ \tilde{\lambda} $ multiplier.}
\label{fig:TrackingDAG3}
\end{figure*}

The obtained dynamic behaviors of the total consumed energy $ \widetilde{\mathcal{E}}_{TOT} $, network energy $\widetilde{\mathcal{E}}_{NET} $ and \textit{lambda} multipliers $ \tilde{\lambda} $ returned by the \textit{RAP} solution are reported in Figs.\ \ref{fig:TrackingDAG1}, \ref{fig:TrackingDAG2} and \ref{fig:TrackingDAG3} for three test values of the clipping factor $ a_{MAX} $ and under \textit{DAG1}, \textit{DAG2} and \textit{DAG3}, respectively. 

An examination of the reported time-plots leads to three main insights. First, even in the presence of the (aforementioned) abrupt changes of the environmental setup, the corresponding \textit{lambda} multipliers remain almost surely vanishing (see the bottom parts of Figs.\ \ref{fig:TrackingDAG1}, \ref{fig:TrackingDAG2} and \ref{fig:TrackingDAG3}). This supports the conclusion that all the resource allocations computed by the RAP solution are, indeed, \textit{feasible} (i.e., they meet the \textit{RAP} feasibility constraint in \eqref{eq:TH_0mintimesT_DAGinequality}). Second, the abrupt step-like jumps of the plots of $ \widetilde{\mathcal{E}}_{NET} $ in the middle parts of Figs.\ \ref{fig:TrackingDAG1}, \ref{fig:TrackingDAG2} and \ref{fig:TrackingDAG3} are due to the combined effects of \textit{both} the changes of the availability of the WiFi connection and the re-allocation of the Cellular up/down throughput triggered by the underlying execution of the \textit{RAP} iterations. Third, a comparative examination of the red-green-blue colored plots of the upper parts in Figs.\ \ref{fig:TrackingDAG1}, \ref{fig:TrackingDAG2}, and \ref{fig:TrackingDAG3} confirms that bigger values of the clipping factor $ a_{MAX} $ of \eqref{eq:psi_l,m} and \eqref{eq:xi,m} speed up the convergence to the corresponding steady-states, but also tend to introduce larger steady-state oscillations.

Overall, two final insights stem from the carried out tracking analysis. First, at least in the carried out tests, values of $ a_{MAX} $ ranging over the interval: $ 1.0 \times 10^{-7} $ --- $ 2.5 \times 10^{-7} $ guarantee good trade-offs among the contrasting requirements of quick reaction to mobility-induced variations of the operative environment and stable behavior in the steady-state. Second, a number of the primal-dual iterations $ I_{MAX} $ limited up to $ 450 $ --- $ 600 $ suffices, in order to reach stable resource allocations in the presence of abrupt environmental changes.


\subsection{Tuning the energy performance-vs.-computational complexity trade-off of the proposed A-GTA strategy }
\label{ssec:TuningTheEnergy}

\begin{figure*}[htb]
\centering
\includegraphics[width=0.70\textwidth]{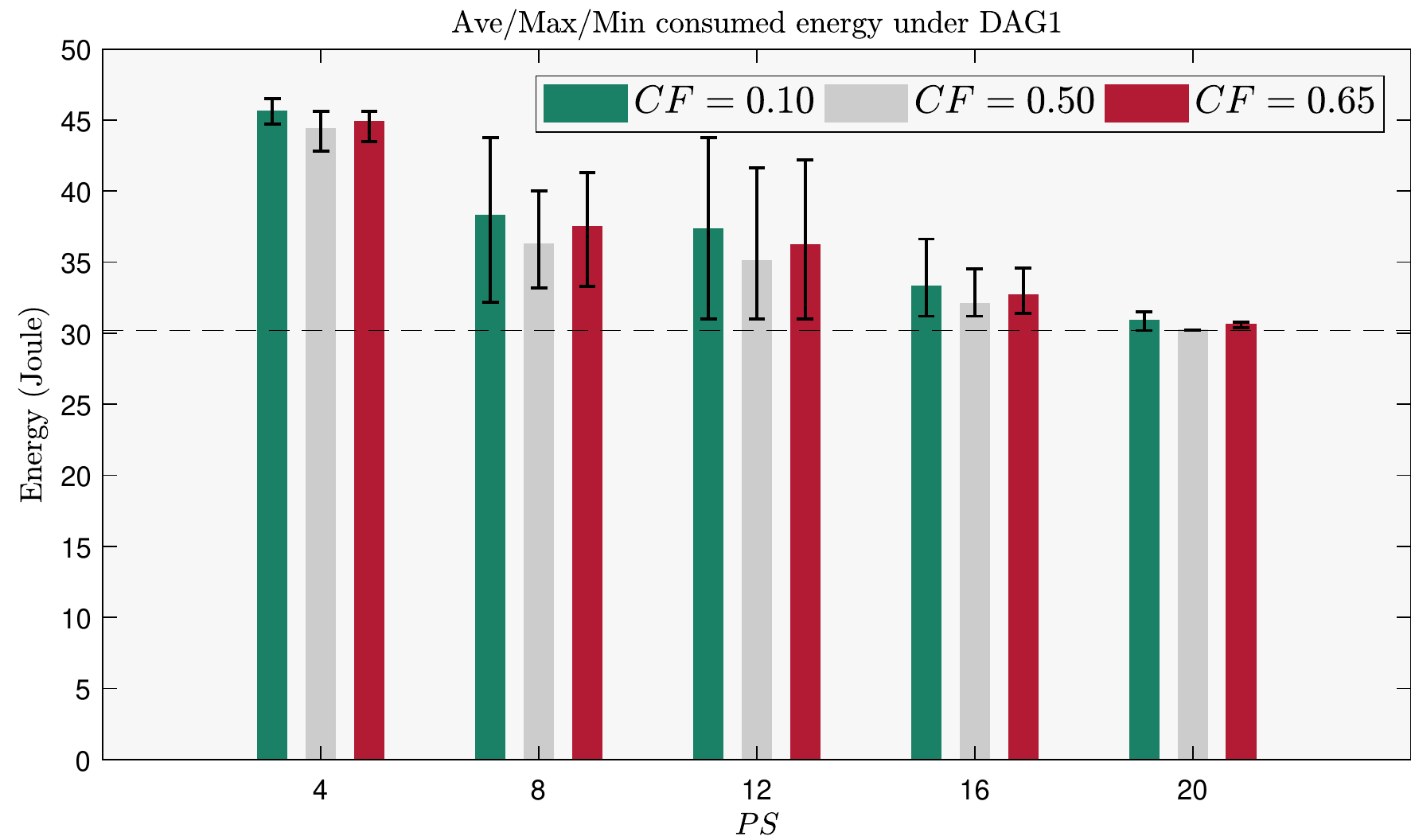}
\caption{Bar plots of the average energy and energy jitters of the proposed \textit{A-GTA} strategy for various values of the population size $ PS $ and crossover fraction $ CF $ under DAG1 at $ T_{DAG}^{\left(MAX\right)} = 0.3 $ (s), $ G_{MAX}=10 $, and $ MN = \mathrm{round}  \left( \left( V-2 \right) / 2 \right) $. As ultimate benchmark, the (horizontal) dashed line reports the corresponding energy consumed by the \textit{A-ES} strategy.}
\label{fig:AGTAstrategyDAG1}
\end{figure*}

\begin{figure*}[htb]
\centering
\includegraphics[width=0.70\textwidth]{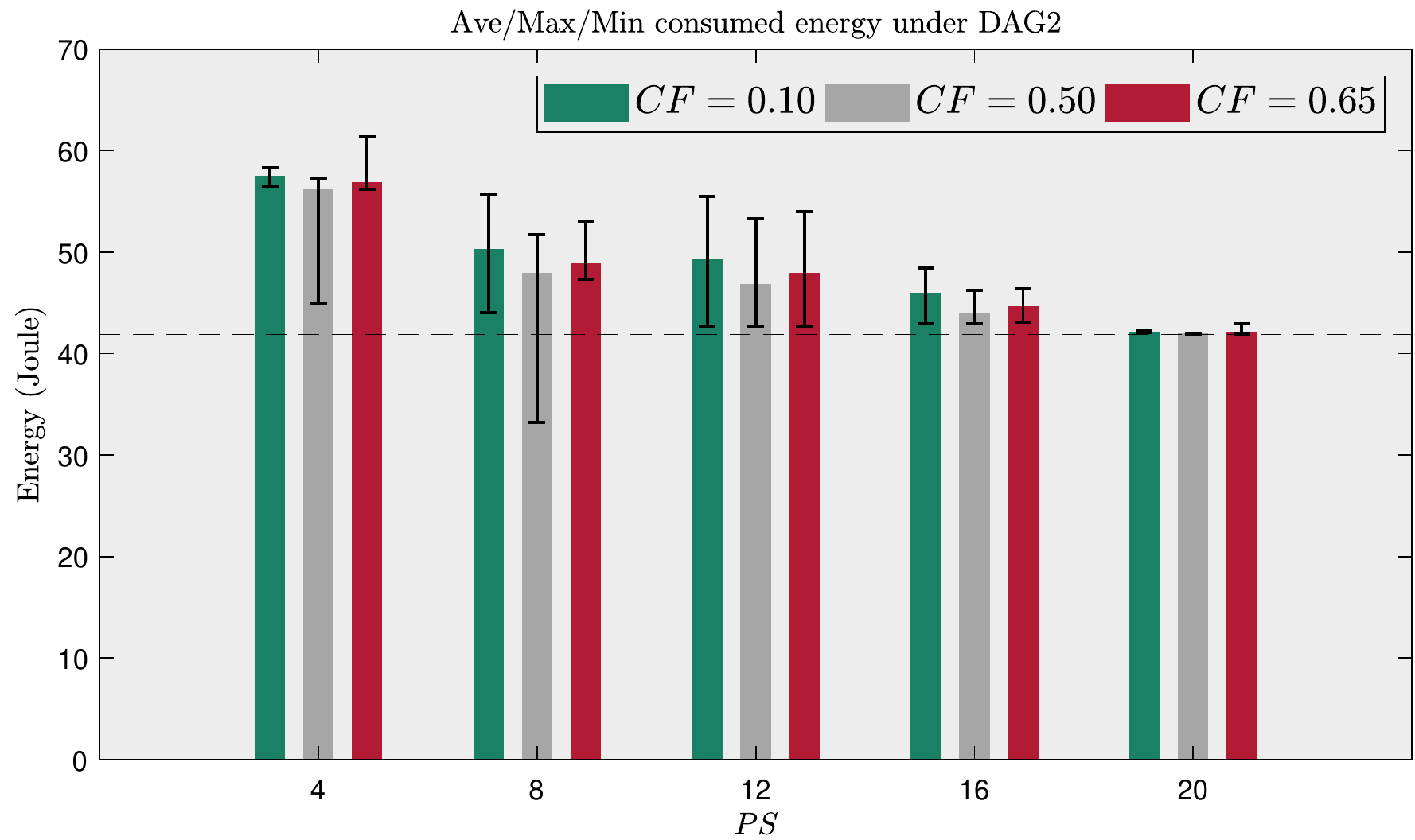}
\caption{Bar plots of the average energy and energy jitters of the proposed \textit{A-GTA} strategy for various values of the population size $ PS $ and crossover fraction $ CF $ under DAG2 at $ T_{DAG}^{\left(MAX\right)} = 0.3 $ (s), $ G_{MAX}=10 $, and $ MN = \mathrm{round}  \left( \left( V-2 \right) / 2 \right) $. As ultimate benchmark, the (horizontal) dashed line reports the corresponding energy consumed by the \textit{A-ES} strategy.}
\label{fig:AGTAstrategyDAG2}
\end{figure*}

\begin{figure*}[htb]
\centering
\includegraphics[width=0.70\textwidth]{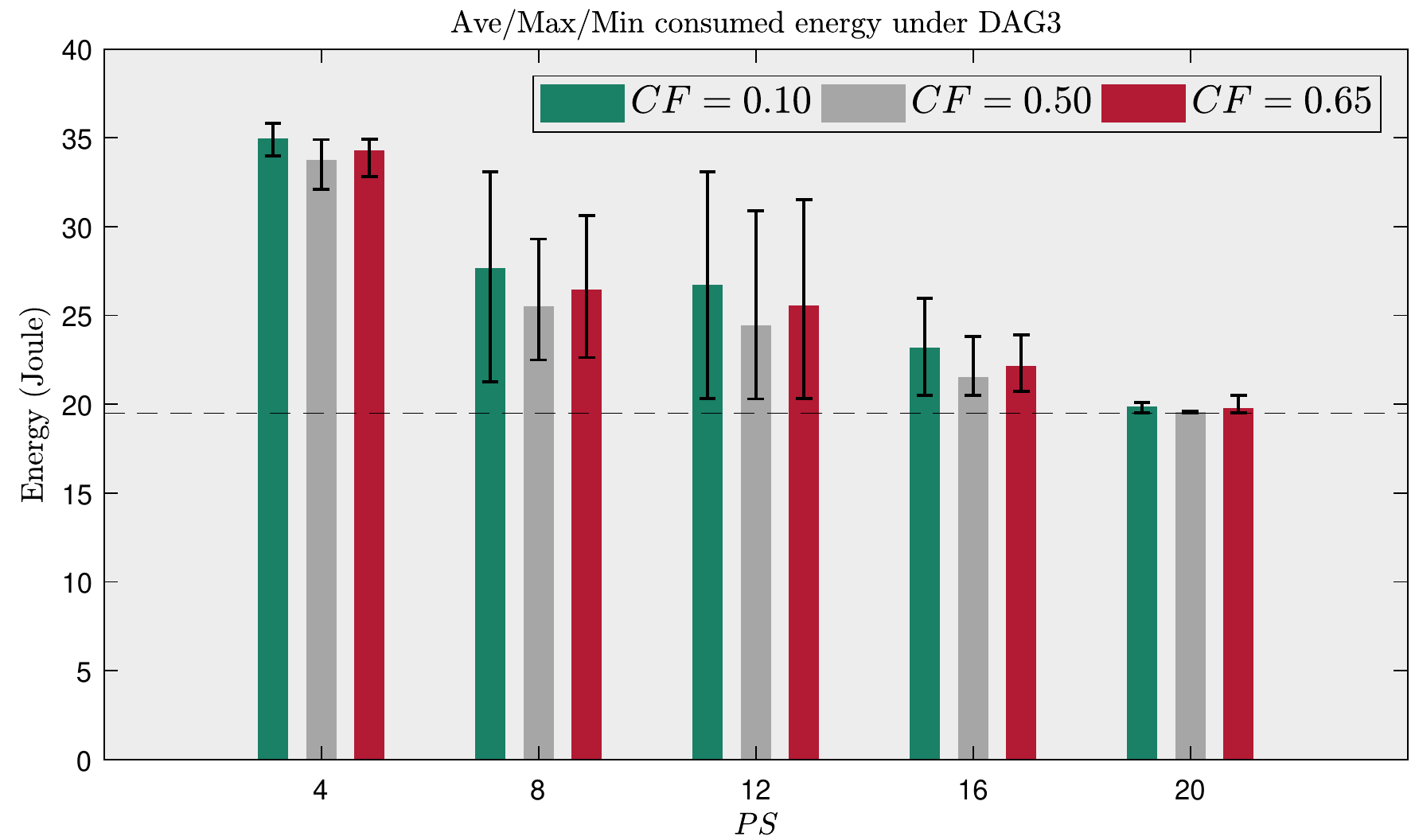}
\caption{Bar plots of the average energy and energy jitters of the proposed \textit{A-GTA} strategy for various values of the population size $ PS $ and crossover fraction $ CF $ under DAG3 at $ T_{DAG}^{\left(MAX\right)} = 0.3 $ (s), $ G_{MAX}=10 $, and $ MN = \mathrm{round}  \left( \left( V-2 \right) / 2 \right) $. As ultimate benchmark, the (horizontal) dashed line reports the corresponding energy consumed by the \textit{A-ES} strategy.}
\label{fig:AGTAstrategyDAG3}
\end{figure*}

The goal of this section is to check the sensitivity of the energy performance of the proposed \textit{A-GTA-S} on the input parameters $ PS $, $ CF $ and $ G_{MAX} $ of Algorithm \ref{algorithm:alg_2}, that specify the population size, fraction of crossed over population and number of performed generations of the underlying GA. The obtained numerical results are reported by the bar plots of Figs.\ \ref{fig:AGTAstrategyDAG1}, \ref{fig:AGTAstrategyDAG2} and \ref{fig:AGTAstrategyDAG3} under \textit{DAG1}, \textit{DAG2}, and \textit{DAG3}, respectively. In order to put the reported results under a right perspective, we note that: (i) since we have numerically ascertained that the obtained energy performance mainly depends on the product population size by generation number: $ PS \times G_{MAX} $, the reported plots refer to the cases of $ PS=4,8,12,16 $, and $ 20 $ at fixed $ G_{MAX}=10 $; (ii) each bar plot reports the average, maximum and minimum energy consumption of the \textit{A-GTA-S} over $ 50 $ independent trials; (iii) as ultimate benchmark, the corresponding energy $ \mathcal{E}_{A-ES-S} $ returned by the \textit{A-ES-S} are also reported in Figs.\ \ref{fig:AGTAstrategyDAG1}, \ref{fig:AGTAstrategyDAG2} and \ref{fig:AGTAstrategyDAG3}.

A comparative examination of these bar plots lead to three main insights. 

First, at fixed $ {CF} $ and for increasing values of $ {PS} $, the average energy consumed by the proposed \textit{A-GTA-S} monotonically decreases and reaches the benchmark ones of the corresponding \textit{A-ES-S} at $ PS=20 $. At the same time, the associated energy jitters decrease for increasing $ PS $ and tend to vanish at $ PS=20 $. These monotonic behaviors are compliant with the (previously remarked) elitary nature of the implemented GA and support its actual effectiveness in the considered application scenarios.

Second, at a fixed $ PS $, both the average energy and energy jitters of the A-GTA-S tend to increase for $ CF < 0.5 $ and $ CF > 0.5 $, while they tend to reach their respective minima at $ CF \cong 0.5 $.

Third, the above two trends are the same under all three test DAGs (i.e., they seem not to be so sensitive on the topologies of the considered DAGs) and we have numerically ascertained that they occur under the overall tested spectrum of values of $ T_{DAG}^{\left(MAX\right)}$.

Overall, the final insight stemming from the analysis of the bar plots of Figs.\ \ref{fig:AGTAstrategyDAG1}, \ref{fig:AGTAstrategyDAG2} and \ref{fig:AGTAstrategyDAG3} is that, at least in the carried out tests, the setting: $ PS=20 $ and $ CF=0.5 $ is the most energy performing one. Under this setting, the ratio between the implementation complexities of the benchmark \textit{E-ES-S} and the proposed \textit{A-GTA-S} remains quite high and of the order of (see Table \ref{table:03}): $ 3^7 / (20 \times 10) \cong 11 $.

\subsection{Task and resource allocation performance of the proposed A-GTA-S}
\label{ssec:TaskResourceAllocation}

In this section, we test the task placement, resource allocation and energy performance of the proposed \textit{A-GTA-S} under the considered test DAGs of Fig.\ \ref{fig:DAGs}. This is done for values of the maximum allowed execution time $T_{DAG}^{\left(MAX\right)}$ of $ 0.3,0.6,1.2 $, and $ 2.4 \mathrm{\left(s\right)} $. The final goal is to acquire insights about the effects of the DAG topology on the task allocation patterns and energy consumption featuring the proposed \textit{A-GTA-S}. In this regard, three main sets of conclusions stem from the numerical results reported in Tables \ref{table:04}, \ref{table:05} and \ref{table:06}. 

\medskip\noindent\textit{Sensitivity of the A-GTA-S performance to the allowed maximum execution times} -- A comparative examination of the numerical values reported in the $ 13 $-th columns of Tables \ref{table:04}, \ref{table:05}, and \ref{table:06} points out that, in all simulated cases, the total energy $ \mathcal{E}_{TOT} $ consumed by the proposed\textit{ A-GTA-S} remains limited up to $ 42 $ (Joule). Since all the test DAGs of Fig.\ \ref{fig:DAGs} share the same sum values of the task workloads and edge weights, this support the conclusion that these factors play the major role in dictating the energy efficiency of the performed task and resource allocations. However, a more detailed examination of Table \ref{table:04}, \ref{table:05} and \ref{table:06} also unveils two interesting trends.

First, under a fixed DAG, the consumed energy tends to decrease for increasing values of $T_{DAG}^{\left(MAX\right)}$. Roughly speaking, this first trend is due to the fact that larger values of $T_{DAG}^{\left(MAX\right)}$ allow the \textit{RAP} solution of Algorithm \ref{algorithm:alg_1} to lower the steady-state computing frequencies and/or the corresponding per-connection throughput (see the numerical values reported by the corresponding columns of Tables \ref{table:04}, \ref{table:05} and \ref{table:06}). This reduces, in turn, the dynamic (i.e., resource-depending) components of the total consumed energy (see the energy models of Section \ref{sec:ModelingEnergyConsumption}).  

The second trend arises from the observation that, in general, at a fixed $T_{DAG}^{\left(MAX\right)}$, the energy consumption returned by the proposed \textit{A-GTA-S} tends to be larger under the tree topology of \textit{DAG2}, and, then, it somewhat decreases under the mesh topology of \textit{DAG1} and the hybrid topology of \textit{DAG3}. Intuitively, this trend is caused by the behavior of the corresponding network energy $ \mathcal{E}_{NET} $. In fact, a comparative examination of the results reported in the last columns of Tables \ref{table:04}, \ref{table:05} and \ref{table:06} points out that the network energy consumed by the tree topology are larger than the corresponding ones of the mesh and hybrid topologies. This behavior is, indeed, compliant with the observation that the tree topology offers, by design, the smallest number of input-output paths. This forces the \textit{A-GTA-S} to increase, in turn, the data traffic allocated to each input-output path, so that the dynamic (i.e., throughput-depending) components of the per-connection network energy increase too (see \eqref{eq:E_N1toN2dynamic} and \eqref{eq:P_MtoNdynamic}).

\medskip\noindent\textit{Features of the task allocation patterns returned by A-GTA-S} -- Columns $ 2 $, $ 3 $, and $ 4 $ of Tables \ref{table:04}, \ref{table:05} and \ref{table:06} report the ID numbers of the tasks allocated by the \textit{A-GTA-S} to the Mobile, Fog and Cloud nodes under \textit{DAG1}, \textit{DAG2}, and \textit{DAG3}, respectively. Although the reported allocation patterns may strongly depend on the specifically considered DAGs, two main trends may be detected. First, at low values of $T_{DAG}^{\left(MAX\right)}$, medium-size communication-intensive tasks are typically allocated to the Fog node, while large-size communication-light tasks are assigned by \textit{A-GTA-S} to the Cloud node. The Mobile device typically executes small-size communication-intensive tasks. Second, an increasing number of tasks are shifted from the Cloud node to the Fog node and/or to the Mobile device when $T_{DAG}^{\left(MAX\right)}$ decreases more and more.

\medskip\noindent\textit{How the A-GTA-S exploits the Fog-Cloud backhaul connection} -- A native feature of the three-tier platform of Fig.\ \ref{fig:Mobile-Fog-CloudPlatform} is the presence of a (possibly, multi-hop and/or wired) two-way backhaul connection, that interconnects the Fog and Cloud nodes. Hence, it may be of interest to attain insight about how the proposed \textit{A-GTA-S} exploits this auxiliary connection. Intuitively, we expect that the backhaul connection is utilized when there are large-size tasks to be allocated to the Cloud and the volumes of data output by the execution of these tasks are also large. Therefore, since the up/down Fog-Mobile WiFi connections of Fig. 4 are more energy efficient than the corresponding Cloud-Mobile cellular ones, it may be energy-saving to transport the processed data from/to the Cloud to/from the Fog over the two-way backhaul connection. So doing, the Fog node of Fig.\ \ref{fig:Mobile-Fog-CloudPlatform} acts as \textit{relay} node by forwarding the needed data over the WiFi up/down connections of Fig.\ \ref{fig:Mobile-Fog-CloudPlatform}. This is, indeed, the general strategy followed by the \textit{A-GTA-S}, in order to allow energy-efficient executions of \textit{DAG1} and \textit{DAG3} under all the considered spectrum of allowed maximum DAG execution times $T_{DAG}^{\left(MAX\right)}$. In fact, an examination of the corresponding Tables \ref{table:04} and \ref{table:06} points out that the optimized execution strategy returned by \textit{A-GTA-S} utilizes: (i) the Cellular/Wifi up-connections of Fig.\ \ref{fig:Mobile-Fog-CloudPlatform} for uploading the data to be processed by the Cloud/Fog nodes; (ii) the two-way backhaul connection: $ C \leftrightarrow F $, in order to allow the Cloud and Fog nodes to exchange partially processed data; and, (iii) the WiFi down connection: $ F\to M $ for the final delivering of the processed data to the Mobile device.

The (somewhat unexpected) final lesson is that, at least in the described operating scenarios, the utilization of (single-hop) Cloud-Mobile and/or Mobil-Fog links are \textit{less} energy-efficient than the exploitation of the (multi-hop) Cloud-Fog-Mobile path.

\begin{table*}[htb]
\caption{Task allocation, resource allocation and energy consumption of the proposed \textit{A-GTA} strategy under \textit{DAG1}. $T_{DAG}^{(MAX)}$ is measured in (s), all the resources are measured in (Mb/s) while the energy is measured in (Joule).}
\label{table:04}
\centering
\small\resizebox{\textwidth}{!}{
\setlength{\extrarowheight}{1ex}
\renewcommand{\arraystretch}{1.5}
\Huge{
\begin{tabular}{*{14}{c}}
\toprule
\rowcolor[HTML]{BABFC6}
$ \bm{T_{DAG}^{\left(MAX\right)}} $ & \textbf{Mobile Tasks} & \textbf{Fog Tasks} & \textbf{Cloud Tasks} & $ \bm{f_M} $ & $ \bm{f_F} $ & $ \bm{f_C} $ & $ \bm{f_{M \to F}} $ & $ \bm{f_{F \to M}} $ & $ \bm{f_{M \to C}} $ & $ \bm{f_{C \to M}} $ & $ \bm{f_{C \leftrightarrow F}} $ & $ \bm{\mathcal{E}_{TOT}} $ & $ \bm{\mathcal{E}_{NET}} $ \\[1ex]
\midrule
\rowcolor[HTML]{F2F5F9}
0.3 & $\left\{1,9\right\}$ & $\left\{2,4,5,7,8\right\}$ & $\left\{3,6\right\}$ & 11.99 & 4.44 & 2.47 & 7.46 & 8.24 & 0.00 & 0.00 & 3.70 & 30.19 & 13.37 \\ [1ex]
\rowcolor[HTML]{DCE1E8}
0.6 & $\left\{1,9\right\}$ & $\left\{2,4,8\right\}$ & $\left\{3,5,6,7\right\}$ & 11.88 & 3.24 & 2.67 & 6.78 & 7.28 & 0.00 & 0.00 & 3.70 & 27.10 & 11.84 \\ [1ex]
\rowcolor[HTML]{F2F5F9}
1.2 & $\left\{1,9\right\}$ & $\left\{2,4,8\right\}$ & $\left\{3,5,6,7\right\}$ & 11.56 & 3.04 & 2.65 & 6.58 & 7.18 & 0.00 & 0.00 & 3.70 & 26.35 & 10.82 \\ [1ex]
\rowcolor[HTML]{DCE1E8}
2.4 & $\left\{1,8,9\right\}$ & $\left\{2,4\right\}$ & $\left\{3,5,6,7\right\}$ & 11.78 & 2.54 & 2.55 & 6.26 & 7.05 & 0.00 & 0.00 & 3.70 & 25.84 & 10.64\\ [1ex]
\bottomrule
\end{tabular}}}
\end{table*}
%
%
%
%
\begin{table*}
\caption{Task allocation, resource allocation and energy consumption of the proposed \textit{A-GTA} strategy under \textit{DAG2}. $T_{DAG}^{(MAX)}$ is measured in (s), all the resources are measured in (Mb/s) while the energy is measured in (Joule).}
\label{table:05}
\centering
\small\resizebox{\textwidth}{!}{
\setlength{\extrarowheight}{1ex}
\renewcommand{\arraystretch}{1.5}
\Huge{
\begin{tabular}{*{14}{c}}
\toprule
\rowcolor[HTML]{BABFC6}
$ \bm{T_{DAG}^{\left(MAX\right)}} $ & \textbf{Mobile Tasks} & \textbf{Fog Tasks} & \textbf{Cloud Tasks} & $ \bm{f_M} $ & $ \bm{f_F} $ & $ \bm{f_C} $ & $ \bm{f_{M \to F}} $ & $ \bm{f_{F \to M}} $ & $ \bm{f_{M \to C}} $ & $ \bm{f_{C \to M}} $ & $ \bm{f_{C \leftrightarrow F}} $ & $ \bm{\mathcal{E}_{TOT}} $ & $ \bm{\mathcal{E}_{NET}} $ \\[1ex]
\midrule
\rowcolor[HTML]{F2F5F9}
0.3 & $\left\{1,2,9\right\}$ & $\left\{4,6,7,8\right\}$ & $\left\{3,5\right\}$ & 11.97 & 3.76 & 2.97 & 7.77 & 8.16 & 4.12 & 1.45 & 3.70 & 41.92 & 20.47 \\[1ex]
\rowcolor[HTML]{DCE1E8}
0.6 & $\left\{1,2,9\right\}$ & $\left\{3,8\right\}$ & $\left\{-\right\}$ & 11.92 & 1.85 & 0.00 & 7.01 & 5.96 & 0.00 & 0.00 & 0.00 & 28.60 & 11.12 \\[1ex]
\rowcolor[HTML]{F2F5F9}
1.2 & $\left\{1,2,9\right\}$ & $\left\{3,8\right\}$ & $\left\{-\right\}$ & 11.79 & 1.83 & 0.00 & 6.95 & 5.87 & 0.00 & 0.00 & 0.00 & 27.30 & 10.81 \\[1ex]
\rowcolor[HTML]{DCE1E8}
2.4 & $\left\{1,4,9\right\}$ & $\left\{5,8\right\}$ & $\left\{-\right\}$ & 11.96 & 1.54 & 0.00 & 6.75 & 5.65 & 0.00 & 0.00 & 0.00 & 27.03 & 10.62 \\[1ex]
\bottomrule
\end{tabular}}}
\end{table*}
%
%
%
%
\begin{table*}
\caption{Task allocation, resource allocation and energy consumption of the proposed \textit{A-GTA} strategy under \textit{DAG3}. $T_{DAG}^{(MAX)}$ is measured in (s), all the resources are measured in (Mb/s) while the energy is measured in (Joule).}
\label{table:06}
\centering
\small\resizebox{\textwidth}{!}{
\setlength{\extrarowheight}{1ex}
\renewcommand{\arraystretch}{1.5}
\Huge{
\begin{tabular}{*{14}{c}}
\toprule
\rowcolor[HTML]{BABFC6}
$ \bm{T_{DAG}^{\left(MAX\right)}} $ & \textbf{Mobile Tasks} & \textbf{Fog Tasks} & \textbf{Cloud Tasks} & $ \bm{f_M} $ & $ \bm{f_F} $ & $ \bm{f_C} $ & $ \bm{f_{M \to F}} $ & $ \bm{f_{F \to M}} $ & $ \bm{f_{M \to C}} $ & $ \bm{f_{C \to M}} $ & $ \bm{f_{C \leftrightarrow F}} $ & $ \bm{\mathcal{E}_{TOT}} $ & $ \bm{\mathcal{E}_{NET}} $ \\[1ex]
\midrule
\rowcolor[HTML]{F2F5F9}
0.3 & $\left\{1,9\right\}$ & $\left\{5,8\right\}$ & $\left\{2,4\right\}$ & 11.95 & 2.36 & 2.48 & 0.00 & 5.91 & 1.35 & 0.00 & 3.70 & 19.51 & 5.93 \\ [1ex]
\rowcolor[HTML]{DCE1E8}
0.6 & $\left\{1,9\right\}$ & $\left\{5,8\right\}$ & $\left\{2,4\right\}$ & 11.91 & 2.28 & 2.37 & 0.00 & 5.82 & 1.31 & 0.00 & 3.70 & 19.37 & 5.43 \\ [1ex]
\rowcolor[HTML]{F2F5F9}
1.2 & $\left\{1,9\right\}$ & $\left\{5,8\right\}$ & $\left\{2,3\right\}$ & 11.81 & 2.36 & 2.15 & 0.00 & 5.61 & 1.24 & 0.00 & 3.70 & 19.21 & 5.28 \\ [1ex]
\rowcolor[HTML]{DCE1E8}
2.4 & $\left\{1,2,9\right\}$ & $\left\{4,8\right\}$ & $\left\{3\right\}$ & 11.94 & 2.16 & 2.01 & 0.00 & 5.22 & 1.19 & 0.00 & 3.70 & 18.13 & 5.01\\ [1ex]
\specialrule{0.5pt}{0pc}{0pc}
\bottomrule
\end{tabular}}}
\end{table*}

\subsection{Performance comparisons against the benchmark strategies}
\label{ssec:PerformanceComparisons}

In this section, we compare the energy performance of the proposed \textit{A-GTA-S} against the corresponding ones of the five benchmark strategies of Section \ref{ssec:SimulatedPlatform}. The pursued threefold goal is to acquire some insight about: (i) the energy reduction stemming from the \textit{dynamic} optimization of the computing-networking resources versus the corresponding case of \textit{static} resource usage; (ii) the performance gap between the proposed \textit{A-GTA-S} and the exhaustive search-based \textit{A-ES-S}; and, (iii) the energy-saving capability offered by the Mobile-Fog-Cloud \textit{three-tier} computing platform of Fig.\ \ref{fig:Mobile-Fog-CloudPlatform} versus the only Mobile, Mobile-Cloud and Mobile-Fog corresponding ones.  

\begin{figure*}[ht!]
\centering
\includegraphics[width=0.67\textwidth]{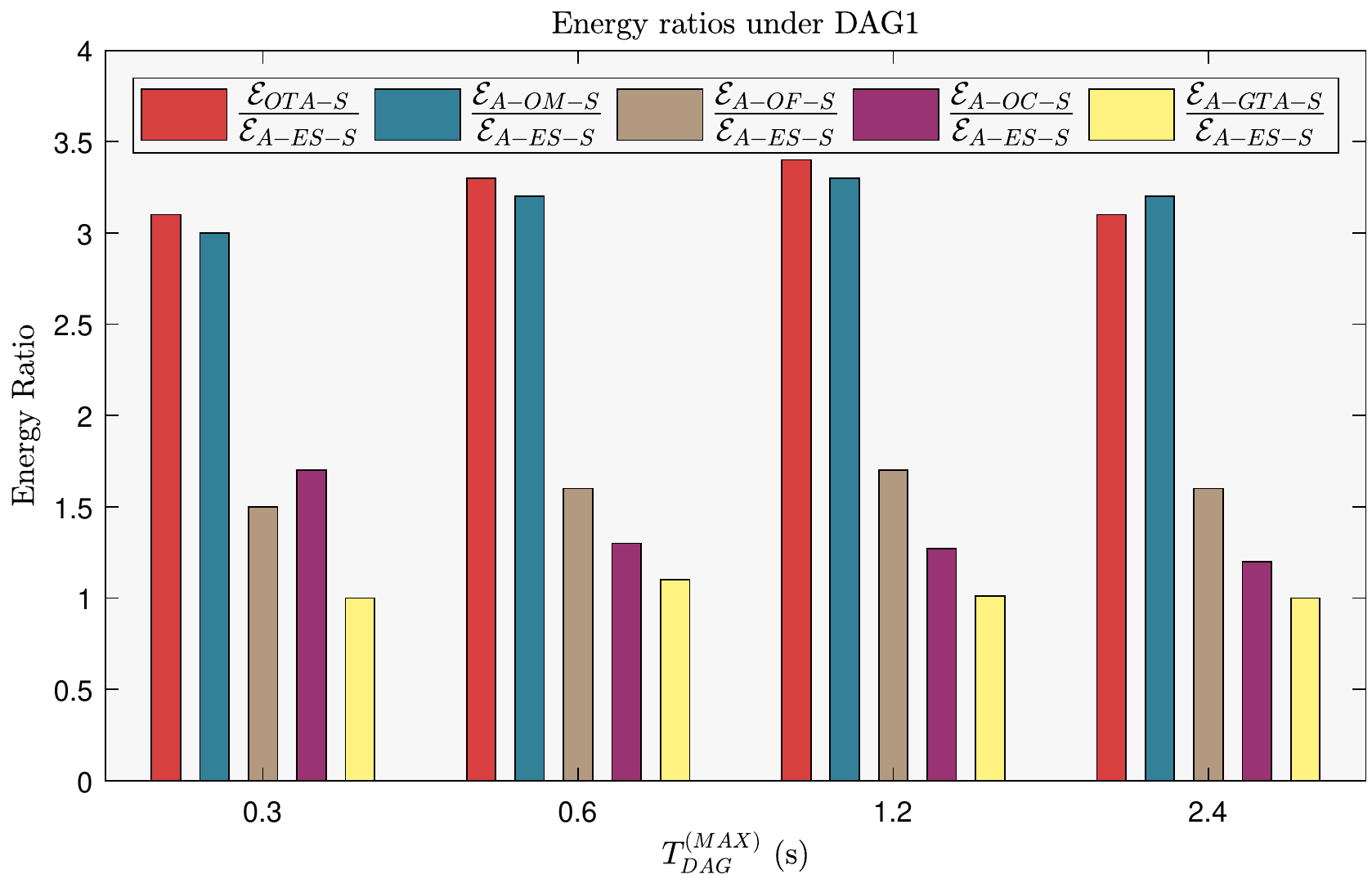}
\caption{Bar plots of the energy ratios under DAG1. All the reported ratios are normalized with respect to the corresponding total energy consumed by \textit{A-ES-S}.}
\label{fig:EnergyRatiosDAG1}
\end{figure*}

\begin{figure*}[ht!]
\centering
\includegraphics[width=0.65\textwidth]{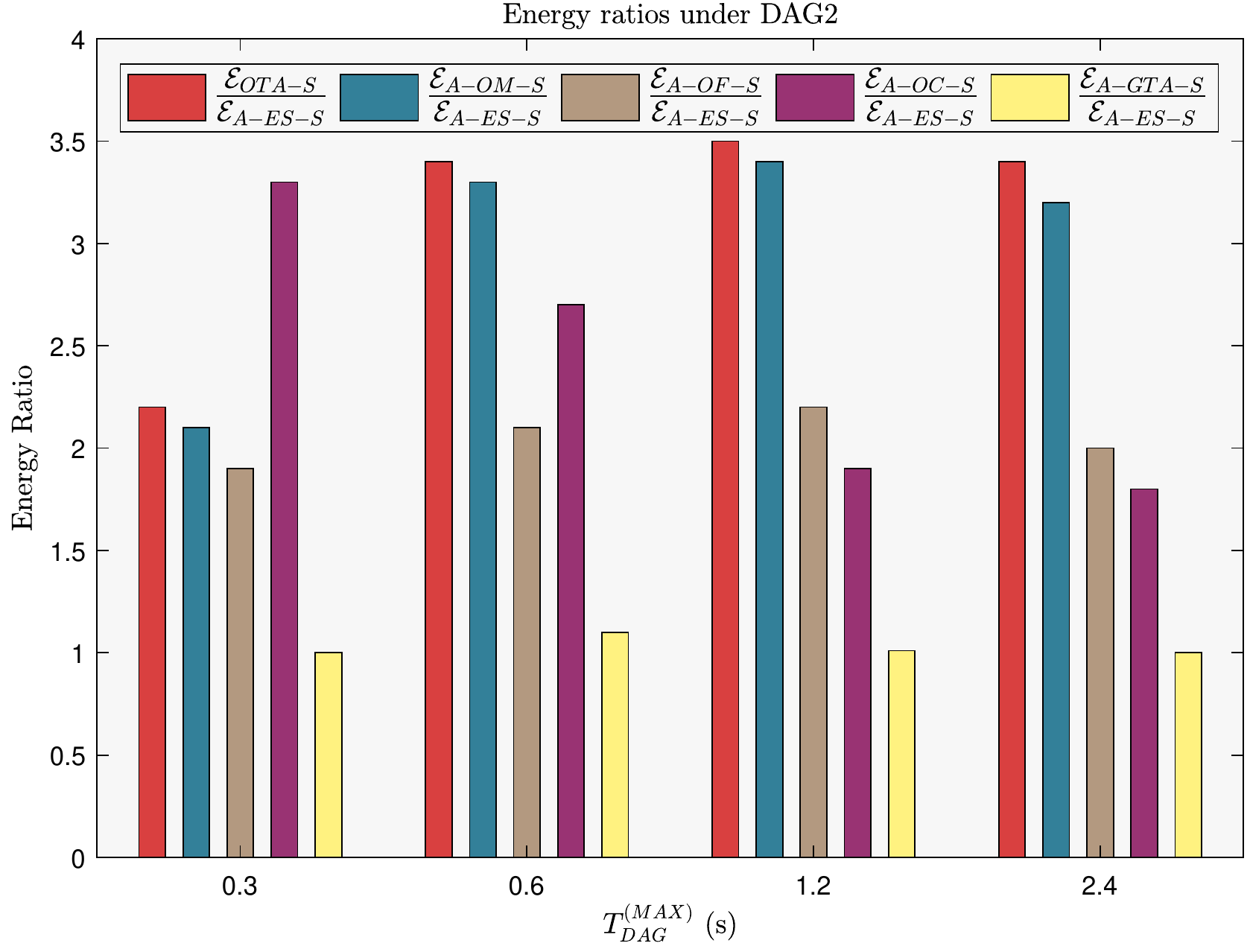}
\caption{Bar plots of the energy ratios under DAG2. All the reported ratios are normalized with respect to the corresponding total energy consumed by \textit{A-ES-S}.}
\label{fig:EnergyRatiosDAG2}
\end{figure*}

\begin{figure*}[ht!]
\centering
\includegraphics[width=0.65\textwidth]{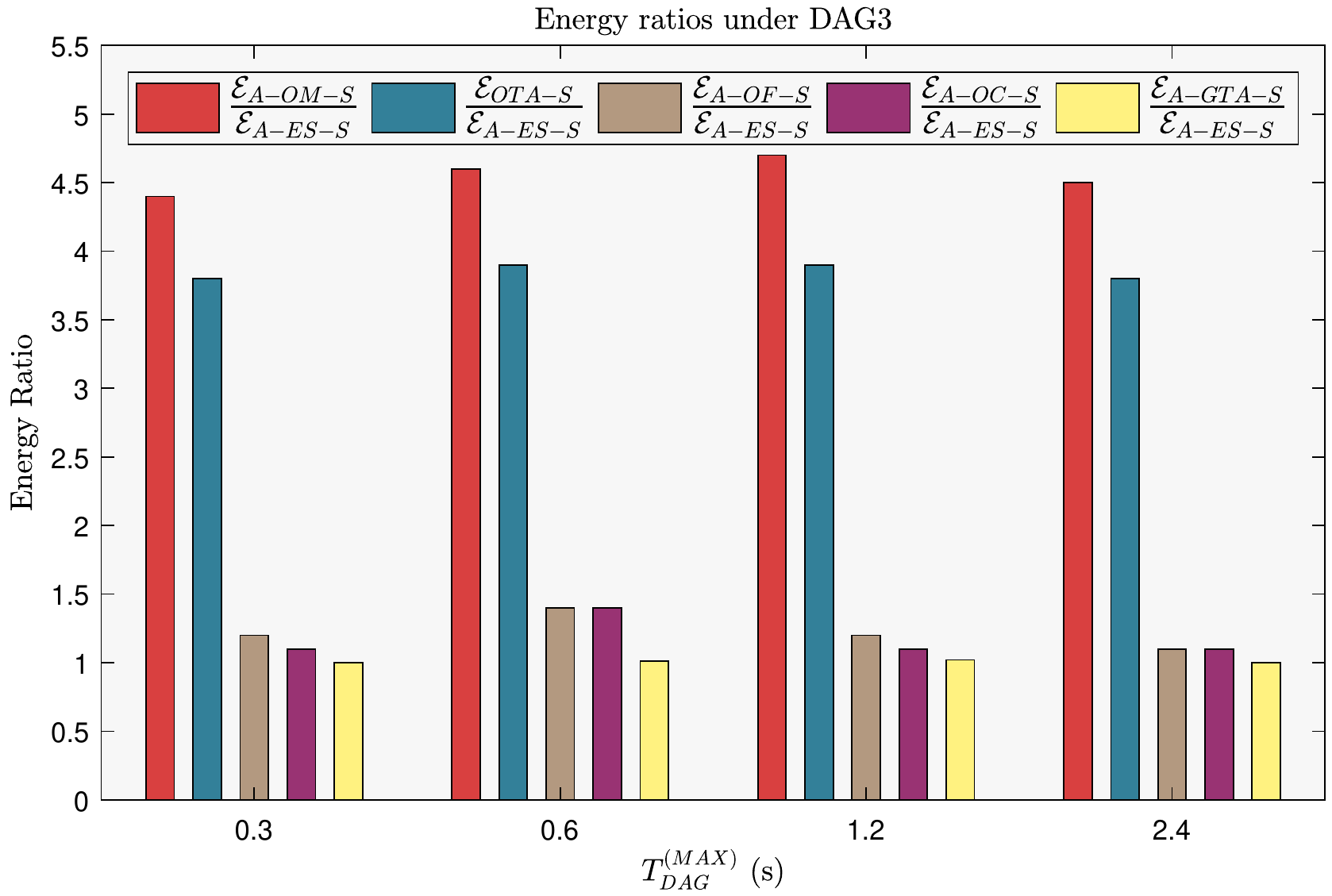}
\caption{Bar plots of the energy ratios under DAG3. All the reported ratios are normalized with respect to the corresponding total energy consumed by \textit{A-ES-S}.}
\label{fig:EnergyRatiosDAG3}
\end{figure*}

The obtained numerical results are summarized by the bar plots of Figs.\ \ref{fig:EnergyRatiosDAG1}, \ref{fig:EnergyRatiosDAG2}, and \ref{fig:EnergyRatiosDAG3} under \textit{DAG1}, \textit{DAG2}, and \textit{DAG3}, respectively. Their examination gives arise to the following three main sets of remarks.

\medskip\noindent\textit{A-GTA-S versus O-TA-S} -- A number of seminal (even quite recent) contributions \cite{cuervo2010,yang2013,mahmoodi2016,chun2011} tackles with the problem of the resource augmentation of mobile devices by developing various heuristic/meta-heuristic/optimal solutions for energy-efficient task offloading. However, they \textit{neglect} to consider, indeed, the companion problem of the dynamic scaling of the computing and/or network resources. Hence, a key (still open) question concerns how much energy may be actually saved by \textit{jointly} performing task and dynamic resource allocation. By design, a direct comparison of the energy consumed by the \textit{A-GTA-S} and \textit{O-TA-S} provides the response to this question. In this regard, a comparative examination of the red and yellow-colored bars of Figs.\ \ref{fig:EnergyRatiosDAG1}, \ref{fig:EnergyRatiosDAG2} and \ref{fig:EnergyRatiosDAG3} leads to three main insights. First, the energy ratio $\mathcal{E}_{O-TA-S} / \mathcal{E}_{A-GTA-S}$ ranges over the intervals: $ 3.0 $ -- $ 3.1 $, $ 2.2 $ -- $ 3.4 $, and: $ 3.8 $ -- $ 3.9 $ under \textit{DAG1}, \textit{DAG2} and \textit{DAG3}, respectively. Second, under a fixed DAG, the energy savings stemming from performing dynamic resource allocation reach their maxima at values of $T_{DAG}^{\left(MAX\right)}$ of the order of $ 0.6 $ -- $ 1.2 $ (s), while tend to somewhat decrease at smaller and higher execution delays. Third, the average energy saving stemming from dynamic optimization is somewhat more relevant under \textit{DAG3}. 

Overall, the key lesson stemming from these considerations is that the dynamic optimization of the allocated computing-networking resources plays, indeed, a \textit{major} role in reducing the energy consumption of the simulated platform of Fig.\ \ref{fig:Mobile-Fog-CloudPlatform}.        

\medskip\noindent\textit{A-GTA-S versus A-ES-S} -- We pass now to focus on the trade-offs among the energy performance and computational complexity that are attained by the proposed (meta-heuristic) \textit{A-GTA} and the benchmark (optimal) \textit{A-ES} strategies. In this regard, we recall that, in the carried out simulations, the computing complexity of the benchmark \textit{A-ES-S} is about $ 11 $ times larger that the corresponding one of the proposed \textit{A-GTA-S} (see the last part of Section \ref{ssec:TuningTheEnergy}). At the same time, a direct inspection of the yellow-colored bars of Figs. \ref{fig:EnergyRatiosDAG1}, \ref{fig:EnergyRatiosDAG2} and \ref{fig:EnergyRatiosDAG3} unveils that the energy gaps between the proposed \textit{A-GTA-S} and the benchmark \textit{A-ES-S} maintain below $ 2 \% $ over the full spectrum of the considered maximum DAG execution times. These considerations lead to the conclusion that the tested implementation of the proposed \textit{A-GTA-S} retains, indeed, good performance-vs.-complexity trade-offs against the benchmark \textit{A-ES-S} one.  

\medskip\noindent\textit{Three-tier versus single/two-tier execution platforms} -- A potential drawback of multi-tier distributed computing platforms is that the number of involved network connections tend to grow with the number of inter-connected tiers, and this could increase the network component of the overall consumed energy. In this regard, we recall that the (previously defined) \textit{A-OM}, \textit{A-OF} and \textit{A-OC} benchmark strategies utilize, by design, only the Mobile device and the two-tier Fog-Mobile and Cloud-Mobile platforms for the execution of the application DAGs. Furthermore, all these benchmark strategies perform dynamic scaling of the utilized computing frequencies and wireless network throughput (see their definitions of Section \ref{ssec:SimulatedPlatform}). Hence, in order to attain insight about the net trade-off among the reduction of the computing energy arising from the utilization of multi-tier computing platforms and the corresponding increment of the network energy needed for their inter-connection, it suffices to compare the blue, cyan, magenta, and yellow-colored bars of Figs.\ \ref{fig:EnergyRatiosDAG1}, \ref{fig:EnergyRatiosDAG2} and \ref{fig:EnergyRatiosDAG3}. Their comparison leads to two main conclusions. 

First, the energy ratio $\mathcal{E}_{A-OM-S} / \mathcal{E}_{A-GTA-S}$ takes values over the intervals: $ 3.0 $ -- $ 3.3 $, $ 2.1 $ -- $ 3.4 $, and $ 4.4 $ -- $ 4.7 $, under \textit{DAG1}, \textit{DAG2} and \textit{DAG3}, respectively. The corresponding intervals of the energy ratios $\mathcal{E}_{A-OF-S} / \mathcal{E}_{A-GTA-S}$ , and $\mathcal{E}_{A-OC-S} / \mathcal{E}_{A-GTA-S}$ are: $ 1.5 $ -- $ 1.7 $,  $ 1.9 $ -- $ 2.1 $,  $ 1.1 $ -- $ 1.4 $,  and: $ 1.2 $ -- $ 1.7 $,  $ 1.8 $ -- $ 3.3 $,  $ 1.2 $ -- $ 1.4 $, respectively. Hence, in the carried out simulations, the minimum (i.e., worst case) energy-savings guaranteed by the three-tier Mobile-Fog-Cloud platform of Fig.\ \ref{fig:Mobile-Fog-CloudPlatform} over the Mobile, Mobile-Fog and Mobile-Cloud ones are of the order of: $ 110\% $, $ 10\% $  and $ 20\% $, while the corresponding maximum values are around: $ 370\% $, $ 110\% $ and $ 230\% $. 

Second, at fixed $T_{DAG}^{\left(MAX\right)}$, the average energy savings offered by the Mobile-Fog-Cloud platform over the considered benchmark ones tend to be somewhat more substantial under \textit{DAG3}. Intuitively, this is due to the fact that \textit{DAG3} is the hybrid combination of the basic mesh and tree topologies, so that its energy-saving executions tend to take more advantage from the \textit{simultaneous} utilization of all the available Mobile, Fog, and Cloud computing nodes of Fig.\ \ref{fig:Mobile-Fog-CloudPlatform}.



\subsection{Sensitivity of the A-GTA-S energy performance on the DAG computing-to-communication ratios}
\label{ssec:SensitivityA-GTA-S}

The goal of this section is to test the sensitivity of the average energy performance of the proposed \textit{A-GTA-S} on the Computing-to-Communication Ratio (CCR) of the benchmark DAGs of Fig.\ \ref{fig:DAGs}. Formally speaking, the CCR of an application DAG is defined as the ratio between the corresponding per-task average workload and per-edge average weight \cite{andrade2014}. Hence, in order to carry out fair energy comparisons, all the tests of this section have been performed by taking the summation of the task workloads and edge weights of each DAG fixed at $ 4.98 $ (Mbit), regardless of the actual value assumed by the corresponding CCR. 

The average total and network energy consumption obtained by running the proposed \textit{A-GTA-S} of Algorithm \ref{algorithm:alg_2} are reported in Table \ref{table:07} for values of CCR ranging from $ 4 $ (case of computing-intensive DAGs) to $ 0.5 $ (case of communication-intensive DAGs). All these results refer to the Eco-centric service model of \eqref{eq:EcoCentric} at $T_{DAG}^{\left(MAX\right)} = 1.0 $ (s).

An examination of these results unveils that, in all carried out tests, both the total energy $ \mathcal{E}_{TOT} $ and the network energy $ \mathcal{E}_{NET} $ consumed by the \textit{A-GTA-S} attain their maxima at $ CCR=1 $ (i.e., in the case of balanced computing and communication loads), while they decrease at lower and higher CCR values.  We have numerically ascertained that this (seemingly unexpected) behavior is, indeed, induced by the considered Eco-centric service scenario. In fact, under this service model, the \textit{JOP} objective function in \eqref{eq:JOP1} accounts for the computing energy of all  Mobile-Fog-Cloud nodes of the simulated system of Fig.\ \ref{fig:Mobile-Fog-CloudPlatform} (see the expression of $ \mathcal{E}_{TOT} $ in \eqref{eq:E_Tot} at  $ \theta_M = \theta_F = \theta_C =1 $). This triggers the allocation policy followed by the \textit{A-GTA-S} to scatter the DAG workload over \textit{all} the available computing nodes when the CCR values are high, so to reduce the resulting total computing energy. However, decreasing values of CCR increase the consumed network energy, so that, as it could be expected, the increment of the network energy balances the corresponding reduction of the computing one under balanced operating conditions (i.e., at $ CCR=1 $). Further decrements of the CCR values induce the allocation policy applied by the \textit{A-GTA-S} to reduce the number of utilized computing nodes, in order to save network energy and, then, lower the resulting total energy.  

Overall, two conclusions stem from the carried out analysis. First, under the Eco-centric service model, the most energy demanding operating conditions take place for CCR around the unit, while less energy is wasted at higher and lower values of CCR. Second, a comparison of the first and last rows of Table \ref{table:07} also points out that computing-intensive operating conditions consume more energy than communication intensive ones under all considered DAGs.

However, we anticipate that these conclusions, could be, indeed, no longer true when the Mobile-centric service model of the next Section \ref{ssec:PerformanceSensitivity} is considered.

\begin{table*}
\caption{Total and network energy consumption (Joule) of the proposed \textit{A-GTA-S} at $ CCR = 4, 2, 1$ and $ 0.5 $. Case of $ \theta_M = \theta_F = \theta_C = 1 $ at $ T_{DAG}^{\left(MAX\right)} = 1.0 $ (s). Each reported energy value is averaged over $ 20 $ independent runs of Algorithm \ref{algorithm:alg_2}.}
\label{table:07}
\centering
\renewcommand{\arraystretch}{1.5}
\begin{tabular}{
>{\columncolor[HTML]{DEE5E4}}l
>{\columncolor[HTML]{80DBFC}}c
>{\columncolor[HTML]{80DBFC}}c
>{\columncolor[HTML]{F5E1BE}}c
>{\columncolor[HTML]{F5E1BE}}c
>{\columncolor[HTML]{9FDDD3}}c
>{\columncolor[HTML]{9FDDD3}}c
}
%
%
\rowcolor[HTML]{7BB6F2}
\toprule
\bfseries CCR & \multicolumn{2}{c}{\bfseries DAG1} & \multicolumn{2}{c}{\bfseries DAG2} & \multicolumn{2}{c}{\bfseries DAG3} \\
\arrayrulecolor{white}
\midrule
& $ \bm{\mathcal{E}_{TOT}} $ & $ \bm{\mathcal{E}_{NET}} $ & $ \bm{\mathcal{E}_{TOT}} $ & $ \bm{\mathcal{E}_{NET}} $ & $ \bm{\mathcal{E}_{TOT}} $ & $ \bm{\mathcal{E}_{NET}} $ \\ 
\midrule
4.0  & 22.54 & 6.47  & 22.91 & 8.10  & 16.00 &  5.96 \\ 
2.0  & 26.94 & 10.48 & 27.31 & 12.64 & 19.05 &  6.55 \\ 
1.0  & 27.31 & 11.12 & 27.94 & 13.10 & 19.51 &  6.93 \\ 
0.5  & 19.64 & 5.11  & 21.04 & 7.55  & 14.38 &  3.63 \\ [0.5ex]
\arrayrulecolor{black}
\bottomrule
\end{tabular}
\renewcommand{\arraystretch}{1.0}
\end{table*}

\subsection{Performance sensitivity on the adopted service model}
\label{ssec:PerformanceSensitivity}

The conclusions of the last section trigger us to further investigate the performance sensitivity of the proposed \textit{A-GTA-S} on the actually adopted service model.

\begin{figure*}[ht!]
\centering
\includegraphics[width=0.8\textwidth]{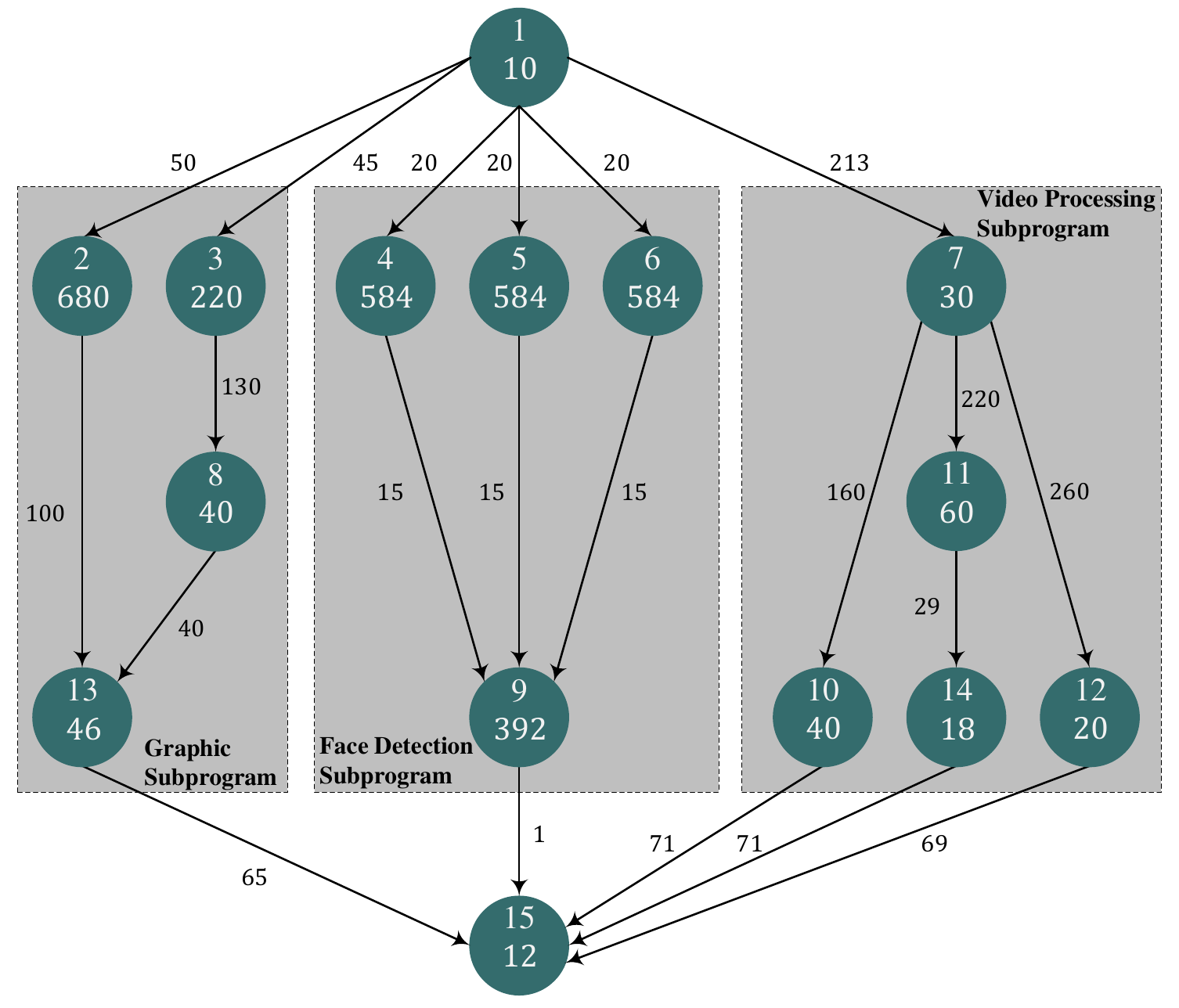}
\caption{Test \textit{DAG4} describing a real-world stream application of radio-navigation. Task workloads and edge weights are in (kbit).}
\label{fig:DAG4}
\end{figure*}

For this purpose, we have considered the test \textit{DAG4} reported in Fig.\ \ref{fig:DAG4}, that is typically considered in the literature for comparing task allocation policies under different service models \cite{mahmoodi2016}. Specifically, \textit{DAG4} details the workflow of a real-world video navigation program for mobile stream application. It involves the parallel execution of three sub-programs, namely a graphic sub-program (left section of Fig.\ \ref{fig:DAG4}), a subprogram for face detection (middle section of Fig.\ \ref{fig:DAG4}), and a video-processing subprogram (right section of Fig.\ \ref{fig:DAG4}). All these sub-programs share the same input and output nodes (i.e., nodes $ 1 $ and $ 15 $ in Fig.\ \ref{fig:DAG4}), that implement data-rendering functionalities and, then, are executed by the Mobile device. \textit{DAG4} is a quite large-size DAG composed of $ 15 $ tasks and $ 21 $ edges. Its topology retains the following features that make it a challenging test DAG: (i) it is asymmetric; (ii) it is composed of the parallel combination of three heterogeneous sub-DAGs, that exhibit fork, parallel and tree-shaped topologies; and, (iii) the face detection and video processing subprograms are computing and communication-intensive, while the graphic subprogram is of mixed type.   

Tables \ref{table:08} and \ref{table:09} report the simulated performance of the proposed \textit{A-GTA-S} under the Eco-centric and Mobile-centric service models of \eqref{eq:EcoCentric} and \eqref{eq:MobileCentric}, respectively. All the reported numerical results have been obtained by running Algorithm \ref{algorithm:alg_2} at $ PS=120 $ and $ G_{MAX}=20 $, and each one refers to the best (i.e., minimum-energy) outcome obtained over $ 10 $ independent runs. An examination of these results leads to three main insights. 

\begin{table*}
\caption{Task allocations and energy consumption of the proposed {\small A-GTA} strategy under \textbf{\small DAG4}. \textbf{\small Eco-centric} case of $ \theta_M = \theta_F = \theta_C = 1 $. $T_{DAG}^{(MAX)}$ is measured in (s) while energy is measured in (Joule).}
\label{table:08}
\centering
\resizebox{0.90\textwidth}{!}{
\setlength{\extrarowheight}{1ex}
\renewcommand{\arraystretch}{1.4}
\begin{tabular}{*{6}{c}}
%
\toprule
\rowcolor[HTML]{BABFC6}
$ \bm{T_{DAG}^{\left(MAX\right)}} $ & \textbf{Mobile Tasks IDs} & \textbf{Fog Tasks IDs} & \textbf{Cloud Tasks IDs} & $ \bm{\mathcal{E}_{A-GTA-MOB}} $ & $ \bm{\mathcal{E}_{A-GTA-TOT}} $ \\[1ex]
\midrule
\rowcolor[HTML]{F2F5F9}
0.3 & $\left\{ 1,7,10-12,14,15 \right\}$ & $\left\{ 2-6,8,9,13 \right\}$ & $\left\{-\right\}$ & 10.33 & 27.10 \\ [0.5ex]
\rowcolor[HTML]{DCE1E8}
0.6 & $\left\{ 1,7,10-12,14,15 \right\}$ &  $\left\{ 2,3,8,13 \right\}$  & $\left\{ 4-6,9 \right\}$ &  9.24 & 23.63 \\ [0.5ex]
\rowcolor[HTML]{F2F5F9}
1.2 & $\left\{ 1,7,10-12,14,15 \right\}$ & $\left\{-\right\}$ & $\left\{ 2-6,8,9,13 \right\}$ &  8.12 & 21.73 \\ [0.5ex]
\rowcolor[HTML]{DCE1E8}
2.4 & $\left\{ 1,7,9-15 \right\}$        & $\left\{-\right\}$ &    $\left\{ 2-6,8 \right\}$   &  7.41 & 21.10 \\ [0.5ex]
\bottomrule
\end{tabular}}
\renewcommand{\arraystretch}{1.0}
\end{table*}
\begin{table*}[ht!]
\caption{Task allocations and energy consumption of the proposed {\small A-GTA} strategy under \textbf{\small DAG4}. \textbf{\small Mobile-centric} case of $ \theta_M = 1 $ and $ \theta_F = \theta_C = 0 $. $T_{DAG}^{(MAX)}$ is measured in (s) while energy is measured in (Joule).}
\label{table:09}
\centering
\resizebox{0.90\textwidth}{!}{
\setlength{\extrarowheight}{1ex}
\renewcommand{\arraystretch}{1.4}
\begin{tabular}{*{6}{c}}
\arrayrulecolor{black}
\toprule
\rowcolor[HTML]{BABFC6}
$ \bm{T_{DAG}^{\left(MAX\right)}} $ & \textbf{Mobile Tasks IDs} & \textbf{Fog Tasks IDs} & \textbf{Cloud Tasks IDs} & $ \bm{\mathcal{E}_{A-GTA-MOB}} $ & $ \bm{\mathcal{E}_{A-GTA-TOT}} $ \\[1ex]
\midrule
\rowcolor[HTML]{F2F5F9}
0.3 & $\left\{ 1,14,15 \right\}$ & $\left\{ 2-8,10-13 \right\}$ &  $\left\{ 9 \right\}$    &  7.38 & 42.66 \\ [0.5ex]
\rowcolor[HTML]{DCE1E8}	
0.6 & $\left\{ 1,14,15 \right\}$ & $\left\{ 3-8,10-13 \right\}$ &  $\left\{ 2,9 \right\}$  &  7.05 & 38.61 \\ [0.5ex]
\rowcolor[HTML]{F2F5F9}
1.2 & $\left\{ 1,14,15 \right\}$ &$\left\{ 3,5-8,10-13 \right\}$& $\left\{ 2,4,9 \right\}$ &  6.65 & 30.01 \\ [0.5ex]
\rowcolor[HTML]{DCE1E8}
2.4 & $\left\{ 1,12,14,15 \right\}$ & $\left\{-\right\}$ &$\left\{ 2-11,13 \right\}$&  6.33 & 26.63 \\ [0.5ex]
\bottomrule
\end{tabular}}
\renewcommand{\arraystretch}{1.0}
\end{table*}

First, a comparison of the task allocation patterns reported in the second columns of Tables \ref{table:08} and \ref{table:09} unveils that, in average, the workload allocated to the Mobile device under the Eco-centric service model is about $ 2.5 $ times larger than the corresponding one under the Mobile-centric case. This is a (first) direct consequence of the fact that both the computing and network resources are made available for-free under the Mobile-centric framework.

Second, a comparison of the numerical values reported in the last columns of Tables \ref{table:08} and \ref{table:09} points out that the total energy $ \mathcal{E}_{A-GTA-TOT} $ consumed by the overall platform of Fig.\ \ref{fig:Mobile-Fog-CloudPlatform} under the Mobile-centric framework is about $ 58.0\% $, $ 45.0\% $, $ 38.0\% $ and $ 26.0\% $ larger than the corresponding one of the Eco-centric case at $ T_{DAG}^{\left(MAX\right)} = 0.3 , 0.6 , 1.2$, and $ 2.4 $ (s), respectively.

Third, the computing-plus-networking energy $ \mathcal{E}_{\scalebox{0.5}{${A\!-\!GTA\!-\!MOB}$}} $ consumed by the Mobile device under the Eco-centric framework is about $ 40.0\% $, $ 31.0\% $, $ 22.0\% $ and $ 17.0\% $ larger than the corresponding one under the Mobile-centric case. In this regard, we have also numerically ascertained that, in the Mobile-centric case, the network component of the overall profiled energy $ \mathcal{E}_{A-GTA-MOB} $ is substantial, and of the order of about: $ 86.8\% $, $ 85.0\% $, $ 72.2\% $ and $ 61.0\% $ at $ T_{DAG}^{\left(MAX\right)} = 0.3 , 0.6 , 1.2$, and $ 2.4 $ (s), respectively.

Overall, the ultimate lesson, which stems from the above discussion, is that both the energy consumption and the task allocation patterns strongly depend on the actually adopted service model.

\subsection{Average energy performance of multi/single-tier ecosystems under randomly time-varying WiFi connectivity}
\label{ssec:AverageEnergy}

The goal of this section is twofold. First, we aim at indagating on the sensitivity of the average energy performance of the simulated ecosystem of Fig.\ \ref{fig:Mobile-Fog-CloudPlatform} when, due to device mobility and limited coverage of the Fog node, the availability of the Mobile-Fog WiFi connection alternates ON-OFF periods in a random way. A related consideration is that the number of involved communication links tend to grow with the number of inter-connected tiers, and this may lead, in turn, to an increment of the network component of the overall consumed energy. Therefore, a second goal of this section is to investigate about the net trade-off among the reduction of the computing energy arising from the utilization of multi-tier computing nodes and the corresponding increment of the network energy needed for their inter-connection.

In order to meet this twofold goal, we have numerically evaluated the energy performance of the ecosystem of Fig.\ \ref{fig:Mobile-Fog-CloudPlatform} under \textit{DAG4} of Fig.\ \ref{fig:DAG4} at $ CCR = 2 $, $ \theta_M = \theta_F = \theta_C = 1 $, and $ T_{DAG}^{\left(MAX\right)} = 1.0 $ (s).

Specifically, in the simulated scenario considered here:
\begin{enumerate}[i.]
\item the up/down Mobile-Cloud cellular connection is permanently ON; 
\item the up/down Mobile-Fog WiFi connection is available only during a fraction $ AV_{WiFi} \in \left[0,1\right]$ of the DAG execution time; and, 
\item the corresponding up/down WiFi throughput $ R_{M \to F} $ and $ R_{F \to C} $ are modeled as two unit-correlated random variables, whose probability density functions (PDFs) are uniform over the corresponding allowed intervals: $ \left[ 0,R_{M \to F}^{MAX} \right] $ and $ \left[ 0,R_{F \to M}^{MAX} \right] $, and present two Dirac's spikes of areas: $ \left( 1 - AV_{WiFi} \right) $ at the origin.
\end{enumerate}

The second column of Table \ref{table:10} reports the energy consumption of the Mobile device, while the third and fourth columns give the total energy wasted by the overall simulated ecosystem of Fig.\ \ref{fig:Mobile-Fog-CloudPlatform}, together with corresponding network energy. In order to account for the random nature of the simulated WiFi connections, each value of Table \ref{table:10} is the average over $ 50 $ independent runs of the proposed \textit{A-GTA-S}.

\begin{table*}
\caption{Average energy consumption (Joule) of the proposed \textit{A-GTA-S} strategy under randomly time-variant availability of the up/down WiFi connections. Case of \textit{DAG4} at $ \theta_M = \theta_F = \theta_C = 1 $ and $ T_{DAG}^{\left(MAX\right)} = 1.0 \, \mathrm{\left(s\right)} $.}
\label{table:10}
\centering
\resizebox{0.8\textwidth}{!}{
\setlength{\extrarowheight}{1.0ex}
\renewcommand{\arraystretch}{1.1}
\begin{tabular}{*{4}{c}}
%
\toprule
\rowcolor[HTML]{BABFC6}
$ \bm{AV_{WiFi}} $ & \textbf{Average} $ \bm{\mathcal{E}_{A-GTA-MOB}} $ & \textbf{Average} $ \bm{\mathcal{E}_{A-GTA-TOT}} $ & \textbf{Average} $ \bm{\mathcal{E}_{A-GTA-NET}} $ \\[1ex]
\midrule
\rowcolor[HTML]{F2F5F9}
1.00 & 7.83  &  22.17 & 7.51 \\[0.5ex]
\rowcolor[HTML]{DCE1E8}
0.75 & 8.40  &  23.70 & 6.10 \\[0.5ex]
\rowcolor[HTML]{F2F5F9}
0.50 & 9.01  &  25.10 & 4.82 \\[0.5ex]
\rowcolor[HTML]{DCE1E8}
0.25 & 10.91 &  26.50 & 3.41 \\[0.5ex]
\rowcolor[HTML]{F2F5F9}
0.00 & 14.33 &  28.98 & 2.51 \\[0.5ex]
\bottomrule
\end{tabular}}
\renewcommand{\arraystretch}{1.0}
\end{table*}

\medskip\noindent\textit{Effects of the intermittent WiFi availability} -- A comparative examination of the columns Table \ref{table:10} points out that the effects of the availability $ AV_{WiFi} $ of the WiFi connection on the reported energy are substantially different. Specifically, by passing from $ AV_{WiFi}=1 $ (i.e., both Fog and Cloud nodes are permanently available for DAG execution) to $ AV_{WiFi}=0 $ (i.e., only the Cloud node is available for DAG execution), we experience that: (i) the average computing-plus-network energy $ \mathcal{E}_{A-GTA-MOB} $ consumed by the Mobile device increases of about $ 83.0\% $ (see the second column of Table \ref{table:10}); (ii) the average total computing-plus-communication energy $ \mathcal{E}_{A-GTA-TOT} $ of the overall ecosystem of Fig.\ \ref{fig:Mobile-Fog-CloudPlatform} increases by about $ 30.7\% $ (see the third column of Table \ref{table:10}); and, (iii) the corresponding average network energy $ \mathcal{E}_{A-GTA-NET} $ consumed by the overall ecosystem decreases by about $ 199.0\% $ (see the last column of Table \ref{table:10}). The common rationale behind these trends is that the execution of more and more tasks are shifted from the Fog node to the Mobile and Cloud ones for decreasing values of $ AV_{WiFi} $. As a matter of this trend, the volume of the total inter-node traffic decreases, but the computing components of the energy wasted by both the Mobile device and the overall ecosystem increase.

\medskip\noindent\textit{Single-tier versus multi-tier ecosystems} -- In order to further corroborate this trend, we have also tested that the corresponding average energy consumption of the (previously introduced) \textit{A-OM} strategy is $ \mathcal{E}_{A-OM} = 89.91 $ (Joule) under the same simulated setting. In this regard, we point out that:
\begin{enumerate}[i.]
\item since, by design, the \textit{A-OM} strategy utilizes only the Mobile device for task execution, this strategy features the performance of a \textit{single-tier} execution platform, whose total energy consumption $ \mathcal{E}_{A-OM} $ equates to the corresponding computing energy (i.e., by design, $ \mathcal{E}_{A-OM-NET} $ is vanishing);
\item since the proposed \textit{A-GTA} strategy exploits, by design, all the actually available computing nodes for task placement, the returned energy $ \mathcal{E}_{A-GTA-TOT} $ of Table \ref{table:10} at $ AV_{WiFi} = 0 $ (resp., $ AV_{WiFi} = 1 $) captures the energy consumption of the two-tier Mobile-Cloud (resp., three-tier Mobile-Fog-Cloud) execution platform embedded in Fig.\ \ref{fig:Mobile-Fog-CloudPlatform}.
\end{enumerate}

Overall, on the basis of these remarks, we conclude that the average computing-plus-networking total energy consumption of the simulated ecosystem of Fig.\ \ref{fig:Mobile-Fog-CloudPlatform} increases for decreasing number of the exploited tiers, and equates to, indeed, $ 22.17 $, $ 28.98 $, and $ 89.91 $ (Juole) when three tiers, two tiers and a single tier are activated, respectively.

This final conclusion provides further full-fledged support both for the multi-tier networked design approach and the Eco-centric perspective pursued by our work.

\section{Conclusion and hints for future research}
\label{sec:Conclusion}

It is expected that the convergence of Fog Computing, Cloud Computing and multi-radio 5G technology allows resource-limited smartphones to support throughput-sensitive mobile stream applications in an energy efficient way. Motivated by this expectation, in this paper, we develop and discuss the main implementation aspects of \textit{EcoMobiFog}, a technological platform for the adaptive joint optimization of the resource allocation and task offloading in 5G-networked virtualized ecosystems composed by an arbitrary number of Fog/Cloud nodes. The energy-delay performance of the solving framework implemented by \textit{EcoMobiFog} is numerically evaluated and compared with respect to the corresponding ones of some state-of-the-art benchmark solutions under a number of operative scenarios that embrace both Eco and Mobile-centric service models.

Being the overall afforded topic still in its infancy, we believe that the presented results could be extended along (at least) four main research directions.

First, the developed solving approach reflects the basic features of the current Middleware management platforms, in which the task scheduling discipline is statically assigned at the compiling time. Hence, including in the afforded \textit{JOP} formulation also the dynamic optimization of the task-execution ordering followed by the computing nodes may be a first research direction of potential interest. The main expected challenge stems from the fact that the dynamic optimization of the task scheduling discipline has been recently proved to be an NP-hard integer-valued problem, even in the basic case of fixed resource allocation \cite{mahmoodi2016}.

A second hint for future research arises from the consideration that 5G technology adopts, by design, massive numbers of transmit/receive antennas at the terminals \cite{Baccarelli2007}. Hence, including the effects of space-time coding and spatial multiplexing \cite{baccarelli2004b} in the energy models of Section \ref{ssec:ModelingComputingEnergy} may be valuable.

A third future research line moves from the consideration that the adaptive framework of Sections \ref{ssec:AdaptiveSolving} and \ref{ssec:DesignTimeVarying} for the dynamic adjustments of the utilized networking-plus-computing resources is purely reactive, i.e., it does not exploit any form of forecasting of the mobility-affected environmental conditions. Including in the solving framework pro-active optimization tools that are capable of predicting future resource utilization \cite{baccarelli1996} could be a further research line of potential interest.

Finally, it could be worthwhile to carry out the implementation of a (small-scale) test-bed of the overall proposed \textit{EcoMobiFog} technological platform of section \ref{sec:Technologicalplatform}, in order to check its performance through real-world field-trials. This is, indeed, the ultimate goal of the (ongoing) \textit{GAUChO} research project (see \url{https://www.gaucho.unifi.it}), that provides the reference framework of this work. 

\section*{Acknowledgment}
\label{sec:ack}

This work was supported in part by the project: ``GAUChO -- A Green Adaptive Fog Computing and networking Architecture'', through the
MIUR Progetti di Ricerca di Rilevante Interesse Nazionale (PRIN) Bando 2015 under Grant 2015YPXH4W\_004, and in part by the
projects: ``Vehicular Fog: energy-efficient QoS mining and dissemination of multimedia Big Data streams'' (V-Fog and V-Fog2),
and: ``SoFT: Fog of Social IoT'', through the Sapienza University of Rome, Bandi 2016, 2017, and 2018.

\appendices

\section{Main taxonomy and simulated setup}
\label{appendix:A}

The following Table \ref{table:11} reports the main symbols used in this paper, their meaning/role, measuring units and simulated values.

\small{
\begin{strip}
\begin{center}
\begin{longtab}
\begin{longtable}{*{4}{l}}
\caption{List of the main parameters, their meaning/role, measuring units and simulated values.}\label{table:11}\\
%
%
\toprule
\rowcolor[HTML]{BABFC6} 
\tone{\textbf{Parameter}} & \ttwo{\textbf{Meaning/Role}} & \tthree{\textbf{Measuring Units}} & \tfour{\textbf{Simulated Settings}} \\ 
%
%
\midrule
%
\rowcolor[HTML]{F2F5F9}
\tone{$Q$} & \ttwo{Number of Fog nodes} & \tthree{Dimensionless} &  \tfour{$Q = 1$} \\
%
\tone{\makecell[cl]{$ \mathcal{A} \overset{\underset{\mathrm{def}}{}}{=}$ \\ $ \left\{ M,F_1,\dots,F_Q,C \right\}$ }} & \ttwo{Set of the available computing nodes} &  \tthree{Dimensionless}  & \tfour{$ \mathcal{A} = \left\{ M,F,C \right\}$}  \\
%
\rowcolor[HTML]{F2F5F9}
\tone{\makecell[cl]{$ {B\!H\!S} \overset{\underset{\mathrm{def}}{}}{=} $ \\ $ \left\{F_1,\dots,F_Q,C \right\} $ }} & \ttwo{Set of nodes of the Backhaul network}  &  \tthree{Dimensionless}  & \tfour{$ {B\!H\!S} = \left\{ F,C \right\}$} \\ 
%
\tone{$n_{_N}, N \in \mathcal{A} $} & \ttwo{Number of the (virtual) computing cores equipping the computing node $ N $}  & \tthree{Dimensionless} & \tfour{$ n_M=1, n_F=4, n_C=12 $} \\ 
%
\rowcolor[HTML]{F2F5F9}
\tone{$ \overline{NF}_{{N}' \to {N}''} $} & \ttwo{Average number of failures of the connection from computing node $ {N}' $ to computing node $ {N}'' $ } & \tthree{Dimensionless} & \tfour{\makecell[cl]{ $ \overline{NF}_{M \to F} = \overline{NF}_{F \to M} = 1.1 $ \\ \\ $ \overline{NF}_{M \to C} = \overline{NF}_{C \to M} = 0.1 $ \\ \\ $ \overline{NF}_{F \to C} = \overline{NF}_{C \to F} = 0.01 $}}  \\
%
\tone{$V$} & \ttwo{Number of tasks of the application DAG}  & \tthree{Dimensionless} & \tfour{$ V \geq 9 $} \\
%
\rowcolor[HTML]{F2F5F9}
\tone{$ \vec{x} = \left[x_1,\dots,x_V\right] $} & \ttwo{Vector of task allocation, with component $ x_i \in \mathcal{A} $} & \tthree{Dimensionless} & \tfour{Optimization variable} \\ 
%
\tone{$f_{N}, N \in \mathcal{A} $} & \ttwo{Per-core computing frequency at the computing node $ N $}  & \tthree{$ \mathrm{bit/s} $} & \tfour{Optimization variable} \\ 
%
%
\rowcolor[HTML]{F2F5F9}
\tone{$f_{N}^{\left(M\!A\!X\right)}, N \in \mathcal{A} $} & \ttwo{Per-core maximum computing frequency at the computing node $ N $}  & \tthree{$ \mathrm{bit/s} $} & \tfour{\makecell[cl]{ $ f_{M}^{\left(MAX\right)} = 12 \times 10^{6}$ \\ \\  $ f_{F}^{\left(MAX\right)} = 12 \times 10^{6}$ \\ \\  $ f_{C}^{\left(MAX\right)} = 12 \times 10^{6}$}} \\ 
%
\tone{$\mathcal{E}_{N}, N \in \mathcal{A} $} & \ttwo{Computing energy consumed by the device clone at node $ N $}  & \tthree{$ \mathrm{Joule} $} & \tfour{To be optimized} \\ 
\rowcolor[HTML]{F2F5F9}
\tone{$ R_{M \to N} $} & \ttwo{Up throughput of the TCP/IP connection from the Mobile to the Cloud/Fog node $ N \in {B\!H\!S} $} &  \tthree{$ \mathrm{bit/s} $}  & \tfour{Optimization variable} \\ 
%
\tone{$ R_{N \to M} $} & \ttwo{Down throughput of the TCP/IP connection from the Cloud/Fog node $ N \in {B\!H\!S} $ to the Mobile} &  \tthree{$ \mathrm{bit/s} $} & \tfour{Optimization variable} \\ 
%
\rowcolor[HTML]{F2F5F9}
\tone{$ \mathcal{E}_{M \to N}, \mathcal{E}_{N \to M} $} & \ttwo{Energy consumed by the one-way wireless connections: $ {M \to N} $, and $ {N \to M} $, with $ N \in {B\!H\!S} $} &  \tthree{$ \mathrm{Joule} $} & \tfour{To be optimized} \\ 
%
\tone{$ R_{M \to N}^{\left(MAX\right)} $} & \ttwo{Maximum throughput of the TCP/IP connection from the Mobile to the Cloud/Fog node $ N \in {B\!H\!S} $} &  \tthree{$ \mathrm{bit/s} $} & \tfour{\makecell[cl]{ $ R_{M \to F}^{\left(MAX\right)} = 8.0 \times 10^{6}$ \\ \\ $ R_{M \to C}^{\left(MAX\right)} = 6.5 \times 10^{6}$}} \\ 
%
\rowcolor[HTML]{F2F5F9}
\tone{$ R_{N \to M}^{\left(MAX\right)} $} & \ttwo{Maximum throughput of the TCP/IP connections from the Cloud/Fog node $ N \in {B\!H\!S} $ to the Mobile} &  \tthree{$ \mathrm{bit/s} $} & \tfour{\makecell[cl]{ $ R_{F \to M}^{\left(MAX\right)} = 9.0 \times 10^{6}$ \\ \\ $ R_{C \to M}^{\left(MAX\right)} = 7.0 \times 10^{6}$}} \\ 
%
\tone{$ R_{N_1 \leftrightarrow N_2} $} & \ttwo{Throughput of the backhaul TCP/IP two-way connection between nodes $ N_1 $, $ N_2 $, with $ N_1 \ne N_2 $, and $ N_1,N_2 \in {B\!H\!S} $} &  \tthree{$ \mathrm{bit/s} $} & \tfour{$ R_{C \leftrightarrow F} = 3.7 \times 10^{6}$}  \\
%
\rowcolor[HTML]{F2F5F9}
\tone{$ \mathcal{E}_{N_1 \leftrightarrow N_2} $} & \ttwo{Energy consumed by the two-way backhaul connection: $ N_1 \leftrightarrow N_2 $, with $ N_1 \ne N_2 $, and $ N_1,N_2 \in {B\!H\!S} $} &  \tthree{$ \mathrm{Joule} $} & \tfour{To be optimized} \\ 
%
\tone{$ T_{DAG} $} & \ttwo{Total execution time of the considered application DAG} &  \tthree{$ \mathrm{s} $}  & \tfour{To be optimized}  \\
%
\rowcolor[HTML]{F2F5F9}
\tone{$ T_{DAG}^{\left(M\!AX\right)} \overset{\underset{\mathrm{def}}{}}{=} \frac{1}{ T \! H_{0}^{\left(M\!I\! N\right)}}$} & \ttwo{Per-DAG maximum allowed execution time} &  \tthree{$ \mathrm{s} $}  & \tfour{$ 0.3 \leq T_{DAG}^{\left(M\!AX\right)} \leq 2.4 $}  \\
%
\tone{$ T_{N}^{\left(SER\right)} , \left( N \in \mathcal{A} \right) $} & \ttwo{Service time of node $ N $} &  \tthree{$ \mathrm{s} $}  & \tfour{To be optimized}  \\
%
\rowcolor[HTML]{F2F5F9}
\tone{\makecell[cl]{$ T_{i,N}^{\left(EXE\right)} , $ \\ $ N \in \mathcal{A} , i=1,\dots,V $}}  & \ttwo{Execution time of the $ i $-th task at node $ N $} &  \tthree{$ \mathrm{s} $}  & \tfour{To be optimized}  \\
%
\tone{$ \theta_N \in \left\{ 0,1 \right\} $} & \ttwo{Binary parameter. It is zero (resp., unit valued) if the computing-plus-networking energy consumed by the computing node $ N \in \left\{ M,F,C \right\} $ is (resp., is not) for free}   &  \tthree{Dimensionless} & \tfour{\makecell[cl]{ $ \theta_M = \theta_F = \theta_C = 1 $ \\ \\ $ \theta_M =1,  \theta_F = \theta_C = 0 $ }}  \\
%
\rowcolor[HTML]{F2F5F9}
\tone{$ nc_{N} $} & \ttwo{Number of containers simultaneously running atop the CPU at the computing node $ N \in\mathcal{A} $}  &  \tthree{Dimensionless} & \tfour{\makecell[cl]{ $ nc_{M}= 1 $ \\ $ nc_{F}= 16 $ \\ $ nc_{C}= 24 $ }}  \\
\tone{$ \gamma_{_N} $} & \ttwo{Positive exponent of the dynamic power consumption of the CPU at the computing node $ N \in\mathcal{A} $}  &  \tthree{Dimensionless} & \tfour{\makecell[cl]{ $ \gamma_{_M}= 3.2 $ \\ $ \gamma_{_F}= 3.1 $ \\ $ \gamma_{_C}= 3.0 $ }}  \\
%
\rowcolor[HTML]{F2F5F9}
\tone{$ k_{N} $} & \ttwo{Positive scaling factor profiling the dynamic power consumption of the CPU at the computing node $ N \in \mathcal{A} $} &  \tthree{$ \displaystyle \frac{\mathrm{Watt}}{{\mathrm{(bit/s)}}^{^{\gamma_{_N}}}} $} & \tfour{\makecell[cl]{ $ k_{M} = 7.50 \times 10^{-21} $ \\ $ k_{F} = 9.78 \times 10^{-20} $ \\ $ k_{C} = 1.14 \times 10^{-19} $ }}  \\
%
\tone{$ r_{N} $} & \ttwo{Fraction of the overall computing power shared by the cores at the computing node $ N \in \mathcal{A} $} &  \tthree{Dimensionless} & \tfour{\makecell[cl]{ $ r_{M} = 0.0  $ \\ $ k_{F} = 0.2 $ \\ $ k_{C}  = 0.1 $ }} \\
%
\rowcolor[HTML]{F2F5F9}
\tone{$ \mathcal{P}^{\left( IDLE \right)}_{CPU-N} $} & \ttwo{Power consumed in the idle state by the physical CPU at the computing node $ N \in \mathcal{A} $} &  \tthree{$ \mathrm{Watt} $} & \tfour{\makecell[cl]{ $ \mathcal{P}^{\left( IDLE \right)}_{CPU-M} = 1.2 $ \\ \\ $ \mathcal{P}^{\left( IDLE \right)}_{CPU-F} = 220 $ \\ \\ $ \mathcal{P}^{\left( IDLE \right)}_{CPU-C} = 440  $ }} \\ 
%
\tone{$ \mathcal{P}^{\left( IDLE \right)}_{BHNET-N} $} & \ttwo{Power consumed in the idle state by each physical Ethernet NIC at the Fog and Cloud nodes} &  \tthree{$ \mathrm{Watt} $} & \begingroup\makeatletter\def\f@size{7}\check@mathfonts $ \mathcal{P}^{\left( IDLE \right)}_{BHNET-F} = \mathcal{P}^{\left( IDLE \right)}_{BHNET-C} =10^{-14} $ \endgroup  \\
%
\rowcolor[HTML]{F2F5F9}
\tone{$ \mathcal{P}^{\left( IDLE \right)}_{SRNET-N-k} $} & \ttwo{Power consumed in the idle state by the $ k $-th  short-range physical NIC at node $ N $, with $ N \in \left\{ M,F_1,\dots,F_Q \right\} $ and $ k = 1, \dots, Q $} &  \tthree{$ \mathrm{Watt} $} & \begingroup\makeatletter\def\f@size{7}\check@mathfonts $ \mathcal{P}^{\left( IDLE \right)}_{SRNET-M} = \mathcal{P}^{\left( IDLE \right)}_{SRNET-F} = 1.3 $ \endgroup  \\
%
\tone{$ \mathcal{P}^{\left( IDLE \right)}_{LRNET} $} & \ttwo{Power consumed in the idle state by each physical long-range NIC at the Mobile device and  Cloud node}  &  \tthree{$ \mathrm{Watt} $} & \tfour{$ \mathcal{P}^{\left( IDLE \right)}_{LRNET} = 0.82 $}  \\
%
\rowcolor[HTML]{F2F5F9}
\tone{$\xi^{\left(Tx\right)}_{\left(N_1,N_2\right)}$} & \ttwo{Positive exponent of the dynamic power consumption of the wireless NIC connecting the computing nodes $ N_1 $ and $ N_2 $ and operating in the transmit mode} & \tthree{Dimensionless} & \tfour{\makecell[cl]{$\xi^{\left(Tx\right)}_{\left(M,F\right)} = 2.40 $ \\ \\ $\xi^{\left(Tx\right)}_{\left(M,C\right)} = 2.45 $ }} \\
%
\tone{$\xi^{\left(Rx\right)}_{\left(N_1,N_2\right)}$} & \ttwo{Positive exponent of the dynamic power consumption of the wireless NIC connecting the computing nodes $ N_1 $ and $ N_2 $ and operating in the receive mode} & \tthree{Dimensionless} & \tfour{\makecell[cl]{$\xi^{\left(Rx\right)}_{\left(M,F\right)} = 2.20 $ \\ \\ $\xi^{\left(Rx\right)}_{\left(M,C\right)} = 2.34$}} \\
%
\rowcolor[HTML]{F2F5F9}
\tone{$\eta$} & \ttwo{Positive exponent of the RTTs of the short and long-range TCP/IP wireless connections} & \tthree{Dimensionless} & \tfour{$\eta = 0.6$} \\
%
\tone{$ RTT_{\left( M,N \right)} $} & \ttwo{Average RTT of the TCP/IP short-range connection between the Mobile device and Fog node $ N \in \left\{ F_1,\dots,F_Q \right\} $} & \tthree{$ \mathrm{s} $} & \tfour{$ RTT_{\left( M,F \right)}  = 1.0 \times 10^{-3}$} \\
%
\rowcolor[HTML]{F2F5F9}
\tone{$ RTT_{\left( M,C \right)} $} & \ttwo{Average RTT of the Mobile-Cloud TCP/IP long-range cellular connection} & \tthree{$ \mathrm{s} $} & \tfour{$ RTT_{\left( M,C \right)} = 1.0 \times 10^{-2}$} \\
%
\tone{$ RTT_{\left( N_1,N_2 \right)} $} & \ttwo{Average RTT of the (possibly, multi-hop) two-way TCP/IP backhaul connection 
between nodes $ N_1 $ and $ N_2 $, with $ N_1 \ne N_2 $ and $ N_1,N_2 \in {B\!H\!S} $} & \tthree{$ \mathrm{s} $} & \tfour{$ RTT_{\left( N_1,N_2 \right)} $} \\
%
\rowcolor[HTML]{F2F5F9}
\tone{$ MSS^{\left( N_1,N_2 \right)} $} & \ttwo{Maximum size of a TCP segment of the TCP/IP connection between the computing nodes $ N_1 $ and $ N_2 $, with $ N_1 \ne N_2 $ and $ N_1,N_2 \in \mathcal{A} $} & \tthree{$ \mathrm{bit} $} & \tfour{$ MSS= 12.0 \times 10^{3} $} \\
%
\tone{$Pr_{_{LOSS}}^{\left( N_1,N_2 \right)}$} & \ttwo{Loss probability of the TCP/IP backhaul connection between nodes $ N_1 $ and $ N_2 $, with $ N_1 \ne N_2 $ and $ N_1,N_2 \in {B\!H\!S} $} & \tthree{Dimensionless} & \tfour{$ Pr_{_{LOSS}}^{\left( N_1,N_2 \right)} = 1.56 \times 10^{-5}$} \\
%
\rowcolor[HTML]{F2F5F9}
\tone{$ \Omega ^{\left(Tx\right)}_{\left(N_1,N_2\right)}$} & \ttwo{Power profile of the wireless NIC connecting the computing nodes $ N_1 $, $ N_2 $ in the transmit mode} & \tthree{$ \displaystyle \frac{\mathrm{Watt}}{\mathrm{(bit/s)}^ {\xi^{\left(Tx\right)}_{\left(N_1,N_2\right)} } \! \!\times \! \! \mathrm{(s)}^{\eta}} $} & \tfour{\makecell[cl]{$ \Omega ^{\left(Tx\right)}_{\left(M,F\right)} = 5.00 \times 10^{-14}$ \\ \\ $ \Omega ^{\left(Tx\right)}_{\left(M,C\right)} = 2.31 \times 10^{-13}$ }} \\
%
\tone{$ \Omega ^{\left(Rx\right)}_{\left(N_1,N_2\right)}$} & \ttwo{Power profile of the wireless NIC connecting the computing nodes $ N_1 $, $ N_2 $ in the receive mode} & \tthree{$ \displaystyle \frac{\mathrm{Watt}}{\mathrm{(bit/s)}^ {\xi^{\left(Rx\right)}_{\left(N_1,N_2\right)} } \! \! \times \! \! \mathrm{(s)}^{\eta}} $} & \tfour{\makecell[cl]{$ \Omega ^{\left(Rx\right)}_{\left(M,F\right)} = 1.40 \times 10^{-14}$ \\ \\ $ \Omega ^{\left(Rx\right)}_{\left(M,C\right)} = 8.10 \times 10^{-15}$ }} \\
%
\rowcolor[HTML]{F2F5F9}
\tone{$ I_{MAX} $} & \ttwo{Maximum number of primal-dual iterations performed by the \textit{RAP}} & \tthree{Dimensionless} & \tfour{$ 500 \leq I_{MAX} \leq 700 $} \\ 
%
\tone{$ a_{MAX} $} & \ttwo{Clipping factor of the \textit{RAP} iterations} & \tthree{Dimensionless} & \tfour{$ 10^{-7} \leq a_{MAX} \leq 8.0 \times 10^{-7} $} \\ 
%
\rowcolor[HTML]{F2F5F9}
\tone{$no_{HOP}^{\left( N_1,N_2 \right)}$} & \ttwo{Number of hops of the backhaul connection between nodes $ N_1 $ and $ N_2 $, with $ N_1 \ne N_2 $ and $ N_1,N_2 \in {B\!H\!S} $} & \tthree{Dimensionless} & \tfour{$ no_{HOP}^{\left( C,F \right)} = 4 $} \\
%
\tone{$\mathcal{P}_{HOP}^{\left( N_1 \to N_2 \right)}$} & \ttwo{One-way per-hop average power consumed by the backhaul connection between nodes $ N_1 $ and $ N_2 $, with $ N_1 \ne N_2 $, and $ N_1,N_2 \in {B\!H\!S} $} & \tthree{$ \mathrm{Watt} $} & \tfour{$ \mathcal{P}_{HOP}^{\left( C \to F \right)} = 2.85 \times 10^{-1} $} \\
%
\rowcolor[HTML]{F2F5F9}
\tone{$ PS $} & \ttwo{Population size of the Genetic algorithms} & \tthree{Dimensionless} & \tfour{$ 4 \leq PS \leq 120 $} \\
%
\tone{$ CF $} & \ttwo{Fraction of the population size that undergoes Genetic crossover} & \tthree{Dimensionless} & \tfour{$ CF = 0.5 $} \\
%
\rowcolor[HTML]{F2F5F9}
\tone{$ G_{MAX} $} & \ttwo{Number of generations run by the Genetic algorithms} & \tthree{Dimensionless} & \tfour{\makecell[bl]{$ 10 \le G_{MAX} \leq 20 $ }} \\
%
\tone{$ MN $} & \ttwo{Number of elements of each task allocation vector that undergo Genetic mutation} & \tthree{Dimensionless} & \tfour{$ MN = \text{round}  \left( \left( V-2 \right) / 2 \right) $} \\
\specialrule{0.7pt}{0pc}{0pc}
%
%
\end{longtable}
\end{longtab}
\end{center}
\end{strip}
}

\normalsize

\begin{figure*}[!t]
\normalsize
\setcounter{tempcount}{\value{equation}}
\setcounter{equation}{\value{tempcount}}
\begin{align}
T_{DAG}^{\left(UP\right)}  \equiv
\sum_{i=1}^{V}  {T_{EXE}^{\left(MAX\right)}}  
=
V \left(
\frac{s^{\left(MAX\right)}}{\underset{N \in \mathcal{A}}{\min}\left\{n_N f_N^{\left(MAX\right)}\right\}}
+
\frac{w_{IN}^{\left(MAX\right)}\left(1+\underset{N_1,N_2 \in \mathcal{A}}{\max}\left\{\overline{NF}_{N_1 \to N_2}\right\}\right)}{\underset{N_1,N_2 \in \mathcal{A}}{\min}\left\{R_{N_1 \to N_2}^{\left(MAX\right)}\right\}}
\right), \qquad \text{\textit{SEQ -- STS}}
\label{eq:appT_DAGUP-Sequential}
\end{align}
	\begin{align}
	T_{DAG}^{\left(UP\right)} & \equiv 
	\underset{1 \leq i \leq V}{\max}
	\left\{ T_{EXE}^{\left(MAX\right)} \right\}  \nonumber \\
	& =
	\left(
	\frac{s^{\left(MAX\right)}}{ \left( \frac{\underset{1 \leq i \leq V}{\min}\left\{\phi_i\right\}}{\sum_{j=1}^{V}{\phi_{j}}} \right)  \times \underset{N \in \mathcal{A}}{\min}\left\{n_N f_N^{\left(MAX\right)}\right\}}
	+
	\frac{w_{IN}^{\left(MAX\right)}  \left( 1 + \underset{N_1,N_2 \in \mathcal{A}}{\max}\left\{\overline{NF}_{N_1 \to N_2}\right\}\right)}{\underset{N_1,N_2 \in \mathcal{A}}{\min}\left\{R_{N_1 \to N_2}^{\left(MAX\right)}\right\}}
	\right), \qquad \text{\textit{WPS -- PTS}}
	\label{eq:appT_DAGUP-Parallel}
	\end{align}
	\setcounter{equation}{\value{tempcount}}
	\addtocounter{equation}{2}
	\hrulefill
	\vspace*{4pt}
\end{figure*}

\section{ Proof of the condition for the JOP feasibility}
\label{appendix:B}

The proof of Proposition \ref{proposition:proposition1} exploits some basic formal properties of $ T_{DAG} $ that are reported in the following Lemma \ref{lemma:lem1}:

\begin{lemma}[\textit{Formal properties of $ T_{DAG} $}]\label{lemma:lem1}
$ $
Let the assumptions on $ T_{DAG} $ of Section \ref{ssec:Per-DAGexecutiontimes} be met. Then, we have that:
\begin{enumerate}[a.]
\item $ T_{DAG} $ is a jointly convex function of the $ \left(3Q+4\right) $ scalar optimization variables gathered by the resource vector $ \overrightarrow{RS} $ in \eqref{eq:RS};
\item $ T_{DAG} $ is a non-decreasing function of the task sizes: $ \left\{ s_i , i=1,\dots,V \right\} $ and edge weights $ \left\{ d_{ij} \left(i,j\right) \in E \right\} $ of the considered application DAG. Furthermore, $ T_{DAG} $ is a non-increasing function of the processing capacities: $ \left\{ n_N f_N, N \in \mathcal{A} \right\} $ of the computing nodes and the transport throughput: $ \left\{ R_{N_1 \to N_2}, N_1 \ne N_2 ; N_1,N_2 \in \mathcal{A} \right\} $ of the underlying network connections.
\end{enumerate}
\hfill\ensuremath{\blacksquare}
\end{lemma}

\begin{proof}
$ $
\begin{enumerate}[a)]
\item According to the assumptions reported in Sections \ref{ssec:Servicedisciplines} and \ref{ssec:executiontimes}, each per-task service time $ T_{N,i}^{\left(SER\right)} $ (resp., per-task network delay $ T_{N,i}^{\left(NET\right)} $) is convex with respect to the corresponding computing frequency $ f_N $ (resp., the connection throughput $ R_{N_1 \to N} $), because, by design, it scales as $ 1/f_N $   (resp., $ 1/R_{N_1 \to N} $). Hence, each per-task execution time $ T_{N,i}^{\left(EXE\right)} $ in \eqref{eq:T_i,N-EXE} is also convex in the involved optimization variables, because it is the summation of two convex functions. As a consequence, since $ T_{DAG} $  is, by assumption, a jointly convex and non-decreasing composition of convex functions (see Section \ref{ssec:Per-DAGexecutiontimes}), it is jointly convex in the optimization variables gathered by the resource vector $ \overrightarrow{RS} $ in \eqref{eq:RS}.
\item By design, each per-task service time $ T_{N,i}^{\left(SER\right)} $ (resp., each per-task network delay $ T_{N,i}^{\left(NET\right)} $) does not decrease for increasing task sizes  $ \left\{ s_i , i=1,\dots,V \right\} $ (resp., edge weights $ \left\{ d_{ij} \left(i,j\right) \in E \right\} $), while it does not increase for increasing processing capacities $ \left\{ n_N f_N, N \in \mathcal{A} \right\} $ (resp., transport throughput $ \left\{ R_{N_1 \to N_2}, N_1 \ne N_2 ; N_1,N_2 \in \mathcal{A} \right\} $) (see the assumptions of Sections \ref{ssec:Servicedisciplines} and \ref{ssec:executiontimes}). Hence, being the summation of the corresponding per-task service time and network delay, the same monotonic properties are also retained by the resulting per-task execution time $ T_{N,i}^{\left(EXE\right)} $ in \eqref{eq:T_i,N-EXE}. As a consequence, by assumption, $ T_{DAG} $ is not decreasing with respect to each per-task execution time (see Section \ref{ssec:Per-DAGexecutiontimes}), the validity of the stated monotonic properties directly follows.
\end{enumerate}
\end{proof}

By leveraging the stated formal properties of $ T_{DAG} $, we note that:
\begin{enumerate}[i.]
\item $ Tmax^{\left(SER\right)} $ in \eqref{eq:Tmax_ser} is a feasible upper bound on all task execution times. This is due to the fact that $ Tmax^{\left(SER\right)} $ is computed by jointly considering the maximum task size (see the numerator of \eqref{eq:R_N1N2}), together with the minimum per-task fraction of the per-node computing frequency and the minimum of the allowed per-node maximum processing frequencies (see the product at the denominator of \eqref{eq:Tmax_ser}); and,
\item $ Tmax^{\left(NET\right)} $ in \eqref{eq:Tmax_net} is a feasible upper bound on all network times. This is due to the fact that $ Tmax^{\left(NET\right)} $ is computed by jointly considering the maximum volume of the per-task input data and the maximum network failure factor (see the product at the numerator of \eqref{eq:Tmax_net}), together with the minimum of the per-connection maximum throughput (see the denominator of \eqref{eq:Tmax_net}).
\end{enumerate}

As a consequence, the resulting $ T_{EXE}^{\left(MAX\right)} $ in \eqref{eq:T_ExeMax} constitutes, by design, a feasible upper bound on the set of the per-task execution times. Hence, the validity of the feasibility condition in \eqref{eq:T_DAGup} directly arises from the not decreasing behavior of $ T_{DAG} $ with respect to the per-task execution times (see Section \ref{ssec:Per-DAGexecutiontimes}).

Before proceeding, two explicative remarks about the meaning/role of the reported feasibility condition are in order.

First, the sufficiency of this condition stems from the fact that it considers the \textit{worst} case in which the task of maximum size is also the task whose execution requires the maximum volume of input data (see $ s^{\left(MAX\right)} $ and $ w_{IN}^{\left(MAX\right)} $ at the numerators of \eqref{eq:Tmax_ser} and \eqref{eq:Tmax_net}, respectively). Second, the evaluation of $ T_{DAG}^{\left(UP\right)} $ in \eqref{eq:T_DAGup} may be carried out in closed-form by exploiting only the defining parameters of the considered \textit{JOP}. Just as application examples, in the case of the (previously introduced) Sequential service and scheduling disciplines, $ T_{DAG}^{\left(UP\right)} $ is computed as in \eqref{eq:appT_DAGUP-Sequential}, while, in the case of intra-node WPS service discipline and inter-node Parallel Task Scheduling discipline, \eqref{eq:appT_DAGUP-Parallel} holds.

\section{ Proof of the RAP convexity}
\label{appendix:C}

The proof of the \textit{RAP} convexity relies on the formal properties of the per-node computing energy and per-connection wireless network energy proved in the following Lemma \ref{lemma:lem2} and Lemma \ref{lemma:lem3}, respectively.

\begin{lemma}[\textit{On the convexity of the per-node computing energy}]\label{lemma:lem2}
$ $
Let the assumptions on $ T_{DAG} $ of Section \ref{ssec:Per-DAGexecutiontimes} be met. Furthermore, let the exponents of the computing energy of \eqref{eq:E_N} meet the following inequality:
\begin{equation}
\gamma_N  \geq  2, \quad  N \in \mathcal{A} ,
\label{eq:appgamma_Ninequality}
\end{equation}	
Then, each computing energy $ \mathcal{E}_N , N \in \mathcal{A} $, is jointly convex in the $ \left(3Q+4\right) $ scalar optimization variables gathered by the resource vector $ \overrightarrow{RS} $ in \eqref{eq:RS}.
\hfill\ensuremath{\blacksquare}
\end{lemma}


\begin{proof}
$ $
Since $ \mathcal{E}_N $ is the summation of a static part $ \mathcal{E}_N^{\left(STA\right)} $ and a dynamic one $ \mathcal{E}_N^{\left(DYN\right)} $, it suffices to separately prove the convexity of $ \mathcal{E}_N^{\left(STA\right)} $ and $ \mathcal{E}_N^{\left(DYN\right)} $. In this regard, we note that:
\begin{enumerate}[i.]
\item an inspection of the first term on the RHS of \eqref{eq:E_N} points out that $ \mathcal{E}_N^{\left(STA\right)} $ depends on the involved optimization variables only through the DAG execution time $ T_{DAG} $, whose convexity has been already proved by Lemma \ref{lemma:lem1};
\item $ \mathcal{E}_N^{\left(DYN\right)} $ depends on the computing frequency $ f_N $ through the product: $ \left(f_N\right)^{\gamma_N} T_N^{\left(SER\right)} $. Hence, since $ T_N^{\left(SER\right)} $ scales, by assumption, as: $ 1/f_N $ (see Section \ref{ssec:executiontimes}), the above product scales as:  $ \left(f_N\right)^{\gamma_N -1} $, which, in turn, is a convex function in $ f_N $ for $ \gamma_N \geq 2 $.
\end{enumerate}
This completes the proof.
\end{proof}

\begin{lemma}[\textit{On the convexity of the network energy of the wireless connections}]\label{lemma:lem3}
$ $
Let the assumptions on $ T_{DAG} $ of Section \ref{ssec:Per-DAGexecutiontimes} be met. Furthermore, let the exponents of the wireless network power of \eqref{eq:P_MtoNdynamic} meet the following inequalities:
\begin{equation}
\xi_{\left(M,N\right)}^{\left(Tx\right)}  \geq 2, \,
\text{ and }
\xi_{\left(M,N\right)}^{\left(Rx\right)}  \geq 2, \,
\quad N \in {BHS} .
\label{eq:appxi_MNinequality}
\end{equation}	
Then, the network energy $ \mathcal{E}_{M \to N} $ and $ \mathcal{E}_{N \to M} $, $ N \in {BHS} $, of each up/down wireless connection are jointly convex in the $ \left(3Q+4\right) $ scalar optimization variables gathered by the resource vector $ \overrightarrow{RS} $ in \eqref{eq:RS}.
\hfill\ensuremath{\blacksquare}
\end{lemma}


\begin{proof}
$ $
After noting that each wireless network energy is still the summation of a static and dynamic component (see Section \ref{ssec:ModelingComputingEnergy}), the proof can be carried out by replicating the same steps already reported for the proof of Lemma \ref{lemma:lem2}.
\end{proof}

About the backhaul network of Fig.\ \ref{fig:multi-tier}, we point out that the total energy $ \mathcal{E}_{BH-NET} $ in \eqref{eq:E_BHNET} consumed by the (two-way) backhaul connections: $ \left\{ N_1 \leftrightarrow N_2 , N_1 \ne N_2, \, N_1,N_2 \in BHS \right\} $ depends on the involved optimization variables only through $ T_{DAG} $ (see the last part of Section \ref{ssec:ModelingComputingEnergy}), that, in turn, is guaranteed to be convex by Lemma \ref{lemma:lem1}.  


In order to formally prove the convexity of the \textit{RAP}, it suffices to note that:
\begin{enumerate}[i.]
\item under any given task allocation vector $ \vec{x} $, the objective function: $ \mathcal{E}_{TOT} $ in \eqref{eq:JOP1} of the \textit{RAP} is a linear superposition (with positive coefficients) of energy terms, whose convexity are guaranteed by the results of Lemmas \ref{lemma:lem1}, \ref{lemma:lem2} and \ref{lemma:lem3}; and,
\item the convexity of $ T_{DAG} $ assures that the inequality-type constraint in \eqref{eq:JOP2} on the required application throughput is convex.
\end{enumerate}
The proof of Proposition \ref{proposition:proposition3} is now complete.


\section{ Proof of the condition for the RAP feasibility}
\label{appendix:D}

Let us proceed to prove the sufficient and necessary parts of the \textit{RAP} feasibility condition of Proposition \ref{proposition:proposition4}.


\hspace*{\parindent}\textit{Sufficient part} -- Let us assume that the condition in \eqref{eq:TH_0mintimesT_DAGinequality} is met under the assigned task allocation vector $ \vec{x} $. This means, in turn, that $ \overrightarrow{RS}^{\left(MAX\right)} $ also meets this condition. Since, by design, $ \overrightarrow{RS}^{\left(MAX\right)} $ also meets all the box constraints in \eqref{eq:JOP3} -- \eqref{eq:JOP5} on the maximum allowed resources, it is a feasible solution of the \textit{RAP}, and, then, the \textit{RAP} is feasible.

\textit{Necessary part} -- The proof is by contradiction. Hence, let us assume that the condition in \eqref{eq:TH_0mintimesT_DAGinequality} is not met under the assigned task allocation vector $ \vec{x} $. Since Lemma \ref{lemma:lem1} guarantees that $ T_{DAG} $ is a non-increasing function of each component of the resource allocation vector, we would increase at least a component of $ \overrightarrow{RS}^{\left(MAX\right)} $, in order to decrease the value of $ T_{DAG} $ and, then, attempt to meet the inequality in \eqref{eq:TH_0mintimesT_DAGinequality}. However, so doing, at least one of the frequency/throughput resources would violate its upper bound and this would give rise to an infeasible resource allocation vector. This proves, in turn, that the condition in \eqref{eq:TH_0mintimesT_DAGinequality} is necessary for the \textit{RAP} feasibility.


\section{ Proof of the satisfaction of the Slater's qualification condition}
\label{appendix:E}

Let us assume that the conditions of Proposition \ref{proposition:proposition3} are met, so that the \textit{RAP} is a convex optimization problem in the $ \overrightarrow{RS} $ optimization vector variables. Hence, the Slater's qualification condition requires that there exists at least a \textit{feasible} resource allocation vector that meets the convex constraint in \eqref{eq:JOP2} with the strict inequality (see, for example, \cite[Section 5.3]{bazaraa2017}). However, if the feasibility condition in \eqref{eq:TH_0mintimesT_DAGinequality} is met with the strict inequality, the vector $ \overrightarrow{RS}^{\left(MAX\right)} $ of the maximal resources is, by design, feasible, and satisfies the convex constraint in \eqref{eq:JOP2} with the strict inequality. This proves, in turn, that the Slater's qualification condition holds.

\section{ Expressions of the derivatives of the Lagrangian function}
\label{appendix:F}

From the definition of Lagrangian function in \eqref{eq:Lagrangian}, the corresponding (scalar) gradients with respect to the per-node computing frequencies, per-connection throughput and Lagrange multiplier read as follows: 
\begin{align}
\frac{\partial \mathcal{L}}{\partial f_N} = &
\sum_{{N}' \in \mathcal{A}}{ \theta_{{N}'}  \left( \frac{\partial \mathcal{E}_{{N}'}}{\partial f_N} \right)   }  +
\sum_{{N}' \in \mathcal{A}}{
\sum_{\begin{subarray}{l} {N}'' \in \mathcal{A} \\ {N}' \ne {N}'' \end{subarray}}{
\frac{\partial \mathcal{E}_{{N}' \leftrightarrow {N}''}}{\partial f_N}
}} \nonumber \\
& + \lambda \times {T\!H}_{0}^{\left(MIN\right)} \left( \frac{\partial T_{DAG}}{\partial f_N} \right)
, \; N \in \mathcal{A} ,
\label{eq:appDelLagDelf}
\end{align}
\begin{align}
\frac{\partial \mathcal{L}}{\partial R_{N_1 \to N_2}} \! = \! \! &
\sum_{{N}' \in \mathcal{A}}{ \!\! \theta_{{N}'}  \left( \frac{\partial \mathcal{E}_{{N}'}}{\partial R_{N_1 \to N_2}} \right) }
+ \!\!
\sum_{{N}' \in \mathcal{A}}{
\sum_{\begin{subarray}{l} {N}'' \in \mathcal{A} \\ {N}' \ne {N}'' \end{subarray}}{
\frac{\partial \mathcal{E}_{{N}' \leftrightarrow {N}''}}{\partial R_{N_1 \to N_2}}
}} \nonumber \\
& + \lambda \times {T\!H}_{0}^{\left(MIN\right)} \left( \frac{\partial T_{DAG}}{\partial R_{N_1 \to N_2}} \right) ,
\label{eq:appDelLagDelR}
\end{align}
%
for $ N_1 = M $, $ N_2 \in {B\!H\!S} $ and $ N_1 \in {B\!H\!S} $, $ N_2 = M $; and:
%
\begin{equation}
\frac{\partial \mathcal{L}}{\partial \lambda} =
\left({T\!H}_{0}^{\left(MIN\right)} \times T_{DAG}\right) - 1  .
\label{eq:appDelLagDelLambda}
\end{equation}
%
Passing to consider the involved derivatives of the per-node computing energy, from the model relationships of Section \ref{ssec:ModelingComputingEnergy}, we obtain:
\begin{align}
\frac{ \partial \mathcal{E}_{N} }{ \partial f_{N'} }  =  &
\left( \frac{ \mathcal{P}_{CPU-N}^{\left(IDLE\right)} }{ {nc}_{N} } \right)
\times
\left( \frac{ \partial T_{DAG} }{ \partial f_{N'} } \right) \nonumber \\
& +
\delta \left( N - {N}' \right) \left( \gamma_{N} -1 \right) k_{N} \left(1 - r_{N}\right) C_{N} \left(f_{N}\right)^{\gamma_{N}-2},
\label{eq:appDelEDelf}
\end{align}
%
for $ N,{N}' \in \mathcal{A} $, and:
%
\begin{equation}
\frac{ \partial \mathcal{E}_{N} }{ \partial R_{N_1 \to N_2} }  =
\left( \frac{ \mathcal{P}_{CPU-N}^{\left(IDLE\right)} }{ {nc}_{N} } \right)
\times
\left( \frac{ \partial T_{DAG} }{ \partial R_{N_1 \to N_2} } \right)
\label{eq:appDelEDelR}
\end{equation}
%
for $ N_1 = M $, $ N_2 \in {B\!H\!S} $ and $ N_1\in {B\!H\!S} $, $ N_2 = M $, with the dummy position:
%
\begin{equation}
C_N
\overset{\mathrm{def}}{=}
\begin{cases}
\displaystyle \sum_{i=1}^{V} {\delta \left( x_i - N \right) s_i}   ,
& \scalebox{0.9}{\text{\textit{SEQ} service discipline}},\\
\underset{1 \leq i \leq V}{\max}
\left\{
\displaystyle \frac{\delta \left( x_i - N \right) s_i}
{\frac{\phi_{i}}{\sum_{j=1}^{V}{ \delta \left( x_j - N \right) s_j }}}
\right\} ,
& \scalebox{0.9}{\text{\textit{WPS} service discipline}} .
\end{cases}
\label{eq:appC_N}
\end{equation}

\begin{figure*}[!t]
\normalsize
\setcounter{tempcount}{\value{equation}}
\setcounter{equation}{\value{tempcount}}
\begin{flalign}
\frac{ \partial \mathcal{E}_{{N}' \leftrightarrow {N}''} }{\partial f_{N}}  =
A_{{N}' \leftrightarrow {N}''} \times 
\left( \frac{ \partial T_{DAG} }{ \partial f_{N} } \right) ,
\, \text{for }  N,{N}',{N}'' \in \mathcal{A} , \,   {N}' \ne {N}'' . & &
\label{eq:appDelE_NNDelf}
\end{flalign}
\begin{flalign}
\frac{ \partial \mathcal{E}_{{N}' \leftrightarrow {N}''} }{\partial R_{N_1 \to N_2}}  = & 
A_{{N}' \leftrightarrow {N}''} \times
\left( \frac{ \partial T_{DAG} }{ \partial R_{N_1 \to N_2} } \right) +
{Vl_{N_1 \to N_2} \times \left[ \delta \left(N_1-{N}'\right) \delta \left(N_2-{N}''\right) + \delta \left(N_1-{N}''\right) \delta \left(N_2-{N}'\right) \right]} &
\nonumber \\
& \! \times \! \!
{\left[
\theta_{N_{1}} \left( \xi_{\left({N}',{N}''\right)}^{\left(Tx\right)}-1 \right)
\! \Omega_{\left({N}',{N}''\right)}^{\left(Tx\right)}
\left( R_{N_1 \to N_2} \right)^{\xi_{\left({N}',{N}''\right)}^{\left(Tx\right)}-2}
\! +
\theta_{N_{2}} \left( \xi_{\left({N}',{N}''\right)}^{\left(Rx\right)}-1 \right)
\! \Omega_{\left({N}',{N}''\right)}^{\left(Rx\right)}
\left( R_{N_1 \to N_2} \right)^{\xi_{\left({N}',{N}''\right)}^{\left(Rx\right)}-2}
\right]} \! , & 
\nonumber \\
&\text{for } {N}',{N}'' \in \mathcal{A}, \,  {N}' \ne {N}''; \, \text{and } N_1 =M, N_2 \in {B\!H\!S} ; \, \text{and } N_1 \in {B\!H\!S}, N_2 =M .
\label{eq:appDelE_NNDelR}
\end{flalign}
\begin{equation}
\begingroup\makeatletter\def\f@size{8.8}\check@mathfonts
A_{{N}' \leftrightarrow {N}''} \overset{\mathrm{def}}{=}
\begin{cases}
\left(  \theta_F \mathcal{P}_{BHNET-F_{l}}^{\left(IDLE\right)}  +  \theta_C \mathcal{P}_{BHNET-C}^{\left(IDLE\right)}  \right)
\times u_{-1}
\left(
\displaystyle\sum_{i=1}^{V}{
\displaystyle\sum_{j=1}^{V}{
a_{ij} \left( \delta \left(x_i-{N}'\right) \delta \left(x_j-{N}''\right) + \delta \left(x_j-{N}'\right) \delta \left(x_i-{N}''\right) \right)
}}
\right)  \! ,
& 
\\
\left( \theta_M + \theta_F \right) \times  \mathcal{P}_{SRNET-M-F_{l}}^{\left(IDLE\right)}
\times u_{-1}
\left(
\displaystyle\sum_{i=1}^{V}{
\displaystyle\sum_{j=1}^{V}{
a_{ij} \left( \delta \left(x_i-M\right) \delta \left(x_j-F_l\right) + \delta \left(x_j-M\right) \delta \left(x_i-F_l\right) \right)
}}
\right)  \! ,
& 
\\
\left( \theta_M + \theta_C \right) \times  \mathcal{P}_{LRNET}^{\left(IDLE\right)}
\times u_{-1}
\left(
\displaystyle\sum_{i=1}^{V}{
\displaystyle\sum_{j=1}^{V}{
a_{ij} \left( \delta \left(x_i-M\right) \delta \left(x_j-C\right) + \delta \left(x_j-M\right) \delta \left(x_i-C\right) \right)
}}
\right)  \! .
& 
\end{cases}
\label{eq:appA_NN}
\endgroup
\end{equation}
\setcounter{equation}{\value{tempcount}}
\addtocounter{equation}{3}
\hrulefill
\vspace*{4pt}
\end{figure*}

Furthermore, by leveraging the defining relationships of Section \ref{ssec:ModelingEnergyInternode}, we arrive at the formulas in \eqref{eq:appDelE_NNDelf} and \eqref{eq:appDelE_NNDelR} (at the top of the next page) for the evaluation of the gradients of the per-connection network energy, with the dummy position in \eqref{eq:appA_NN}.

Finally, from the general expression of $ T_{DAG} $ in \eqref{eq:T_DAG}, the related derivatives read as follows:
\begin{align}
\frac{ \partial T_{DAG} }{ \partial R_{N_1 \to N_2} }    &  =
\left( \frac{ \partial T_{DAG} }{ \partial T_{N_{2}}^{\left(EXE\right)} } \right)
\times
\left( \frac{ \partial T_{N_{2}}^{\left(EXE\right)} }{ \partial R_{N_1 \to N_2} } \right) \nonumber \\
& \equiv 
- \left( \frac{ \partial \mathcal{X} }{ \partial T_{N_{2}}^{\left(EXE\right)} } \right)
\times
\left( \frac{ Vl_{N_1 \to N_2} }{ \left( R_{N_1 \to N_2} \right)^{2}  } \right),
\label{eq:appDelT_DAGDelR}
\end{align}
%
for $ N_1 = M $, $ N_2  \in {B\!H\!S} $, and $ N_1  \in {B\!H\!S} $, $ N_2 = M $; and:
%
\begin{align}
\frac{ \partial T_{DAG} }{ \partial f_N }    &  =
\left( \frac{ \partial T_{DAG} }{ \partial T_{N}^{\left(EXE\right)} } \right)
\times
\left( \frac{ \partial T_{N}^{\left(EXE\right)} }{ \partial f_N } \right) \nonumber \\
& \equiv 
- \left( \frac{ \partial \mathcal{X} }{ \partial T_{N}^{\left(EXE\right)} } \right)
\times
\left( \frac{ C_N }{ n_N \left( f_N \right)^{2}  } \right), \quad N \in \mathcal{A} .
\label{eq:appDelT_DAGDelf}
\end{align}

In the above expressions, we recall that: (i) $ T_{N_2}^{\left(EXE\right)} $ and $ T_{N}^{\left(EXE\right)} $ are the total execution times at nodes $ N_2 $  and $ N $, respectively (see Section \ref{ssec:executiontimes}); (ii) $ Vl_{N_1 \to N_2} $ is the total volume of data transported from node $ N_1 $ to node $ N_2 $ (see Section \ref{ssec:ModelingComputingEnergy}); (iii) $ C_N $ is the dummy constant in \eqref{eq:appC_N}; and, (iv) $ \mathcal{X} \left(.\right) $ is the application-depending function in \eqref{eq:T_DAG} that is adopted for the formal description of the considered $ T_{DAG} $.

In this regard, we observe that the $ \mathcal{X} $-function encloses the $ {\max} $ function when the adopted task scheduling discipline is the \textit{PTS} one (see Section \ref{ssec:Per-DAGexecutiontimes}). Hence, in order to carry out the derivatives in \eqref{eq:appDelT_DAGDelR} and \eqref{eq:appDelT_DAGDelf}, we pursue the (quite usual) approach to approximate the $ {\max} $ function with the following upper bound \cite{bazaraa2017}:
\begin{equation}
\underset{1 \leq l \leq V}{\max} \left\{b_l\right\}  \leq
\left( \sum_{l=1}^{V} { \left(b_l\right)^{r} } \right) ^{1/r} ,  \quad r \geq 1 .
\label{eq:appMaxb_lLessSum}
\end{equation}

This approach is, indeed, supported by the following three considerations. First, at any given $ r $, the above upper bound admits continuous derivatives with respect to its arguments. Second, since the following limit expression holds:
\begin{equation}
\underset{1 \leq l \leq V}{\max} \left\{b_l\right\}  =
\lim\limits_{r \to + \infty}
{\left( \sum_{l=1}^{V} { \left(b_l\right)^{r} } \right) ^{1/r}} , 
\label{eq:appMaxb_lEqualLim}
\end{equation}
the approximation error incurred by using the upper bound in \eqref{eq:appMaxb_lLessSum} in place of the actual $ \max $ function vanishes for large values of the $ r $ exponent. Third, at least in the carried out numerical tests, values of $ r $ of the order of $20$ -- $25$ suffice, in order to limit the incurred approximation errors within $ 1 \% $.

\balance


\end{document}